\documentclass{article}

\usepackage{amsmath}
\usepackage{amssymb}
\usepackage{authblk}
\usepackage{hyperref}
\usepackage{graphicx}
\usepackage{framed}
\usepackage{setspace}
\usepackage{fullpage}

\title{General Relativistic Hydrodynamics in Discrete Spacetime: Perfect Fluid Accretion onto Static and Spinning Black Holes}
\author[1]{Jonathan Gorard}
\affil[1]{Princeton University\protect\\
Princeton, NJ, United States\footnote{\href{mailto:gorard@princeton.edu}{gorard@princeton.edu}}\footnote{Current affiliation. Research and development work was performed at, and funded by, the Wolfram Institute.}}

\begin{document}

\maketitle

\begin{abstract}
We study the problem of a spherically-symmetric distribution of a perfect relativistic fluid accreting onto a (potentially spinning) black hole within a fully discrete spacetime setting. This problem has previously been studied extensively in the context of continuum spacetimes, beginning with the purely analytic work of Bondi in the spherically-symmetric Newtonian case, Michel in the spherically-symmetric general relativistic case, and Petrich, Shapiro and Teukolsky in the axially-symmetric general relativistic case relevant for spinning black holes. However, the purpose of the present work is to determine the effect of discretization of the underlying spacetime upon the mass/energy and momentum accretion rates, the overall morphology and characteristics of the accretion flow, and the drag force exerted on the black hole in the case of non-zero spin. In order to achieve this, we first develop a novel formulation of the equations of general relativistic hydrodynamics that is more directly amenable to rigorous analysis within a discrete spacetime setting, and we then proceed to implement this formulation into the \textsc{Gravitas} computational general relativity framework. Through a combination of mathematical analysis and explicit numerical simulation in \textsc{Gravitas}, we discover that the mass/energy and momentum accretion rates both decrease monotonically as functions of the underlying spacetime discretization scale, with this effect becoming more pronounced for higher values of the black hole spin parameter, higher fluid temperatures, and stiffer equation of state parameters. We also find that the exerted drag force is highly sensitive to the value of the underlying discretization scale in the case of spinning black hole spacetimes, with certain instabilities becoming significantly more pronounced at certain critical values of the discretization parameter. We discuss some potentially observable consequences of these results, as well as some directions for future theoretical investigation.
\end{abstract}

\section{Introduction}

The accretion of an idealized fluid onto a compact object (e.g. a neutron star or a black hole) remains one of the most widely-studied problems in astrophysics and cosmology, as it can be used as a minimal mathematical or numerical model for such a wide variety of phenomena, including the formation and growth of supermassive black holes at the centers of galaxies\cite{salpeter}\cite{volonteri}, the formation and growth of primordial black holes during the early universe\cite{zeldovich}\cite{loraclavijo}, and the dynamics of pulsars, active galactic nuclei and other high-energy astrophysical phenomena\cite{perna}\cite{russell}. Indeed, via modern observational techniques such as reverberation mapping\cite{uttley} and measurements of quasi-periodic oscillations\cite{miyamoto}\cite{vanderklis}, the dynamics of the accretion region very close to a compact object (which are often reflected in the fast variability in the spectrum of the region's X-ray emissions) can be used to provide a powerful and high-precision testbed for general relativity itself, for instance allowing one to test mathematical proposals such as the no-hair theorem experimentally, by determining the degree to which the exterior geometry surrounding a spinning black hole (or other compact object) appears to be well-described by the Kerr metric. This, in turn, presents the exciting possibility that deviations from the predictions of classical general relativity, for instance due to modifications in the microscopic structure of spacetime at or below the Planck scale arising from certain quantum gravity models, may become experimentally verifiable (or falsifiable) via astrophysical observations of such high-energy accretion phenomena in the near future. It would therefore be both useful and instructive to determine a robust and generic set of predictions regarding the effects of spacetime discreteness upon certain relevant accretion parameters, including the accretion rates of mass/energy and momentum onto the compact object; the lift and drag forces exerted upon the compact object (assuming that it possess a non-zero angular momentum value, and/or that the accretion is non-radial) due to the accretion flow; and the morphology, characteristics and dynamics of the accretion flow itself. The purpose of the present article is to commence the lengthy process of deriving such a set of predictions.

One of the prototypical idealized accretion cases conventionally studied is that of a compact object moving at a constant velocity through an ideal gas of uniform density (or, equivalently, an ideal gas with a uniform density and flow velocity accreting onto the compact object), commonly known as Bondi-Hoyle-Lyttleton accretion\cite{bondi}\cite{bondi2}\cite{hoyle}. The standard interpolation formula for the rate of mass accretion in the Bondi-Hoyle-Lyttleton model is, in turn, derived from two important limiting cases: the case where the flow velocity is zero (known as Bondi accretion\cite{bondi2}), and the case where the flow velocity is supersonic (known as Hoyle-Lyttleton accretion\cite{hoyle}). In this article, we shall focus solely upon the case of (radial) Bondi accretion, and we leave the extension of these techniques to the supersonic Hoyle-Lyttleton case, and to the generalized Bondi-Hoyle-Lyttleton case, as an open research problem, ripe for future investigation. Bondi's original analysis\cite{bondi2} considered the case of a spherically-symmetric distribution of ideal gas of initially uniform density (assumed to be of infinite extent), accreting onto a central point mass in pure Newtonian gravity. Michel\cite{michel} later extended Bondi's analytic solution for spherical accretion to the general relativistic case of a static, uncharged, non-rotating black hole (as described by the Schwarzschild metric) as the central compact object, although the polytropic form of the ideal gas equation of state used within Michel's analysis was previously shown by Taub\cite{taub} to be physically reasonable only in the strictly non-relativistic and strictly ultra-relativistic limits (with dimensionless gas temperature much less than unity, and much greater than unity, respectively), and not in the intermediate relativistic case (with dimensionless gas temperature approximately equal to unity). Rather surprisingly, Petrich, Shapiro and Teukolsky\cite{petrich} were even able to extend this analysis beyond the spherically-symmetric spacetimes considered thus far to the axially-symmetric spacetime case, and hence to derive an analytic solution for the accretion of a stiff, ultra-relativistic fluid onto an uncharged but spinning black hole (as described by the Kerr metric) as the central compact object. Although the stiff, ultra-relativistic equation of state used therein is not expected to be physical (since it requires a perfect relativistic fluid whose local sound speed is equal to the speed of light), the exact solution of Petrich, Shapiro and Teukolsky nevertheless provides a useful benchmark for the testing of general relativistic hydrodynamics codes. Font and Ib\'a\~nez\cite{font}\cite{font2}, and later Font, Ib\'a\~nez and Papadopoulos\cite{font3} later performed a detailed and systematic analysis of the (non-radial) case of Bondi-Hoyle-Lyttleton accretion of perfect relativistic fluids, obeying more general forms of the ideal gas equation of state, onto both static and spinning black holes by means of numerical simulations using the so-called ${3 + 1}$ ``Valencia'' formulation of the equations of general relativistic hydrodynamics in hyperbolic conservation law form, due originally to Banyuls, Font, Ib\'a\~nez, Mart\'i and Miralles\cite{banyuls}.

Some of the key insights yielded by this combination of analytical and numerical work included: the discovery of a systematic reduction in the mass accretion rate as a function of the black hole spin (an effect which becomes more significant for higher gas temperatures and higher values of the adiabatic exponent)\cite{font3}; the discovery of a drag force exerted on the black hole, either due to the presence of a downstream region of high fluid density caused by non-radial fluid motion in the case of Bondi-Hoyle-Lyttleton accretion, or redistribution of high fluid pressure regions caused by non-zero black hole spin, or both (which, in turn, result in increased gravitational forces exerted on the black hole by the fluid), with an absence of the ``flip-flop'' fluid instabilities typically seen in purely Newtonian accretion simulations\cite{font}\cite{font2}; and the discovery of a lift force, analogous to the Magnus effect in Newtonian fluid dynamics, exerted on spinning black holes by non-radially accreting fluids due to the asymmetry of the fluid pressure redistribution (with more pressure being redistributed onto the side of the black hole that is counter-rotating with the fluid)\cite{font3}. The primary objective of the present work is to begin the process of determining what kind of effect an underlying discretization of the background spacetime is expected to have on these types of black hole accretion phenomena, as well as on other related astrophysical processes. Since discreteness of the fundamental structure of spacetime is a generic feature of many proposed models of quantum gravity, including casual set theory\cite{bombelli}\cite{bombelli2}\cite{sorkin}\cite{gorard}, causal dynamical triangulations\cite{loll}\cite{ambjorn}, loop quantum gravity\cite{rovelli}\cite{rovelli2}\cite{rovelli3}, and the Wolfram model\cite{gorard2}\cite{gorard3}\cite{gorard4}\cite{gorard5}\cite{gorard6}, it is hoped that such an investigation will eventually enable the observational investigation of certain classes of quantum gravity theories by means of astrophysical probes of near-black hole accretion regions. To this end, we make use of the \textsc{Gravitas} computational general relativity framework\cite{gorard7}\cite{gorard8}, which allows for the configuration, execution, visualization and analysis of complex numerical relativity simulations in both discrete and continuous spacetime settings, by combining a powerful tensor calculus and differential geometry framework on the analytical side, with a sophisticated hypergraph-based adaptive refinement system\cite{gorard9}\cite{gorard10} on the numerical side. Most general relativistic simulations of black hole accretion consider a perfect relativistic fluid evolving on top of a \textit{fixed}, time-independent spacetime metric (typically representing either a Schwarzschild geometry or a Kerr geometry), and thereby neglect any gravitational effects of the fluid density on the black hole itself. Since the mass densities of the fluids in question are usually much smaller than the mass of the central black hole, this is often not an unreasonable simplification to make (this is conventionally referred to as the ``test-fluid'' assumption within the relativistic hydrodynamics literature\cite{font4}). However, since many of the effects in which we are interested for the purposes of this article (such as drag forces exerted by a fluid upon a spinning black hole) depend crucially upon the two-way gravitational interaction between the black hole and the fluid, we do not make this assumption here. Instead, we use \textsc{Gravitas} to configure and run fully general relativistic two-way coupled simulations, evolving the fluid variables and the metric tensor together in parallel.

We begin in Section \ref{sec:Section2} with a brief overview of the purely hyperbolic ${3 + 1}$ ``Valencia'' formalism for general relativistic hydrodynamics of Banyuls, Font, Ib\'a\~nez, Mart\'i and Miralles\cite{banyuls}, together with a description of how a modified version of the formalism can be derived that is specifically adapted for numerical relativistic hydrodynamics in discrete spacetimes, by means of the discrete spacetime ADM formalism already implemented within the \textsc{Gravitas} framework\cite{gorard8}. The final result of this analysis will be the derivation of a complete and fully-coupled system of purely hyperbolic equations for the evolution of the components of the discrete spatial metric tensor and the discrete spacetime fluid variables jointly, together with a set of purely elliptic constraint equations for the discrete spacetime gauge, which can then be implemented directly into \textsc{Gravitas}. We proceed in Section \ref{sec:Section3} to present a weak, integral form of these equations that is amenable to direct numerical solution via finite-volume methods, and we validate the resulting numerical implementation against a standard \textit{special} relativistic hydrodynamics shock tube problem (namely the mildly-relativistic blast wave problem of Donat, Font, Iba\'e\~nez and Marquina\cite{donat}). Particular attention is paid to the validation of the implementation of the conservative-to-primitive variable reconstruction algorithm, which is generally a non-trivial operation in relativistic hydrodynamics and for which we follow the approach of Eulderink and Mellema\cite{eulderink} in deriving a one-dimensional iterative Newton-Raphson solver, which works generically for any ideal gas equation of state. Finally, in Section \ref{sec:Section4}, we show the numerical results of our \textit{general} relativistic hydrodynamics simulations, beginning with a simulation of radial (Bondi-type) accretion onto a static/Schwarzschild black hole, before proceeding to radial (Bondi-type) accretion onto spinning/Kerr black holes, with a variety of spin values, ranging from modest to near-extremal. The broad qualitative features (e.g. the shape of the density profile for the accretion region in the Schwarzschild case, or the splitting of the accretion region into several distinct ``arms'' in the rapidly-spinning Kerr case, etc.) of these simulations appear similar to those obtained from analogous general relativistic hydrodynamics simulations performed in continuous spacetime geometries, although a more rigorous quantitative analysis reveals certain notable discrepancies. In particular, we find that the rates of mass/energy and momentum accretion onto the black hole both appear to be monotonically-decreasing functions of the discretization scale of the underlying spacetime, with increased black hole spin values, higher fluid temperatures and larger values of the adiabatic exponent (i.e. stiffer equations of state) accentuating and amplifying this discretization effect. Moreover, we discover that the drag force exerted on the black hole exhibits a sensitive dependence upon the underlying discretization scale, with certain critical values of the discretization scale resulting in apparent instabilities in the spacetime structure, observable within the feedback effect of the fluid density onto the black hole geometry. All simulation results presented within this (and the previous) section are presented in both ``horizon-adapted'' and ``non-horizon-adapted'' coordinate systems, as proposed by Font, Ib\'a\~nez and Papadopoulos\cite{font5}, so as to eliminate the possibility that any of these effects might simply be a byproduct of unphysical fluid behavior resulting from certain numerical divergences near the black hole horizon. We conclude in Section \ref{sec:Section5} with a brief discussion of potential astrophysical implications of these results, as well as directions for future research and investigation.

Note that all of the \textsc{Gravitas} functionality necessary to reproduce the results presented within this article can be found in the \href{https://github.com/JonathanGorard/Gravitas/}{\textsc{Gravitas} GitHub repository}, with extensive documentation available within both the \textit{Wolfram Function Repository} (e.g. \href{https://resources.wolframcloud.com/FunctionRepository/resources/ADMDecomposition/}{ADMDecomposition} and \href{https://resources.wolframcloud.com/FunctionRepository/resources/StressEnergyTensor/}{StressEnergyTensor}) and within the two previous articles \cite{gorard7} and \cite{gorard8}. This article follows all of the same notational and terminological conventions as these two previous articles; in particular, we assume geometric units with ${c = G = \hbar = 1}$, we employ a metric signature of ${\left( -, +, +, + \right)}$ in all relevant cases, and the Einstein summation convention is assumed throughout (such that all repeated tensor indices are implicitly summed over).

\section{General Relativistic Hydrodynamics in Discrete Spacetime}
\label{sec:Section2}

In order to derive a form of the equations of general relativistic hydrodynamics that is suitable for analysis within a discrete spacetime setting, we begin by considering the ${3 + 1}$ ``Valencia'' formulation of the curved spacetime hydrodynamics equations in conservation law form due to Banyuls, Font, Ib\'a\~nez, Mart\'i and Miralles\cite{banyuls}, which exploits the fundamentally hyperbolic character of the spacetime continuity equations. The equations of general relativistic hydrodynamics represent a mathematical encoding of two distinct physical laws, namely the law of conservation of energy-momentum, and the law of conservation of baryon number. Assuming a spacetime given by a smooth $n$-dimensional Lorentzian manifold ${\left( \mathcal{M}, g \right)}$, the law of conservation of energy-momentum can be represented as a statement that the covariant divergence of the rank-2 stress-energy tensor ${T^{\mu \nu}}$ vanishes identically:

\begin{equation}
\nabla_{\nu} T^{\mu \nu} = \frac{\partial}{\partial x^{\nu}} \left( T^{\mu \nu} \right) + \Gamma_{\nu \sigma}^{\mu} T^{\sigma \nu} + \Gamma_{\nu \sigma}^{\nu} T^{\mu \sigma} = 0,
\end{equation}
while the law of conservation of baryon number can be represented as a statement that the covariant divergence of the rank-1 (rest) mass current vector ${J^{\mu}}$ also vanishes identically:

\begin{equation}
\nabla_{\mu} J^{\mu} = \frac{\partial}{\partial x^{\mu}} \left( J^{\mu} \right) + \Gamma_{\mu \sigma}^{\mu} J^{\sigma} = 0.
\end{equation}
In the above, the spacetime covariant derivative ${\nabla_{\mu}}$ is represented in terms of the coefficients of the Levi-Civita connection ${\nabla}$ on the manifold ${\left( \mathcal{M}, g \right)}$, namely the spacetime Christoffel symbols ${\Gamma_{\mu \nu}^{\rho}}$, themselves represented in terms of partial derivatives of the spacetime metric tensor ${g_{\mu \nu}}$:

\begin{equation}
\Gamma_{\mu \nu}^{\rho} = \frac{1}{2} g^{\rho \sigma} \left( \frac{\partial}{\partial x^{\mu}} \left( g_{\sigma \nu} \right) + \frac{\partial}{\partial x^{\nu}} \left( g_{\mu \sigma} \right) - \frac{\partial}{\partial x^{\sigma}} \left( g_{\mu \nu} \right) \right).
\end{equation}
For the specific case of a \textit{perfect} relativistic fluid in equilibrium, obtained by neglecting all considerations of heat conduction, fluid viscosity and shear stress, the stress-energy tensor ${T^{\mu \nu}}$ and (rest) mass current vector ${J^{\mu}}$ take the forms:

\begin{equation}
T^{\mu \nu}= \rho h u^{\mu} u^{\nu} + P g^{\mu \nu}, \qquad \text{ and } \qquad J^{\mu} = \rho u^{\mu},
\end{equation}
respectively, where ${\rho}$ denotes the (rest) mass density of the fluid, $P$ denotes its hydrostatic pressure, ${u^{\mu}}$ denotes its spacetime velocity, and $h$ denotes its specific relativistic enthalpy:

\begin{equation}
h = 1 + \varepsilon \left( \rho, P \right) + \frac{P}{\rho},
\end{equation}
where ${\varepsilon \left( \rho, P \right)}$ represents the specific internal energy of the fluid. The product ${\rho h}$ of the (rest) mass density ${\rho}$ and the specific relativistic enthalpy $h$ constitutes the total mass-energy density of the fluid. In all of the above, the components ${g^{\mu \nu}}$ are components of the inverse metric tensor ${g^{\mu \nu} = \left( g_{\mu \nu} \right)^{-1}}$, and the tensor indices ${\mu, \nu, \rho, \sigma}$ range across all spacetime coordinate directions ${\left\lbrace 0, \dots, n - 1 \right\rbrace}$ (with ${\left\lbrace x^{\mu} \right\rbrace}$ being a local spacetime coordinate basis), in contrast to the ${3 + 1}$ decomposition formalism discussed below. The resulting system of equations may then be closed by defining an appropriate equation of state, allowing one either to calculate the specific internal energy as a function of the fluid density and hydrostatic pressure ${\varepsilon \left( \rho, P \right)}$, or, equivalently, to calculate the hydrostatic pressure as a function of the fluid density and specific internal energy ${P \left( \rho, \varepsilon \right)}$. The equation of state thus allows one to compute the local sound speed ${c_s}$ of the fluid as:

\begin{equation}
c_s = \frac{1}{\sqrt{h}} \sqrt{\left. \left( \frac{\partial P}{\partial \rho} \right) \right\rvert_{\varepsilon} + \left( \frac{P}{\rho^2} \right) \left. \left( \frac{\partial P}{\partial \varepsilon} \right) \right\rvert_{P}},
\end{equation}
where ${\left. \left( \frac{\partial P}{\partial \rho} \right) \right\rvert_{\varepsilon}}$ and ${\left. \left( \frac{\partial P}{\partial \varepsilon} \right) \right\rvert_{\rho}}$ denote partial derivatives assuming fixed internal energy ${\varepsilon}$ and fixed fluid density ${\rho}$, respectively.

We now proceed to perform a ``${3 + 1}$ decomposition'' (or ``foliation'') of our $n$-dimensional spacetime ${\left( \mathcal{M}, g \right)}$ into a time-ordered sequence of ${\left( n - 1 \right)}$-dimensional spacelike hypersurfaces of Riemannian signature, each with an induced/spatial metric tensor ${\gamma_{\mu \nu}}$, by means of the ADM formalism due originally to Arnowitt, Deser and Misner\cite{arnowitt}\cite{arnowitt2}, and later adapted by York\cite{york} into the form used for the purposes of this article. Within such a decomposition, the overall spacetime line element (or first fundamental form) ${d s^2}$, which normally takes the general form:

\begin{equation}
d s^2 = g_{\mu \nu} d x^{\mu} d x^{\nu},
\end{equation}
with ${\mu, \nu}$ ranging across all spacetime coordinate indices ${\left\lbrace 0, \dots, n - 1 \right\rbrace}$ (and with ${\left\lbrace x^{\mu} \right\rbrace}$ taken to represent a local \textit{spacetime} coordinate basis), can now be written instead as:

\begin{multline}
d s^2 = - \alpha^2 d t^2 + \gamma_{\mu \nu} \left( d x^{\mu} + \beta^{\mu} d t \right) \left( d x^{\nu} + \beta^{\nu} d t \right)\\
= \left( - \alpha^2 + \gamma_{\mu \sigma} \beta^{\sigma} \beta^{\mu} \right) d t^2 + 2 \gamma_{\mu \sigma} \beta^{\sigma} d t d x^{\mu} + \gamma_{\mu \nu} d x^{\mu} d x^{\nu}\\
= \left( - \alpha^2 + \beta_{\mu} \beta^{\mu} \right) d t^2 + 2 \beta_{\mu} d t d x^{\mu} + \gamma_{\mu \nu} d x^{\mu} d x^{\nu},
\end{multline}
with ${\mu, \nu, \sigma}$ ranging across the spatial coordinate indices ${\left\lbrace 0, \dots, n - 2 \right\rbrace}$ only (and with ${\left\lbrace x^{\mu} \right\rbrace}$ now taken to represent a local \textit{spatial} coordinate basis on each hypersurface), and where $t$ designates a distinguished ``time'' coordinate. In the above, the scalar field ${\alpha}$ (known as the \textit{lapse function}) and the ${\left( n - 1 \right)}$-dimensional vector field ${\beta^{\mu}}$ (known as the \textit{shift vector}) correspond to the Lagrange multipliers of the ADM formalism, representing the proper time distance ${d \tau}$ between corresponding points on the neighboring spacelike hypersurfaces labeled by coordinate time values ${t = t_0}$ and ${t = t_0 + d t}$:

\begin{equation}
d \tau \left( t_0, t_0 + d t \right) = \alpha d t,
\end{equation}
as measured in the direction ${\mathbf{n}}$ normal to the ${t = t_0}$ hypersurface, and the relabeling of the spatial coordinate basis ${x^{\mu} \left( t_0 \right)}$ as one moves from the ${t = t_0}$ hypersurface to the neighboring ${t = t_0 + d t}$ hypersurface:

\begin{equation}
x^{\mu} \left( t_0 + d t \right) = x^{\mu} \left( t_0 \right) - \beta^{\mu} d t,
\end{equation}
respectively. The unit vector ${\mathbf{n}}$ that is normal to each spacelike hypersurface is given by the spacetime contravariant derivative ${{}^{\left( 4 \right)} \nabla^{\mu}}$ of the distinguished time coordinate $t$:

\begin{equation}
n^{\mu} = - \alpha {}^{\left( 4 \right)} \nabla^{\mu} t = - \alpha g^{\mu \sigma} {}^{\left( 4 \right)} \nabla_{\sigma} t = - \alpha g^{\mu \sigma} \frac{\partial}{\partial x^{\sigma}} \left( t \right),
\end{equation}
while the ``time vector'' ${\mathbf{t}}$ that determines how points on the ${t = t_0}$ hypersurface map to corresponding points on the ${t = t_0 + d t}$ hypersurface is given by:

\begin{equation}
t^{\mu} = \alpha n^{\left( \mu + 1 \right)} + \beta^{\mu} = - \alpha^2 g^{\mu \sigma} \frac{\partial}{\partial x^{\sigma}} \left( t \right) + \beta^{\mu},
\end{equation}
with ${\mu, \sigma}$ ranging across the spatial coordinate indices ${\left\lbrace 0, \dots, n - 2 \right\rbrace}$ only, and where we have introduced the notational convention of using a bracketed ``4'' to designate spacetime quantities (such that ${{}^{\left( 4 \right)} \nabla_{\mu}}$ denotes the spacetime covariant derivative, as defined above in terms of the spacetime Christoffel symbols ${\Gamma_{\mu \nu}^{\rho}}$, which are henceforth denoted ${{}^{\left( 4 \right)} \Gamma_{\mu \nu}^{\rho}}$), in order to distinguish them from the corresponding spatial quantities, for which we use a bracketed ``3'' instead. Interpreting the ADM formalism as a Hamiltonian formulation of the Einstein field equations, we see that the components ${\gamma_{\mu \nu}}$ of the spatial metric tensor represent the dynamical variables of the theory, with the components ${K_{\mu \nu}}$ of the \textit{extrinsic curvature tensor} (or second fundamental form) representing the corresponding conjugate momenta. These components can be obtained by computing the Lie derivative ${\mathcal{L}}$ of the spatial metric tensor ${\gamma_{\mu \nu}}$ in the direction of the normal vector ${\mathbf{n}}$\cite{alcubierre}:

\begin{equation}
K_{\mu \nu} = - \frac{1}{2} \mathcal{L}_{\mathbf{n}} \gamma_{\mu \nu},
\end{equation}
which expands out to give, explicitly:

\begin{multline}
K_{\mu \nu} = \frac{1}{2 \alpha} \left( {}^{\left( 3 \right)} \nabla_{\nu} \beta_{\mu} + {}^{\left( 3 \right)} \nabla_{\mu} \beta_{\nu} - \frac{\partial}{\partial t} \left( \gamma_{\mu \nu} \right) \right)\\
= \frac{1}{2 \alpha} \left( \frac{\partial}{\partial x^{\nu}} \left( \beta_{\mu} \right) - {}^{\left( 3 \right)} \Gamma_{\nu \mu}^{\sigma} \beta_{\sigma} + \frac{\partial}{\partial x^{\mu}} \left( \beta_{\nu} \right) - {}^{\left( 3 \right)} \Gamma_{\mu \nu}^{\sigma} \beta_{\sigma} - \frac{\partial}{\partial t} \left( \gamma_{\mu \nu} \right) \right),
\end{multline}
where the spatial covariant derivative ${{}^{\left( 3 \right)} \nabla_{\mu}}$ is represented in terms of the coefficients of the induced Levi-Civita connection ${{}^{\left( 3 \right)} \nabla}$ on each spacelike hypersurface, namely the spatial Christoffel symbols ${{}^{\left( 3 \right)} \Gamma_{\mu \nu}^{\rho}}$, themselves represented in terms of partial derivatives of the spatial metric tensor ${\gamma_{\mu \nu}}$:

\begin{equation}
{}^{\left( 3 \right)} \Gamma_{\mu \nu}^{\rho} = \frac{1}{2} \gamma^{\rho \sigma} \left( \frac{\partial}{\partial x^{\mu}} \left( \gamma_{\sigma \nu} \right) + \frac{\partial}{\partial x^{\nu}} \left( \gamma_{\mu \sigma} \right) - \frac{\partial}{\partial x^{\sigma}} \left( \gamma_{\mu \nu} \right) \right).
\end{equation}
Note also that the indices of the shift vector ${\boldsymbol\beta}$ are raised and lowered using the spatial metric tensor ${\gamma_{\mu \nu}}$, and so, in particular, the covector form ${\beta_{\mu}}$ used above is given by:

\begin{equation}
\beta_{\mu} = \gamma_{\mu \sigma} \beta^{\sigma}.
\end{equation}
In all of the above, ${\mu, \nu, \sigma}$ range across the spatial coordinate indices ${\left\lbrace 0, \dots, n - 2 \right\rbrace}$ only.

Just as one can perform a ${3 + 1}$/ADM decomposition of the overall spacetime metric ${g_{\mu \nu}}$ that appears on the left-hand-side of the Einstein field equations, one can equivalently perform a ${3 + 1}$/ADM decomposition of the overall spacetime stress-energy tensor ${T^{\mu \nu}}$ that appears on the right-hand-side of the Einstein field equations\cite{gourgoulhon}. By projecting the continuity equations for the stress-energy tensor ${T^{\mu \nu}}$:

\begin{equation}
{}^{\left( 4 \right)} \nabla_{\nu} T^{\mu \nu} = \frac{\partial}{\partial x^{\nu}} \left( T^{\mu \nu} \right) + {}^{\left( 4 \right)} \Gamma_{\nu \sigma}^{\mu} T^{\sigma \nu} + {}^{\left( 4 \right)} \Gamma_{\nu \sigma}^{\nu} T^{\mu \sigma} = 0,
\end{equation}
with ${\mu, \nu, \sigma}$ ranging across all spacetime coordinate indices ${\left\lbrace 0, \dots, n - 1 \right\rbrace}$, in the purely timelike direction, we obtain the energy conservation equation:

\begin{equation}
\frac{\partial}{\partial t} \left( E \right) - \mathcal{L}_{\boldsymbol\beta} E + \alpha \left( {}^{\left( 3 \right)} \nabla_{\mu} p^{\mu} - K E - K_{\mu \nu} S^{\mu \nu} \right) + 2 p^{\mu} {}^{\left( 3 \right)} \nabla_{\mu} \alpha = 0,
\end{equation}
where the Lie derivative term ${\mathcal{L}_{\boldsymbol\beta} E}$ expands to give:

\begin{multline}
\frac{\partial}{\partial t} \left( E \right) - \beta^{\mu} \frac{\partial}{\partial x^{\mu}} \left( E \right) + \alpha \left( {}^{\left( 3 \right)} \nabla_{\mu} p^{\mu} - K E - K_{\mu \nu} S^{\mu \nu} \right) + 2 p^{\mu} {}^{\left( 3 \right)} \nabla_{\mu} \alpha\\
= \frac{\partial}{\partial t} \left( E \right) - \beta^{\mu} \frac{\partial}{\partial x^{\mu}} \left( E \right) + \alpha \left( \frac{\partial}{\partial x^{\mu}} \left( p^{\mu} \right) + {}^{\left( 3 \right)} \Gamma_{\mu \sigma}^{\mu} p^{\sigma} - K E - K_{\mu \nu} S^{\mu \nu} \right) + 2 p^{\mu} \frac{\partial}{\partial x^{\mu}} \left( \alpha \right) = 0,
\end{multline}
with ${\mu, \nu, \sigma}$ ranging across spatial coordinate indices ${\left\lbrace 0, \dots, n - 2 \right\rbrace}$ only. On the other hand, projecting in the ${\left( n - 1 \right)}$ purely spacelike directions yields the momentum conservation equations:

\begin{equation}
\frac{\partial}{\partial t} \left( p_{\mu} \right) - \mathcal{L}_{\boldsymbol\beta} p_{\mu} + \alpha {}^{\left( 3 \right)} \nabla_{\nu} S_{\mu}^{\nu} + S_{\mu \nu} {}^{\left( 3 \right)} \nabla^{\nu} \alpha - \alpha K p_{\mu} + E {}^{\left( 3 \right)} \nabla_{\mu} \alpha = 0,
\end{equation}
where the Lie derivative term ${\mathcal{L}_{\boldsymbol\beta} p_{\mu}}$ expands, and the contravariant derivative operator ${{}^{\left( 3 \right)} \nabla^{\nu}}$ may be replaced with a corresponding covariant derivative operator ${{}^{\left( 3 \right)} \nabla_{\sigma}}$, to give:

\begin{multline}
\frac{\partial}{\partial t} \left( p_{\mu} \right) - \beta^{\sigma} \frac{\partial}{\partial x^{\sigma}} \left( p_{\mu} \right) - p_{\sigma} \frac{\partial}{\partial x^{\mu}} \left( \beta^{\sigma} \right) + \alpha {}^{\left( 3 \right)} \nabla_{\nu} S_{\mu}^{\nu} + S_{\mu \nu} \gamma^{\nu \sigma} {}^{\left( 3 \right)} \nabla_{\sigma} \alpha - \alpha K p_{\mu} + E {}^{\left( 3 \right)} \nabla_{\mu} \alpha\\
= \frac{\partial}{\partial t} \left( p_{\mu} \right) - \beta^{\sigma} \frac{\partial}{\partial x^{\sigma}} \left( p_{\mu} \right) - p_{\sigma} \frac{\partial}{\partial x^{\mu}} \left( \beta^{\sigma} \right) + \alpha \left( \frac{\partial}{\partial x^{\nu}} \left( S_{\mu}^{\nu} \right) + {}^{\left( 3 \right)} \Gamma_{\nu \sigma}^{\nu} S_{\mu}^{\sigma} - {}^{\left( 3 \right)} \Gamma_{\nu \mu}^{\sigma} S_{\sigma}^{\nu} \right)\\
+ S_{\mu \nu} \gamma^{\nu \sigma} \frac{\partial}{\partial x^{\sigma}} \left( \alpha \right) - \alpha K p_{\mu} + E \frac{\partial}{\partial x^{\mu}} \left( \alpha \right) = 0,
\end{multline}
with ${\mu, \nu, \sigma}$ again ranging across spatial coordinate indices ${\left\lbrace 0, \dots, n - 2 \right\rbrace}$ only. Note that, in the above, $E$, ${p_{\mu}}$ and ${S_{\mu \nu}}$ denote the energy density, the momentum density (in covector form) and the (Cauchy) stress tensor, respectively, of the stress-energy distribution described by ${T^{\mu \nu}}$, as perceived by an observer moving in the direction ${\mathbf{n}}$ normal to the spacelike hypersurfaces, which can be calculated via the componentwise projections:

\begin{equation}
E = T_{\mu \nu} n^{\mu} n^{\nu}, \qquad p_{\alpha} = - T_{\mu \nu} n^{\mu} \bot_{\alpha}^{\nu}, \qquad \text{ and } \qquad S_{\alpha \beta} = T_{\mu \nu} \bot_{\alpha}^{\mu} \bot_{\beta}^{\nu},
\end{equation}
with ${\alpha, \beta}$ ranging across spatial coordinate indices ${\left\lbrace 0, \dots, n - 2 \right\rbrace}$ only, and ${\mu, \nu}$ ranging across all spacetime coordinate indices ${\left\lbrace 0, \dots, n - 1 \right\rbrace}$, respectively. Here, ${\bot_{\mu}^{\nu}}$ are the components of the orthogonal projector (i.e. the projection operator in the normal direction ${\mathbf{n}}$):

\begin{equation}
\bot_{\mu}^{\nu} = \delta_{\left( \mu + 1 \right)}^{\nu} + n_{\left( \mu + 1 \right)} n^{\nu},
\end{equation}
where ${\delta_{\mu}^{\nu}}$ is the identity tensor/Kronecker delta function, and the momentum vector ${\mathbf{p}}$ and (Cauchy) stress tensor ${S_{\mu \nu}}$ are raised and lowered using the spatial metric tensor ${\gamma_{\mu \nu}}$, and so, in particular, for the quantities ${p^{\mu}}$ (in vector form), ${S_{\mu}^{\nu}}$ (in mixed-index form) and ${S^{\mu \nu}}$ (in contravariant form) appearing in the equations above, one has:

\begin{equation}
p^{\mu} = \gamma^{\mu \sigma} p_{\sigma}, \qquad S_{\mu}^{\nu} = \gamma_{\mu \sigma} S^{\sigma \nu} = \gamma_{\mu \sigma} \gamma^{\lambda \nu} S_{\lambda}^{\sigma} = \gamma^{\sigma \nu} S_{\mu \sigma}, \qquad S^{\mu \nu} = \gamma^{\mu \sigma} S_{\sigma}^{\nu} = \gamma^{\sigma \nu} S_{\sigma}^{\mu} = \gamma^{\mu \sigma} \gamma^{\lambda \nu} S_{\sigma \lambda},
\end{equation}
respectively, where ${\mu, \nu, \sigma, \lambda}$ range across spatial coordinate indices ${\left\lbrace 0, \dots, n - 2 \right\rbrace}$ only, and where ${\gamma^{\mu \nu}}$ are components of the inverse spatial metric tensor ${\gamma^{\mu \nu} = \left( \gamma_{\mu \nu} \right)^{-1}}$. Moreover, the indices of the stress-energy tensor ${T^{\mu \nu}}$ and the normal vector ${\mathbf{n}}$ are raised and lowered using the spacetime metric tensor ${g_{\mu \nu}}$, and so, in particular, for the covariant forms ${T_{\mu \nu}}$ and ${n_{\mu}}$ appearing above, one has:

\begin{equation}
T_{\mu \nu} = g_{\mu \sigma} g_{\lambda \nu} T^{\sigma \lambda} = g_{\sigma \nu} T_{\mu}^{\sigma} = g_{\mu \sigma} T_{\nu}^{\sigma}, \qquad n_{\mu} = g_{\mu \sigma} n^{\sigma},
\end{equation}
with ${\mu, \nu, \sigma, \lambda}$ ranging across all spacetime coordinate indices ${\left\lbrace 0, \dots, n - 1 \right\rbrace}$. We have also introduced the notation $K$ to indicate the trace of the extrinsic curvature tensor ${K_{\mu \nu}}$, i.e:

\begin{equation}
K = K_{\mu}^{\mu} = \gamma^{\mu \nu} K_{\mu \nu}.
\end{equation}
Upon comparing the decomposition of the stress-energy tensor ${T^{\mu \nu}}$ to the decomposition of the spacetime metric tensor ${g_{\mu \nu}}$, we see that the energy density $E$ plays the same as the lapse function ${\alpha}$, the momentum density covector ${p_{\mu}}$ plays the same role as the shift vector ${\beta^{\mu}}$, and the (Cauchy) stress tensor ${S_{\mu \nu}}$ plays the same role as the induced/spatial metric tensor ${\gamma_{\mu \nu}}$.

Returning now from considerations of the general ADM formalism to the specific case of general relativistic hydrodynamics, we proceed to consider the (spatial) fluid velocity ${\mathbf{v}}$, as perceived by an observer moving in the direction ${\mathbf{n}}$ normal to the spacelike hypersurfaces, namely:

\begin{equation}
v^{\mu} = \frac{u^{\left( \mu + 1 \right)}}{\alpha u^0} + \frac{\beta^{\mu}}{\alpha},
\end{equation}
where ${\alpha u^0}$ represents the Lorentz factor of the fluid:

\begin{equation}
\alpha u^0 = - u_{\left( \mu + 1 \right)} n^{\left( \mu + 1 \right)} = \frac{1}{\sqrt{1 - \gamma_{\mu \nu} v^{\mu} v^{\nu}}},
\end{equation}
with ${\mu, \nu}$ in all of the above ranging across spatial coordinate indices ${\left\lbrace 0, \dots, n - 2 \right\rbrace}$ only. We shall henceforth treat the fluid (rest) mass density ${\rho}$, the (spatial) fluid velocity components for a normal observer ${v^{\mu}}$, and the fluid pressure $P$, as the primitive variables of our forthcoming system of hyperbolic partial differential equations in conservation law form. For a perfect relativistic fluid, the energy conservation equation obtained from taking a timelike projection of the stress-energy continuity equations becomes:

\begin{multline}
\frac{1}{\sqrt{- \det \left( g_{\mu \nu} \right)}} \left[ \frac{\partial}{\partial t} \left( \sqrt{\det \left( \gamma_{\mu \nu} \right)} \left( \frac{\rho h}{1 - \gamma_{\mu \nu} v^{\mu} v^{\nu}} - P - \frac{\rho}{\sqrt{1 - \gamma_{\mu \nu} v^{\mu} v^{\nu}}} \right) \right) \right.\\
\left. + \frac{\partial}{\partial x^{\rho}} \left( \sqrt{- \det \left( g_{\mu \nu} \right)} \left( \left( \frac{\rho h}{1 - \gamma_{\mu \nu} v^{\mu} v^{\nu}} - P - \frac{\rho}{\sqrt{1 - \gamma_{\mu \nu} v^{\mu} v^{\nu}}} \right) \left( v^{\rho} - \frac{\beta^{\rho}}{\alpha} \right) + P v^{\rho} \right) \right) \right]\\
= \alpha \left( T^{\mu 0} \frac{\partial}{\partial x^{\mu}} \left( \log \left( \alpha \right) \right) - T^{\mu \nu} {}^{\left( 4 \right)} \Gamma_{\nu \mu}^{0} \right),
\end{multline}
with ${\mu, \nu, \rho}$ on the left-hand-side of the equation ranging across spatial coordinate indices ${\left\lbrace 0, \dots, n - 2 \right\rbrace}$ only, and ${\mu, \nu}$ on the right-hand-side of the equation ranging across all spacetime coordinate indices ${\left\lbrace 0, \dots, n - 1 \right\rbrace}$. We have introduced the notation ${\det \left( g_{\mu \nu} \right)}$ and ${\det \left( \gamma_{\mu \nu} \right)}$ in the above to represent determinants of the spacetime and spatial metric tensors ${g_{\mu \nu}}$ and ${\gamma_{\mu \nu}}$, respectively (regarded here as explicit matrices in covariant form); within this notation, the indices ${\mu}$ and ${\nu}$ should therefore be thought of as being purely ``structural''. Likewise, the momentum conservation equations obtained from taking ${\left( n - 1 \right)}$ spacelike projections of the stress-energy continuity equations become:

\begin{multline}
\frac{1}{\sqrt{- \det \left( g_{\mu \nu} \right)}} \left[ \frac{\partial}{\partial t} \left( \sqrt{\det \left( \gamma_{\mu \nu} \right)} \left( \frac{\rho h v_{\sigma}}{1 - \gamma_{\mu \nu} v^{\mu} v^{\nu}} \right) \right) \right.\\
\left. + \frac{\partial}{\partial x^{\rho}} \left( \sqrt{- \det \left( g_{\mu \nu} \right)} \left( \left( \frac{\rho h v_{\sigma}}{1 - \gamma_{\mu \nu} v^{\mu} v^{\nu}} \right) \left( v^{\rho} - \frac{\beta^{\rho}}{\alpha} \right) + P \delta_{\sigma}^{\rho} \right) \right) \right]\\
= T^{\mu \nu} \left( \frac{\partial}{\partial x^{\mu}} \left( g_{\nu \left( \sigma + 1 \right)} \right) - {}^{\left( 4 \right)} \Gamma_{\nu \mu}^{\lambda} g_{\lambda \left( \sigma + 1 \right)} \right),
\end{multline}
with ${\mu, \nu, \rho}$ on the left-hand-side of the equation again ranging across spatial coordinate indices ${\left\lbrace 0, \dots, n - 2 \right\rbrace}$ only, ${\mu, \nu, \lambda}$ on the right-hand-side of the equation ranging across all spacetime coordinate indices ${\left\lbrace 0, \dots, n - 1 \right\rbrace}$, and with ${\sigma}$ on both sides ranging across spatial coordinate indices ${\left\lbrace 0, \dots, n - 2 \right\rbrace}$ only. Finally, the baryon number continuity equation:

\begin{equation}
{}^{\left( 4 \right)} \nabla_{\mu} J^{\mu} = \frac{\partial}{\partial x^{\mu}} \left( J^{\mu} \right) + {}^{\left( 4 \right)} \Gamma_{\mu \sigma}^{\mu} J^{\sigma} = 0,
\end{equation}
yields:

\begin{multline}
\frac{1}{\sqrt{- \det \left( g_{\mu \nu} \right)}} \left[ \frac{\partial}{\partial t} \left( \sqrt{\det \left( \gamma_{\mu \nu} \right)} \left( \frac{\rho}{\sqrt{1 - \gamma_{\mu \nu} v^{\mu} v^{\nu}}} \right) \right) \right.\\
\left. + \frac{\partial}{\partial x^{\rho}} \left( \sqrt{- \det \left( g_{\mu \nu} \right)} \left( \left( \frac{\rho}{\sqrt{1 - \gamma_{\mu \nu} v^{\mu} v^{\nu}}} \right) \left( v^{\rho} - \frac{\beta^{\rho}}{\alpha} \right) \right) \right) \right] = 0,
\end{multline}
with ${\mu, \nu, \rho}$ ranging across spatial coordinate indices ${\left\lbrace 0, \dots, n - 2 \right\rbrace}$ only. In the above, the indices of the (spatial) fluid velocity vector ${\mathbf{v}}$ are raised and lowered using the spatial metric tensor ${\gamma_{\mu \nu}}$, and so, in particular, one has the covector form:

\begin{equation}
v_{\mu} = \gamma_{\mu \sigma} v^{\sigma},
\end{equation}
with ${\mu, \sigma}$ ranging across spatial coordinate indices ${\left\lbrace 0, \dots, n - 2 \right\rbrace}$ only. The conserved quantity appearing within the baryon number continuity equation represents the (rest) mass density $D$ of the fluid as measured by an observer moving in the normal direction ${\mathbf{n}}$:

\begin{equation}
D = \frac{\rho}{\sqrt{1 - \gamma_{\mu \nu} v^{\mu} v^{\nu}}} = - J_{\mu} n^{\mu},
\end{equation}
with ${\mu, \nu}$ on the left-hand-side ranging across spatial coordinate indices ${\left\lbrace 0, \dots, n - 2 \right\rbrace}$ only, and ${\mu}$ on the right-hand-side ranging across all spacetime coordinate indices ${\left\lbrace 0, \dots, n - 1 \right\rbrace}$; the conserved quantity appearing within the energy conservation equation is the difference between the energy density $E$ measured by a normal observer and the (rest) mass density $D$ measured by that same observer:

\begin{equation}
E - D = \frac{\rho h}{1 - \gamma_{\mu \nu} v^{\mu} v^{\nu}} - P - \frac{\rho}{\sqrt{1 - \gamma_{\mu \nu} v^{\mu} v^{\nu}}} = T_{\mu \nu} n^{\mu} n^{\nu} - J_{\mu} n^{\mu},
\end{equation}
with ${\mu, \nu}$ on the left-hand-side ranging across spatial coordinate indices ${\left\lbrace 0, \dots, n - 2 \right\rbrace}$ only, and ${\mu, \nu}$ on the right-hand-side ranging across all spacetime coordinate indices ${\left\lbrace 0, \dots, n - 1 \right\rbrace}$; while, finally, the conserved quantities appearing within the momentum conservation equations are simply the components of the momentum density ${p_{\mu}}$ (represented in covector form) measured by a normal observer:

\begin{equation}
p_{\sigma} = \frac{\rho h v_{\sigma}}{1 - \gamma_{\mu \nu} v^{\mu} v^{\nu}} = - T_{\mu \nu} n^{\mu} \bot_{\sigma}^{\nu},
\end{equation}
with ${\mu, \nu}$ on the left-hand-side ranging across spatial coordinate indices ${\left\lbrace 0, \dots, n - 2 \right\rbrace}$ only, ${\mu, \nu}$ on the right-hand-side ranging across across all spacetime coordinate indices ${\left\lbrace 0, \dots, n - 1 \right\rbrace}$, and with ${\sigma}$ on both sides ranging across spatial coordinate indices ${\left\lbrace 0, \dots, n - 2 \right\rbrace}$ only. Note that the source terms appearing on the right-hand-sides of the energy and momentum conservation equations do not contain any derivatives of the primitive variables ${\rho}$, ${v^{\mu}}$ and $P$, and therefore the hyperbolic character of the overall system of equations is preserved.

However, observe also that the source terms for the energy and momentum conservation equations currently depend upon the overall spacetime metric tensor ${g_{\mu \nu}}$, its partial derivatives and its corresponding Christoffel symbols ${{}^{\left( 4 \right)} \Gamma_{\mu \nu}^{\rho}}$. Moreover, the hyperbolic equations themselves involve a dependence on the spacetime metric determinant ${det \left( g_{\mu \nu} \right)}$. For many simulations in general relativistic hydrodynamics, this does not present a problem, since a time-independent (and often analytic) spacetime metric, such as the Schwarzschild metric for a static black hole or the Kerr metric for a spinning one, is assumed to be fixed in advance, and then a relativistic fluid is simply evolved on top of it\cite{gourgoulhon2}. In such cases, all of the necessary spacetime metric components, spacetime metric derivatives and spacetime Christoffel symbols (along with the spacetime metric determinant) may be precalculated, and their analytical forms (or some appropriate numerical approximations to them) can then be incorporated into the overall simulation code. Although this is an entirely reasonable idealization to use in cases where the gravitational influence of the fluid on the underlying metric may be safely neglected (i.e. the ``test-fluid'' assumption\cite{font4}), which is often true in the case of black hole accretion simulations, this is clearly unsatisfactory for our present purposes, since we intend to evolve the fluid parameters and the spatial metric tensor together in a fully-coupled fashion, in order to determine the effects of spacetime discretization on both the fluid morphology and the resulting spacetime geometry. Eliminating the dependence of the equations on the spacetime metric determinant ${\det \left( g_{\mu \nu} \right)}$ is straightforward since, due to the geometry of the ADM decomposition, this determinant can be directly related to the spatial metric determinant ${\det \left( \gamma_{\mu \nu} \right)}$ by means of the lapse function ${\alpha}$:

\begin{equation}
\sqrt{- \det \left( g_{\mu \nu} \right)} = \alpha \sqrt{\det \left( \gamma_{\mu \nu} \right)}.
\end{equation}
On the other hand, by means of a somewhat more involved calculation, we can rewrite the source terms for the energy and momentum conservation equations purely in terms of components of the stress-energy tensor ${T^{\mu \nu}}$, the primitive variables of the fluid ${\rho}$, ${v^{\mu}}$ (or equivalently ${v_{\mu}}$) and $P$, the ADM gauge variables ${\alpha}$ and ${\beta^{\mu}}$, the spatial metric tensor components ${\gamma_{\mu \nu}}$, and the extrinsic curvature tensor components ${K_{\mu \nu}}$, as follows:

\begin{multline}
\alpha \left( T^{\mu 0} \frac{\partial}{\partial x^{\mu}} \left( \log \left( \alpha \right) \right) - T^{\mu \nu} {}^{\left( 4 \right)} \Gamma_{\nu \mu}^{0} \right) = T^{0 0} \left( \beta^{\mu} \beta^{\nu} K_{\mu \nu} - \beta^{\mu} \frac{\partial}{\partial x^{\mu}} \left( \alpha \right) \right)\\
+ T^{0 \left( \mu + 1 \right)} \left( - \frac{\partial}{\partial x^{\mu}} \left( \alpha \right) + 2 \beta^{\nu} K_{\mu \nu} \right) + T^{\left( \mu + 1 \right) \left( \nu + 1 \right)} K_{\mu \nu},
\end{multline}
with ${\mu, \nu}$ on the left-hand-side ranging across all spacetime coordinate indices ${\left\lbrace 0, \dots, n - 1 \right\rbrace}$ and ${\mu, \nu}$ on the right-hand side ranging across spatial coordinate indices ${\left\lbrace 0, \dots, n - 2 \right\rbrace}$ only, and:

\begin{multline}
T^{\mu \nu} \left( \frac{\partial}{\partial x^{\mu}} \left( g_{\nu \left( \sigma + 1 \right)} \right) - {}^{\left( 4 \right)} \Gamma_{\nu \mu}^{\lambda} g_{\lambda \left( \sigma + 1 \right)} \right) = T^{0 0} \left( \frac{1}{2} \beta^{\mu} \beta^{\nu} \frac{\partial}{\partial x^{\sigma}} \left( \gamma_{\mu \nu} \right) - \alpha \frac{\partial}{\partial x^{\sigma}} \left( \alpha \right) \right) + T^{0 \left( \mu + 1 \right)} \beta^{\nu} \frac{\partial}{\partial x^{\sigma}} \left( \gamma_{\mu \nu} \right)\\
+ \frac{1}{2} T^{\left( \mu + 1 \right) \left( \nu + 1 \right)} \frac{\partial}{\partial x^{\sigma}} \left( \gamma_{\mu \nu} \right) + \frac{\rho h v_{\rho}}{\alpha \left( 1 - \gamma_{\mu \nu} v^{\mu} v^{\nu} \right)} \frac{\partial}{\partial x^{\sigma}} \left( \beta^{\rho} \right),
\end{multline}
with ${\mu, \nu, \lambda}$ on the left-hand-side ranging across all spacetime coordinate indices ${\left\lbrace 0, \dots, n - 1 \right\rbrace}$, ${\mu, \nu, \rho}$ on the right-hand side ranging across spatial coordinate indices ${\left\lbrace 0, \dots, n - 2 \right\rbrace}$ only, and with ${\sigma}$ on both sides ranging across spatial coordinate indices ${\left\lbrace 0, \dots, n - 2 \right\rbrace}$ only.

With these new modifications put in place, our hyperbolic system of equations governing the evolution of a perfect relativistic fluid on an arbitrary (and potentially dynamically-evolving) spacetime now consists of the following form of the energy conservation law:

\begin{multline}
\frac{1}{\alpha \sqrt{\det \left( \gamma_{\mu \nu} \right)}} \left[ \frac{\partial}{\partial t} \left( \sqrt{\det \left( \gamma_{\mu \nu} \right)} \left( \frac{\rho h}{1 - \gamma_{\mu \nu} v^{\mu} v^{\nu}} - P - \frac{\rho}{\sqrt{1 - \gamma_{\mu \nu} v^{\mu} v^{\nu}}} \right) \right) \right.\\
\left. + \frac{\partial}{\partial x^{\rho}} \left( \alpha \sqrt{\det \left( \gamma_{\mu \nu} \right)} \left( \left( \frac{\rho h}{1 - \gamma_{\mu \nu} v^{\mu} v^{\nu}} - P - \frac{\rho}{\sqrt{1 - \gamma_{\mu \nu} v^{\mu} v^{\nu}}} \right) \left( v^{\rho} - \frac{\beta^{\rho}}{\alpha} \right) + P v^{\rho} \right) \right) \right]\\
= T^{0 0} \left( \beta^{\mu} \beta^{\nu} K_{\mu \nu} - \beta^{\mu} \frac{\partial}{\partial x^{\mu}} \left( \alpha \right) \right) + T^{0 \left( \mu + 1 \right)} \left( - \frac{\partial}{\partial x^{\mu}} \left( \alpha \right) + 2 \beta^{\nu} K_{\mu \nu} \right) + T^{\left( \mu + 1 \right) \left( \nu + 1 \right)} K_{\mu \nu},
\end{multline}
with ${\mu, \nu, \rho}$ now ranging across spatial coordinate indices ${\left\lbrace 0, \dots, n - 2 \right\rbrace}$ only for the whole equation, the following form of the momentum conservation law:

\begin{multline}
\frac{1}{\alpha \sqrt{\det \left( \gamma_{\mu \nu} \right)}} \left[ \frac{\partial}{\partial t} \left( \sqrt{\det \left( \gamma_{\mu \nu} \right)} \left( \frac{\rho h v_{\sigma}}{1 - \gamma_{\mu \nu} v^{\mu} v^{\nu}} \right) \right) \right.\\
\left. + \frac{\partial}{\partial x^{\rho}} \left( \alpha \sqrt{\det \left( \gamma_{\mu \nu} \right)} \left( \left( \frac{\rho h v_{\sigma}}{1 - \gamma_{\mu \nu} v^{\mu} v^{\nu}} \right) \left( v^{\rho} - \frac{\beta^{\rho}}{\alpha} \right) + P \delta_{\sigma}^{\rho} \right) \right) \right]\\
= T^{0 0} \left( \frac{1}{2} \beta^{\mu} \beta^{\nu} \frac{\partial}{\partial x^{\sigma}} \left( \gamma_{\mu \nu} \right) - \alpha \frac{\partial}{\partial x^{\sigma}} \left( \alpha \right) \right) + T^{0 \left( \mu + 1 \right)} \beta^{\nu} \frac{\partial}{\partial x^{\sigma}} \left( \gamma_{\mu \nu} \right)\\
+ \frac{1}{2} T^{\left( \mu + 1 \right) \left( \nu + 1 \right)} \frac{\partial}{\partial x^{\sigma}} \left( \gamma_{\mu \nu} \right) + \frac{\rho h v_{\rho}}{\alpha \left( 1 - \gamma_{\mu \nu} v^{\mu} v^{\nu} \right)} \frac{\partial}{\partial x^{\sigma}} \left( \beta^{\rho} \right),
\end{multline}
with ${\mu, \nu, \rho, \sigma}$ now ranging across spatial coordinate indices ${\left\lbrace 0, \dots, n - 2 \right\rbrace}$ only for the whole equation, and the following form of the baryon number conservation law:

\begin{multline}
\frac{1}{\alpha \sqrt{\det \left( \gamma_{\mu \nu} \right)}} \left[ \frac{\partial}{\partial t} \left( \sqrt{\det \left( \gamma_{\mu \nu} \right)} \left( \frac{\rho}{\sqrt{1 - \gamma_{\mu \nu} v^{\mu} v^{\nu}}} \right) \right) \right.\\
\left. + \frac{\partial}{\partial x^{\rho}} \left( \alpha \sqrt{\det \left( \gamma_{\mu \nu} \right)} \left( \left( \frac{\rho}{\sqrt{1 - \gamma_{\mu \nu} v^{\mu} v^{\nu}}} \right) \left( v^{\rho} - \frac{\beta^{\rho}}{\alpha} \right) \right) \right) \right] = 0,
\end{multline}
with ${\mu, \nu, \rho}$ now ranging across spatial coordinate indices ${\left\lbrace 0, \dots, n - 2 \right\rbrace}$ only for the whole equation. The characteristic wave speeds of the fluid system can now be calculated by performing an eigendecomposition of its  corresponding ${\left( n + 1 \right)}$-dimensional Jacobian matrices ${\mathbf{B}^{\rho}}$, namely:

\begin{equation}
\mathbf{B}^{\rho} = \alpha \frac{\partial \begin{bmatrix} 
\left( \frac{\rho h}{1 - \gamma_{\mu \nu} v^{\mu} v^{\nu}} - P - \frac{\rho}{\sqrt{1 - \gamma_{\mu \nu} v^{\mu} v^{\nu}}} \right) \left( v^{\rho} - \frac{\beta^{\rho}}{\alpha} \right) + P v^{\rho}\\
\left( \frac{\rho h v_{\sigma}}{1 - \gamma_{\mu \nu} v^{\mu} v^{\nu}} \right) \left( v^{\rho} - \frac{\beta^{\rho}}{\alpha} \right) + P \delta_{\sigma}^{\rho}\\
\left( \frac{\rho}{\sqrt{1 - \gamma_{\mu \nu} v^{\mu} v^{\nu}}} \right) \left( v^{\rho} - \frac{\beta^{\rho}}{\alpha} \right)
\end{bmatrix}}{\partial \begin{bmatrix}
\frac{\rho h}{1 - \gamma_{\mu \nu} v^{\mu} v^{\nu}} - P - \frac{\rho}{\sqrt{1 - \gamma_{\mu \nu} v^{\mu} v^{\nu}}}\\
\frac{\rho h v_{\sigma}}{1 - \gamma_{\mu \nu} v^{\mu} v^{\nu}}\\
\frac{\rho}{\sqrt{1 - \gamma_{\mu \nu} v^{\mu} v^{\nu}}}
\end{bmatrix}},
\end{equation}
with one Jacobian matrix ${\mathbf{B}^{\rho}}$ associated with each spatial coordinate direction ${x^{\rho}}$. As first calculated by Anile\cite{anile}, Eulderink and Mellema\cite{eulderink}, and later Banyuls et al.\cite{banyuls}, the ${\left( n + 1 \right)}$ eigenvalues of each Jacobian matrix ${\mathbf{B}^{\rho}}$ can be grouped into those corresponding to the \textit{material} wave speeds, i.e:

\begin{equation}
\lambda_{0}^{\rho} = \alpha v^{\rho} - \beta^{\rho},
\end{equation}
which have algebraic multiplicity $n$, and those corresponding to the \textit{acoustic} wave speeds, i.e:

\begin{equation}
\lambda_{\pm}^{\rho} = \frac{\alpha}{1 - \gamma_{\mu \nu} v^{\mu} v^{\nu} c_{s}^{2}} \left[ v^{\rho} \left( 1 - c_{s}^{2} \right) \pm c_s \sqrt{\left( 1 - \gamma_{\mu \nu} v^{\mu} v^{\nu} \right) \left[ \gamma^{\rho \rho} \left( 1 - \gamma_{\mu \nu} v^{\mu} v^{\nu} c_{s}^{2} \right) - v^{\rho} v^{\rho} \left( 1 - c_{s}^{2} \right) \right]} \right] - \beta^{\rho},
\end{equation}
which each have algebraic multiplicity 1.

All that remains for us now is to consider the equations governing the dynamics of the discrete spacetime geometry itself. We start from the full Einstein field equations (including arbitrary stress-energy source terms), which are of a mixed hyperbolic-elliptic character:

\begin{equation}
{}^{\left( 4 \right)} G_{\mu \nu} + \Lambda g_{\mu \nu} = {}^{\left( 4 \right)} R_{\mu \nu} - \frac{1}{2} {}^{\left( 4 \right)} R g_{\mu \nu} + \Lambda g_{\mu \nu} = 8 \pi T_{\mu \nu},
\end{equation}
where ${{}^{\left( 4 \right)} G_{\mu \nu}}$ is the spacetime Einstein tensor:

\begin{equation}
{}^{\left( 4 \right)} G_{\mu \nu} = {}^{\left( 4 \right)} R_{\mu \nu} - \frac{1}{2} {}^{\left( 4 \right)} R g_{\mu \nu},
\end{equation}
${{}^{\left( 4 \right)} R_{\mu \nu}}$ is the spacetime Ricci tensor, obtained by contraction (${{}^{\left( 4 \right)} R_{\mu \nu} = {}^{\left( 4 \right)} R_{\mu \sigma \nu}^{\sigma}}$) of the spacetime Riemann tensor ${{}^{\left( 4 \right)} R_{\sigma \mu \nu}^{\rho}}$:

\begin{equation}
{}^{\left( 4 \right)} R_{\sigma \mu \nu}^{\rho} = \frac{\partial}{\partial x^{\mu}} \left( {}^{\left( 4 \right)} \Gamma_{\sigma \nu}^{\rho} \right) - \frac{\partial}{\partial x^{\nu}} \left( {}^{\left( 4 \right)} \Gamma_{\mu \sigma}^{\rho} \right) + {}^{\left( 4 \right)} \Gamma_{\mu \lambda}^{\rho} {}^{\left( 4 \right)} \Gamma_{\sigma \nu}^{\lambda} - {}^{\left( 4 \right)} \Gamma_{\lambda \nu}^{\rho} {}^{\left( 4 \right)} \Gamma_{\mu \sigma}^{\lambda},
\end{equation}
${{}^{\left( 4 \right)} R}$ is the spacetime Ricci scalar, obtained as the trace (i.e. ${{}^{\left( 4 \right)} R = {}^{\left( 4 \right)} R_{\mu}^{\mu} = g^{\mu \nu} {}^{\left( 4 \right)} R_{\mu \nu}}$) of the spacetime Ricci tensor ${{}^{\left( 4 \right)} R_{\mu \nu}}$, and ${\Lambda}$ is the cosmological constant (essentially taken to be an arbitrary integration constant for our present purposes). We can decompose the ten independent components of the full Einstein field equations (assuming a four-dimensional spacetime manifold ${\left( \mathcal{M}, g \right)}$, otherwise the number of independent components is equal to ${\frac{1}{2} n \left( n + 1 \right)}$ for $n$-dimensional spacetimes) into a system of six purely hyperbolic evolution equations (or ${\frac{1}{2} n \left( n - 1 \right)}$ evolution equations, for $n$-dimensional spacetimes) and a collection of four purely elliptic constraint equations (or $n$ constraint equations, for $n$-dimensional spacetimes), with the latter (constraint) equations arising from the contracted Bianchi identities, which assert that the covariant divergence of the spacetime Einstein tensor ${{}^{\left( 4 \right)} G_{\mu \nu}}$ must vanish identically:

\begin{equation}
{}^{\left( 4 \right)} \nabla_{\nu} {}^{\left( 4 \right)} G^{\mu \nu} = \frac{\partial}{\partial x^{\nu}} \left( {}^{\left( 4 \right)} G^{\mu \nu} \right) + {}^{\left( 4 \right)} \Gamma_{\nu \sigma}^{\mu} {}^{\left( 4 \right)} G^{\sigma \nu} + {}^{\left( 4 \right)} \Gamma_{\nu \sigma}^{\nu} {}^{\left( 4 \right)} G^{\mu \sigma} = 0.
\end{equation}
In all of the above, ${\mu, \nu, \rho, \sigma, \lambda}$ range across all spacetime coordinate indices ${\left\lbrace 0, \dots, n - 1 \right\rbrace}$. Upon performing an ADM decomposition of the spacetime metric, the hyperbolic evolution equations take the form:

\begin{multline}
\frac{\partial}{\partial t} \left( K_{\nu}^{\mu} \right) = \alpha {}^{\left( 3 \right)} R_{\nu}^{\mu} - {}^{\left( 3 \right)} \nabla_{\rho} \left( {}^{\left( 3 \right)} \nabla_{\nu} \alpha \right) \gamma^{\rho \mu} + \alpha K K_{\nu}^{\mu} + \beta^{\rho} {}^{\left( 3 \right)} \nabla_{\rho} K_{\nu}^{\mu}\\
+ K_{\rho}^{\mu} {}^{\left( 3 \right)} \nabla_{\nu} \beta^{\rho} - K_{\nu}^{\rho} {}^{\left( 3 \right)} \nabla_{\rho} \beta^{\mu} - \alpha \left( 8 \pi T_{\left( \rho + 1 \right) \left( \nu + 1 \right)} \gamma^{\rho \mu} - 4 \pi T \delta_{\nu}^{\mu} \right) - \alpha \left( \frac{2 \Lambda}{n - 2} \gamma_{\rho \nu} \right) \gamma^{\rho \mu},
\end{multline}
which then expand out to give:

\begin{multline}
\frac{\partial}{\partial t} \left( K_{\nu}^{\mu} \right) = \alpha {}^{\left( 3 \right)} R_{\nu}^{\mu} - \left( \frac{\partial}{\partial x^{\rho}} \left( \frac{\partial}{\partial x^{\nu}} \left( \alpha \right) \right) - {}^{\left( 3 \right)} \Gamma_{\rho \nu}^{\sigma} \left( \frac{\partial}{\partial x^{\sigma}} \left( \alpha \right) \right) \right) \gamma^{\rho \mu} + \alpha K_{\nu}^{\mu}\\
+ \beta^{\rho} \left( \frac{\partial}{\partial x^{\rho}} \left( K_{\nu}^{\mu} \right) + {}^{\left( 3 \right)} \Gamma_{\rho \sigma}^{\mu} K_{\nu}^{\sigma} - {}^{\left( 3 \right)} \Gamma_{\rho \nu}^{\sigma} K_{\sigma}^{\mu} \right) + K_{\rho}^{\mu} \left( \frac{\partial}{\partial x^{\nu}} \left( \beta^{\rho} \right) + {}^{\left( 3 \right)} \Gamma_{\nu \sigma}^{\rho} \beta^{\sigma} \right)\\
- K_{\nu}^{\rho} \left( \frac{\partial}{\partial x^{\rho}} \left( \beta^{\mu} \right) + {}^{\left( 3 \right)} \Gamma_{\rho \sigma}^{\mu} \beta^{\sigma} \right) - \alpha \left( 8 \pi T_{\left( \rho + 1 \right) \left( \nu + 1 \right)} \gamma^{\rho \mu} - 4 \pi T \delta_{\nu}^{\mu} \right) - \alpha \left( \frac{2 \Lambda}{n - 2} \gamma_{\rho \nu} \right) \gamma^{\rho \mu},
\end{multline}
where ${{}^{\left( 3 \right)} R_{\mu \nu}}$ is the spatial Ricci tensor, obtained by contraction (i.e. ${{}^{\left( 3 \right)} R_{\mu \nu} = {}^{\left( 3 \right)} R_{\mu \sigma \nu}^{\sigma}}$) of the spatial Riemann tensor ${{}^{\left( 3 \right)} R_{\sigma \mu \nu}^{\rho}}$:

\begin{equation}
{}^{\left( 3 \right)} R_{\sigma \mu \nu}^{\rho} = \frac{\partial}{\partial x^{\mu}} \left( {}^{\left( 3 \right)} \Gamma_{\sigma \nu}^{\rho} \right) - \frac{\partial}{\partial x^{\nu}} \left( {}^{\left( 3 \right)} \Gamma_{\mu \sigma}^{\rho} \right) + {}^{\left( 3 \right)} \Gamma_{\mu \lambda}^{\rho} {}^{\left( 3 \right)} \Gamma_{\sigma \nu}^{\lambda} - {}^{\left( 3 \right)} \Gamma_{\lambda \nu}^{\rho} {}^{\left( 3 \right)} \Gamma_{\mu \sigma}^{\lambda},
\end{equation}
$T$ is the trace of the stress-energy tensor (i.e. ${T = g^{\mu \nu} T_{\mu \nu}}$), and the indices of the extrinsic curvature tensor ${K_{\mu \nu}}$ and spatial Ricci tensor ${{}^{\left( 3 \right)} R_{\mu \nu}}$ are raised and lowered using the spatial metric tensor ${\gamma_{\mu \nu}}$, and so, in particular, for the quantities ${K_{\mu}^{\nu}}$ and ${{}^{\left( 3 \right)} R_{\mu}^{\nu}}$ (both in mixed-index form) appearing above, one has:

\begin{equation}
K_{\mu}^{\nu} = \gamma_{\mu \sigma} K^{\sigma \nu} = \gamma_{\mu \sigma} \gamma^{\lambda \nu} K_{\lambda}^{\sigma} = \gamma^{\sigma \nu} K_{\mu \sigma}, \qquad \text{ and } \qquad R_{\mu}^{\nu} = \gamma_{\mu \sigma} R_{\sigma \nu} = \gamma_{\mu \sigma} \gamma^{\lambda \nu} R_{\lambda}^{\sigma} = \gamma^{\sigma \nu} R_{\mu \sigma}.
\end{equation}
Finally, the elliptic constraint equations resulting from the contracted Bianchi identities may be decomposed into a timelike projection, yielding the Hamiltonian constraint equation:

\begin{equation}
\mathcal{H} = {}^{\left( 3 \right)} R + K^2 - K_{\nu}^{\mu} K_{\mu}^{\nu} - 16 \pi \alpha^2 T^{0 0} - 2 \Lambda = 0
\end{equation}
where ${\left( 3 \right) R}$ is the spatial Ricci scalar, obtained as the trace (i.e. ${{}^{\left( 3 \right)} R = {}^{\left( 3 \right)} R_{\mu}^{\mu} = \gamma^{\mu \nu} {}^{\left( 3 \right)} R_{\mu \nu}}$) of the spatial Ricci tensor ${{}^{\left( 3 \right)} R_{\mu \nu}}$, and into a collection of ${\left( n - 1 \right)}$ spacelike projections, yielding the momentum constraint equations:

\begin{equation}
\mathcal{M}_{\mu} = {}^{\left( 3 \right)} \nabla_{\nu} K_{\mu}^{\mu} - {}^{\left( 3 \right)} \nabla_{\mu} K - 8 \pi T_{\left( \mu + 1 \right)}^{0} = 0,
\end{equation}
which then expand out to give:

\begin{equation}
\mathcal{M}_{\mu} = \frac{\partial}{\partial x^{\mu}} \left( K_{\mu}^{\nu} \right) + {}^{\left( 3 \right)} \Gamma_{\nu \sigma}^{\nu} K_{\mu}^{\sigma} - {}^{\left( 3 \right)} \Gamma_{\nu \mu}^{\sigma} K_{\sigma}^{\nu} - \frac{\partial}{\partial x^{\mu}} \left( K \right) - 8 \pi T_{\left( \mu + 1 \right)}^{0} = 0.
\end{equation}
In all of the above, ${\mu, \nu, \rho, \sigma, \lambda}$ range across spatial coordinate indices ${\left\lbrace 0, \dots, n - 2 \right\rbrace}$ only. Although \textsc{Gravitas} solves the elliptic constraint equations automatically when running numerical simulations (typically by means of an iterative solver), we have also validated the algorithms employed within this article by using violations of the constraint equations (and, in particular, the propagation of certain constraint-violating modes) as a means of measuring and quantifying the robustness of the relevant numerical schemes. Note that, analytically, due to the Einstein field equations, the Hamiltonian and momentum constraint equations on the spacetime are satisfied identically whenever the energy and momentum conservation equations on the stress-energy distribution are satisfied, and vice versa.

\section{Numerical Validation: Special Relativistic Hydrodynamics}
\label{sec:Section3}

In order to render the discrete spacetime general relativistic hydrodynamics equations derived within the previous section in a form that is more directly amenable to explicit numerical solution, we begin by subdividing our overall $n$-dimensional spacetime ${\left( \mathcal{M}, g \right)}$ into a collection of simply-connected $n$-dimensional submanifolds ${\Omega \subseteq \mathcal{M}}$ (known as ``control volumes'', ``computational cells'', or ``nodes'' within \textsc{Gravitas}'s terminology), each of which has a closed ${\left( n - 1 \right)}$-dimensional boundary ${\partial \Omega}$. We can then integrate over each of these submanifolds in turn, yielding:

\begin{multline}
\int_{\Omega} \frac{1}{\alpha \sqrt{\det \left( \gamma_{\mu \nu} \right)}} \left[ \frac{\partial}{\partial t} \left( \sqrt{\det \left( \gamma_{\mu \nu} \right)} \left( \frac{\rho h}{1 - \gamma_{\mu \nu} v^{\mu} v^{\nu}} - P - \frac{\rho}{\sqrt{1 - \gamma_{\mu \nu} v^{\mu} v^{\nu}}} \right) \right) \right] d \Omega\\
+ \int_{\Omega} \frac{1}{\alpha \sqrt{\det \left( \gamma_{\mu \nu} \right)}} \left[ \frac{\partial}{\partial x^{\rho}} \left( \alpha \sqrt{\det \left( \gamma_{\mu \nu} \right)} \left( \left( \frac{\rho h}{1 - \gamma_{\mu \nu} v^{\mu} v^{\nu}} - P - \frac{\rho}{\sqrt{1 - \gamma_{\mu \nu} v^{\mu} v^{\nu}}} \right) \left( v^{\rho} - \frac{\beta^{\rho}}{\alpha} \right) + P v^{\rho} \right) \right) \right] d \Omega\\
= \int_{\Omega} \left[ T^{0 0} \left( \beta^{\mu} \beta^{\nu} K_{\mu \nu} - \beta^{\mu} \frac{\partial}{\partial x^{\mu}} \left( \alpha \right) \right) + T^{0 \left( \mu + 1 \right)} \left( - \frac{\partial}{\partial x^{\mu}} \left( \alpha \right) + 2 \beta^{\nu} K_{\mu \nu} \right) + T^{\left( \mu + 1 \right) \left( \nu + 1 \right)} K_{\mu \nu} \right] d \Omega,
\end{multline}
for the energy conservation equation;

\begin{multline}
\int_{\Omega} \frac{1}{\alpha \sqrt{\det \left( \gamma_{\mu \nu} \right)}} \left[ \frac{\partial}{\partial t} \left( \sqrt{\det \left( \gamma_{\mu \nu} \right)} \left( \frac{\rho h v_{\sigma}}{1 - \gamma_{\mu \nu} v^{\mu} v^{\nu}} \right) \right) \right] d \Omega\\
+ \int_{\Omega} \frac{1}{\alpha \sqrt{\det \left( \gamma_{\mu \nu} \right)}} \left[ \frac{\partial}{\partial x^{\rho}} \left( \alpha \sqrt{\det \left( \gamma_{\mu \nu} \right)} \left( \left( \frac{\rho h v_{\sigma}}{1 - \gamma_{\mu \nu} v^{\mu} v^{\nu}} \right) \left( v^{\rho} - \frac{\beta^{\rho}}{\alpha} \right) + P \delta_{\sigma}^{\rho} \right) \right) \right] d \Omega\\
= \int_{\Omega} \left[ T^{0 0} \left( \frac{1}{2} \beta^{\mu} \beta^{\nu} \frac{\partial}{\partial x^{\sigma}} \left( \gamma_{\mu \nu} \right) - \alpha \frac{\partial}{\partial x^{\sigma}} \left( \alpha \right) \right) + T^{0 \left( \mu + 1 \right)} \beta^{\nu} \frac{\partial}{\partial x^{\sigma}} \left( \gamma_{\mu \nu} \right) \right.\\
\left. + \frac{1}{2} T^{\left( \mu + 1 \right) \left( \nu + 1 \right)} \frac{\partial}{\partial x^{\sigma}} \left( \gamma_{\mu \nu} \right) + \frac{\rho h v_{\rho}}{\alpha \left( 1 - \gamma_{\mu \nu} v^{\mu} v^{\nu} \right)} \frac{\partial}{\partial x^{\sigma}} \left( \beta^{\rho} \right) \right] d \Omega,
\end{multline}
for the momentum conservation equations; and:

\begin{multline}
\int_{\Omega} \frac{1}{\alpha \sqrt{\det \left( \gamma_{\mu \nu} \right)}} \left[ \frac{\partial}{\partial t} \left( \sqrt{\det \left( \gamma_{\mu \nu} \right)} \left(  \frac{\rho}{\sqrt{1 - \gamma_{\mu \nu} v^{\mu} v^{\nu}}} \right) \right) \right] d \Omega\\
\int_{\Omega} \frac{1}{\alpha \sqrt{\det \left( \gamma_{\mu \nu} \right)}} \left[ \frac{\partial}{\partial x^{\rho}} \left( \alpha \sqrt{\det \left( \gamma_{\mu \nu} \right)} \left( \left( \frac{\rho}{\sqrt{1 - \gamma_{\mu \nu} v^{\mu} v^{\nu}}} \right) \left( v^{\rho} - \frac{\beta^{\rho}}{\alpha} \right) \right) \right) \right] = 0,
\end{multline}
for the baryon number continuity equation, with ${\mu, \nu, \rho, \sigma}$ in all of the above ranging across spatial coordinate indices ${\left\lbrace 0, \dots, n - 2 \right\rbrace}$ only. This so-called ``weak'' integral form of the conservation equations can then be solved directly using \textsc{Gravitas}'s hypergraph-based finite-volume numerical algorithms\cite{gorard8}\cite{gorard9}. As a means of validating this numerical implementation, we first consider the simplified case of four-dimensional \textit{special} relativistic hydrodynamics, in which the spacetime metric tensor ${g_{\mu \nu}}$ is simply the four-dimensional Minkowski metric ${\eta_{\mu \nu}}$, i.e. ${g_{\mu \nu} = \eta_{\mu \nu} = \mathrm{diag} \left( -1, 1, 1, 1 \right)}$, the spatial metric tensor ${\gamma_{\mu \nu}}$ is the three-dimensional Euclidean metric ${\delta_{\mu \nu}}$, i.e. ${\gamma_{\mu \nu} = \delta_{\mu \nu} = \mathrm{diag} \left( 1, 1, 1 \right)}$, and we select trivial gauge conditions in which the lapse function obeys the geodesic slicing condition (i.e. ${\alpha = 1}$) and the shift vector obeys the normal coordinate conditions (i.e. ${\boldsymbol\beta = \mathbf{0}}$). The energy conservation equation now reduces to:

\begin{equation}
\frac{\partial}{\partial t} \left( \frac{\rho h}{1 - \delta_{\mu \nu} v^{\mu} v^{\nu}} - P - \frac{\rho}{\sqrt{1 - \delta_{\mu \nu} v^{\mu} v^{\nu}}} \right) + \frac{\partial}{\partial x^{\rho}} \left( \left( \frac{\rho h}{1 - \delta_{\mu \nu} v^{\mu} v^{\nu}} - P - \frac{\rho}{\sqrt{1 - \delta_{\mu \nu} v^{\mu} v^{\nu}}} \right) v^{\rho} + P v^{\rho} \right) = 0,
\end{equation}
with the corresponding weak form:

\begin{multline}
\int_{\Omega} \left[ \frac{\partial}{\partial t} \left( \frac{\rho h}{1 - \delta_{\mu \nu} v^{\mu} v^{\nu}} - P - \frac{\rho}{\sqrt{1 - \delta_{\mu \nu} v^{\mu} v^{\nu}}} \right) \right] d \Omega\\
+ \int_{\Omega} \left[ \frac{\partial}{\partial x^{\rho}} \left( \left( \frac{\rho h}{1 - \delta_{\mu \nu} v^{\mu} v^{\nu}} - P - \frac{\rho}{\sqrt{1 - \delta_{\mu \nu} v^{\mu} v^{\nu}}} \right) v^{\rho} + P v^{\rho} \right) \right] d \Omega = 0
\end{multline}
the momentum conservation equations reduce to:

\begin{equation}
\frac{\partial}{\partial t} \left( \frac{\rho h v_{\sigma}}{1 - \delta_{\mu \nu} v^{\mu} v^{\nu}} \right) + \frac{\partial}{\partial x^{\rho}} \left( \left( \frac{\rho h v_{\sigma}}{1 - \delta_{\mu \nu} v^{\mu} v^{\nu}} \right) v^{\rho} + P \delta_{\sigma}^{\rho} \right) = 0,
\end{equation}
with their corresponding weak forms being:

\begin{equation}
\int_{\Omega} \left[ \frac{\partial}{\partial t} \left( \frac{\rho h v_{\sigma}}{1 - \delta_{\mu \nu} v^{\mu} v^{\nu}} \right) \right] d \Omega + \int_{\Omega} \left[ \frac{\partial}{\partial x^{\rho}} \left( \left( \frac{\rho h v_{\sigma}}{1 - \delta_{\mu \nu} v^{\mu} v^{\nu}} \right) v^{\rho} + P \delta_{\sigma}^{\rho} \right) \right] d \Omega = 0
\end{equation}
and the baryon number continuity equation reduces to:

\begin{equation}
\frac{\partial}{\partial t} \left( \frac{\rho}{\sqrt{1 - \delta_{\mu \nu} v^{\mu} v^{\nu}}} \right) + \frac{\partial}{\partial x^{\rho}} \left( \left( \frac{\rho}{\sqrt{1 - \delta_{\mu \nu} v^{\mu} v^{\nu}}} \right) v^{\rho} \right) = 0,
\end{equation}
with its corresponding weak form being:

\begin{equation}
\int_{\Omega} \left[ \frac{\partial}{\partial t} \left( \frac{\rho}{\sqrt{1 - \delta_{\mu \nu} v^{\mu} v^{\nu}}} \right) \right] d \Omega + \int_{\Omega} \left[ \left( \left( \frac{\rho}{\sqrt{1 - \delta_{\mu \nu} v^{\mu} v^{\nu}}} \right) v^{\rho} \right) \right] d \Omega,
\end{equation}
and where ${\mu, \nu, \rho, \sigma}$ in all of the above range, again, across spatial coordinate indices ${\left\lbrace 0, \dots, n - 2 \right\rbrace}$ only.

Our primary objective here is to validate our conservative-to-primitive variable reconstruction algorithm. Since the conservative variables are the quantities that are actually evolved by our numerical algorithm, yet the primitive variables are the quantities that are required for the computation of the flux function on the next time step, it is necessary to perform a conversion from the conservative variables, consisting of the (rest) mass density of the fluid $D$ measured by an observer moving in the normal direction ${\mathbf{n}}$

\begin{equation}
D = \frac{\rho}{\sqrt{1 - \gamma_{\mu \nu} v^{\mu} v^{\nu}}} = -J_{\mu} n^{\mu},
\end{equation}
with ${\mu, \nu}$ on the left-hand-side ranging across spatial coordinate indices ${\left\lbrace 0, \dots, n - 2 \right\rbrace}$ only, and ${\mu}$ on the right-hand-side ranging across all spacetime coordinate indices ${\left\lbrace 0, \dots, n - 1 \right\rbrace}$; the components of the momentum density covector ${p_{\sigma}}$ of the fluid measured by that normal observer:

\begin{equation}
p_{\sigma} = \frac{\rho h v_{\sigma}}{1 - \gamma_{\mu \nu} v^{\mu} v^{\nu}} = - T_{\mu \nu} n^{\mu} \bot_{\sigma}^{\nu},
\end{equation}
with ${\mu, \nu}$ on the left-hand-side ranging across spatial coordinate indices ${\left\lbrace 0, \dots, n - 2 \right\rbrace}$ only, ${\mu, \nu}$ on the right-hand-side ranging across all spacetime coordinate indices ${\left\lbrace 0, \dots, n - 1 \right\rbrace}$, and with ${\sigma}$ on both sides ranging across spatial coordinate indices ${\left\lbrace 0, \dots, n - 2 \right\rbrace}$ only; and the difference between the energy density $E$ and (rest) mass density $D$ of the fluid measured by that same normal observer:

\begin{equation}
E - D = \frac{\rho h}{1 - \gamma_{\mu \nu} v^{\mu} v^{\nu}} - P - \frac{\rho}{\sqrt{1 - \gamma_{\mu \nu} v^{\mu} v^{\nu}}} = T_{\mu \nu} n^{\mu} n^{\nu} - J_{\mu} n^{\mu},
\end{equation}
with ${\mu, \nu}$ on the left-hand-side ranging across spatial coordinate indices ${\left\lbrace 0, \dots, n - 2 \right\rbrace}$ only, and ${\mu, \nu}$ on the right-hand-side ranging across all spacetime coordinate indices ${\left\lbrace 0, \dots, n - 1 \right\rbrace}$, to the primitive variables, consisting of the (rest) mass density of the fluid ${\rho}$, the components of the spatial velocity vector of the fluid measured by a normal observer ${v^{\mu}}$, and the hydrostatic pressure of the fluid $P$. In non-relativistic hydrodynamics, such a conversion can typically be performed purely algebraically, but in both special and general relativistic hydrodynamics (at least assuming a reasonably generic equation of state), the components of the momentum density covector ${p_{\sigma}}$ are not algebraically independent of one other due to the presence of the Lorentz factor ${\sqrt{1 - \gamma_{\mu \nu} v^{\mu} v^{\nu}}}$, and therefore no closed form expression for the primitive variables in terms of the conservative ones is known in general\cite{marti}. One notable exception to this is the case of stiff, ultra-relativistic fluids\cite{petrich2}, in which such a closed form expression \textit{does} exist (and, indeed, the existence of such a simplification is closely related to why Petrich, Shapiro and Teukolsky\cite{petrich} were able to derive an analytic solution for accretion for the accretion of such fluids onto spinning black holes in general axial symmetry), although this particular property of the stiff, ultra-relativistic equation of state certainly does not generalize, as we shall discuss later on in this article.

For this reason, we choose to follow the approach proposed by Eulderink and Mellema\cite{eulderink}, and apply an iterative, non-linear root-finding algorithm (namely the one-dimensional Newton-Raphson method) in order to approximate the roots to the following quartic polynomial in ${\xi}$ numerically:

\begin{equation}
\alpha_4 \xi^3 \left( \xi - \eta \right) + \alpha_2 \xi^2 + \alpha_1 \xi + \alpha_0 = 0,
\end{equation}
where we have defined the variable ${\xi}$ and the constant ${\eta}$ to be given by:

\begin{equation}
\xi = \frac{\sqrt{- g_{\mu \nu} T^{0 \mu} T^{0 \nu}}}{\rho h u^0} = \frac{\left( \sqrt{\left( \frac{\rho h}{1 - \gamma_{\mu \nu} v^{\mu} v^{\nu}} - P \right)^2 - \left( \frac{\rho h v_{\sigma}}{1 - \gamma_{\mu \nu} v^{\mu} v^{\mu}} \right)^2} \right) \left( \sqrt{1 - \gamma_{\mu \nu} v^{\mu} v^{\nu}} \right)}{\rho h},
\end{equation}
and:

\begin{equation}
\eta = \frac{2 \rho u^0 \left( \Gamma - 1 \right)}{\left( \sqrt{- g_{\mu \nu} T^{0 \mu} T^{0 \nu}} \right) \Gamma} = \frac{2 \rho \left( \Gamma - 1 \right)}{\left( \sqrt{\left( \frac{\rho h}{1 - \gamma_{\mu \nu} v^{\mu} v^{\nu}} - P \right)^2 - \left( \frac{\rho h v_{\sigma}}{1 - \gamma_{\mu \nu} v^{\mu} v^{\nu}} \right)^2} \right) \left( \sqrt{1 - \gamma_{\mu \nu} v^{\mu} v^{\nu}} \right) \Gamma},
\end{equation}
respectively, and where the coefficients ${\alpha_4}$, ${\alpha_2}$, ${\alpha_1}$ and ${\alpha_0}$ in front of the terms in the quartic are given by:

\begin{equation}
\alpha_4 = \frac{\left( T^{0 0} \right)^2}{g^{0 0} g_{\mu \nu} T^{0 \mu} T^{0 \nu}} - 1 = \frac{\left( \frac{\rho h}{1 - \gamma_{\mu \nu} v^{\mu} v^{\nu}} - P \right)^2}{\left( \frac{\rho h}{1 - \gamma_{\mu \nu} v^{\mu} v^{\nu}} - P \right)^2 - \left( \frac{\rho h v_{\sigma}}{1 - \gamma_{\mu \nu} v^{\mu} v^{\nu}} \right)^2} - 1,
\end{equation}
\begin{multline}
\alpha_2 = \left( \frac{\Gamma - 2}{\Gamma} \right) \left( \frac{\left( T^{0 0} \right)^2}{g^{0 0} g_{\mu \nu} T^{0 \mu} T^{0 \nu}} - 1 \right) + 1 + \left( \frac{\left( \rho u^0 \right)^2}{g_{\mu \nu} T^{0 \mu} T^{0 \nu}} \right) \left( \frac{\Gamma - 1}{\Gamma} \right)^2\\
= \left( \frac{\Gamma - 2}{\Gamma} \right) \left( \frac{\left( \frac{\rho h}{1 - \gamma_{\mu \nu} v^{\mu} v^{\nu}} - P \right)^2}{\left( \frac{\rho h}{1 - \gamma_{\mu \nu} v^{\mu} v^{\nu}} \right)^2 - \left( \frac{\rho h v_{\sigma}}{1 - \gamma_{\mu \nu} v^{\mu} v^{\nu}} \right)^2} - 1\right) + 1\\
- \left( \frac{\rho^2}{\left( \left( \frac{\rho h}{1 - \gamma_{\mu \nu} v^{\mu} v^{\nu}} - P \right)^2 - \left( \frac{\rho h v_{\sigma}}{1 - \gamma_{\mu \nu} v^{\mu} v^{\nu}} \right)^2 \right) \left( 1 - \gamma_{\mu \nu} v^{\mu} v^{\nu} \right)} \right) \left( \frac{\Gamma - 1}{\Gamma} \right)^2,
\end{multline}
\begin{equation}
\alpha_1 = - \frac{2 \rho u^0 \left( \Gamma - 1 \right)}{\left( \sqrt{- g_{\mu \nu} T^{0 \mu} T^{0 \nu}} \right) \Gamma^2} = - \frac{2 \rho \left( \Gamma - 1 \right)}{\left( \sqrt{\left( \frac{\rho h}{1 - \gamma_{\mu \nu} v^{\mu} v^{\nu}} - P \right)^2 - \left( \frac{\rho h v_{\sigma}}{1 - \gamma_{\mu \nu} v^{\mu} v^{\nu}} \right)^2} \right) \left( \sqrt{1 - \gamma_{\mu \nu} v^{\mu} v^{\nu}} \right) \Gamma^2},
\end{equation}
and:

\begin{equation}
\alpha_0 = - \frac{1}{\Gamma^2},
\end{equation}
respectively. In all of the equations above, ${\mu, \nu}$ on the left-hand-side range across all spacetime coordinate indices ${\left\lbrace 0, \dots, n - 1 \right\rbrace}$, while ${\mu, \nu, \sigma}$ on the right-hand-side range across spatial coordinate indices ${\left\lbrace 0, \dots, n - 2 \right\rbrace}$ only. Since the variable ${\xi}$ itself represents the fluid quantity:

\begin{equation}
\xi = \frac{\left( \sqrt{\left( \frac{\rho h}{1 - \gamma_{\mu \nu} v^{\mu} v^{\nu}} - P \right)^2 - \left( \frac{\rho h v_{\sigma}}{1 - \gamma_{\mu \nu} v^{\mu} v^{\nu}} \right)^2} \right) \left( \sqrt{1 - \gamma_{\mu \nu} v^{\mu} v^{\nu}} \right)}{\rho h},
\end{equation}
we may proceed to use its approximate numerical value as a starting point for computing all of the other fluid quantities of interest, such as the new value of the  Lorentz factor ${W_{new}}$:

\begin{multline}
W_{new} = \frac{1}{2} \left( \frac{\frac{\rho h}{1 - \gamma_{\mu \nu} v^{\mu} v^{\nu}} - P}{\sqrt{\left( \frac{\rho h}{1 - \gamma_{\mu \nu} v^{\mu} v^{\nu}} - P \right)^2 - \left( \frac{\rho h v_{\sigma}}{1 - \gamma_{\mu \nu} v^{\mu} v^{\nu}} \right)^2}} \right) \xi\\
\times \left( 1 + \sqrt{1 + 4 \left( \frac{\Gamma - 1}{\Gamma} \right) \left( \frac{1 - \frac{\rho \xi}{\left( \sqrt{\left( \frac{\rho h}{1 - \gamma_{\mu \nu} v^{\mu} v^{\nu}} - P \right)^2 - \left( \frac{\rho h v_{\sigma}}{1 - \gamma_{\mu \nu} v^{\mu} v^{\nu}} \right)^2} \right) \left( \sqrt{1 - \gamma_{\mu \nu} v^{\mu} v^{\nu}} \right)}}{\frac{\left( \frac{\rho h}{1 - \gamma_{\mu \nu} v^{\mu} v^{\nu}} -P \right)^2 \xi^2}{\left( \frac{\rho h}{1 - \gamma_{\mu \nu} v^{\mu} v^{\nu}} - P \right)^2 - \left( \frac{\rho h v_{\sigma}}{1 - \gamma_{\mu \nu} v^{\mu} v^{\nu}} \right)^2}} \right)} \right),
\end{multline}
and, from it, the new value of the fluid (rest) mass density ${\rho_{new}}$:

\begin{equation}
\rho_{new} = \frac{\rho}{\left( \sqrt{1 - \gamma_{\mu \nu} v^{\mu} v^{\nu}} \right) W_{new}},
\end{equation}
the new value of the specific relativistic enthalpy of the fluid ${h_{new}}$ (from which the new value of its hydrostatic pressure ${P_{new}}$ can then be recovered by means of the equation of state):

\begin{equation}
h_{new} = \frac{\left( \sqrt{\left( \frac{\rho h}{1 - \gamma_{\mu \nu} v^{\mu} v^{\nu}} - P \right)^2 - \left( \frac{\rho h v_{\sigma}}{1 - \gamma_{\mu \nu} v^{\mu} v^{\nu}} \right)^2} \right) \left( \sqrt{1 - \gamma_{\mu \nu} v^{\mu} v^{\nu}} \right)}{\rho \xi^2},
\end{equation}
and finally the components of the new spatial velocity covector of the fluid, as measured by a normal observer, ${v_{\sigma}^{new}}$:

\begin{equation}
v_{\sigma}^{new} = \frac{\rho h v_{\sigma}}{\left( 1 - \gamma_{\mu \nu} v^{\mu} v^{\nu} \right) \rho_{new} h_{new} W_{new}^{2}}.
\end{equation}
In all of the equations above, ${\mu, \nu, \sigma}$ range across spatial coordinate indices ${\left\lbrace 0, \dots, n - 2 \right\rbrace}$ only. As proved by Eulderink and Mellema\cite{eulderink}, this quartic possesses exactly two real roots, with only one obeying the physicality condition ${\xi > 0}$, and the iterative Newton-Raphson method is guaranteed to converge at least quadratically to this solution. In all of the above, we have assumed that the fluid obeys the ideal gas equation of state\cite{courant}, with specific relativistic enthalpy given by

\begin{equation}
h = 1 + \frac{P}{\rho} \left( \frac{\Gamma}{\Gamma - 1} \right),
\end{equation}
where ${\Gamma}$ is the adiabatic exponent of the fluid, such that the local speed of sound ${c_s}$ is simply:

\begin{equation}
c_s = \sqrt{\frac{\Gamma P}{\rho \left( 1 + \left( \frac{P}{\rho} \right) \left( \frac{\Gamma}{\Gamma - 1} \right) \right)}}.
\end{equation}

\begin{figure}[ht]
\centering
\includegraphics[width=0.295\textwidth]{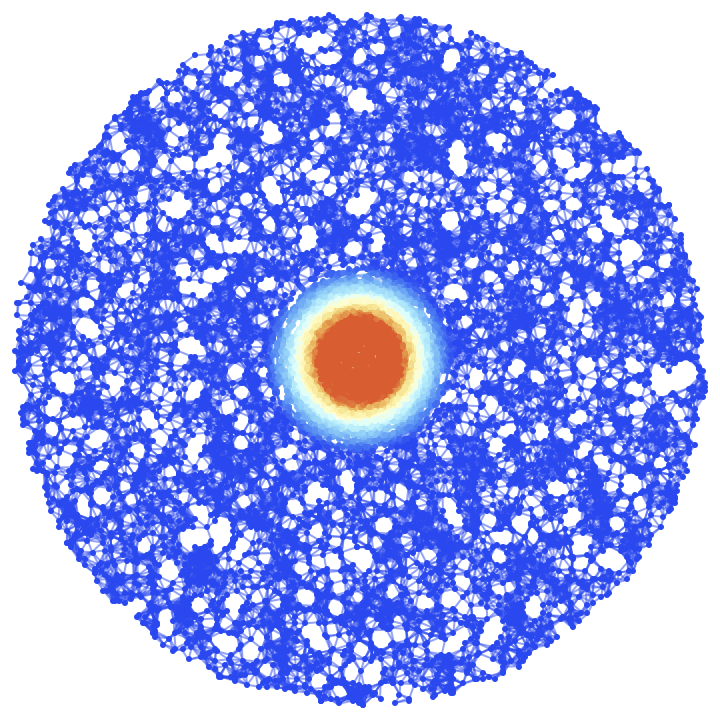}\hspace{0.2\textwidth}
\includegraphics[width=0.295\textwidth]{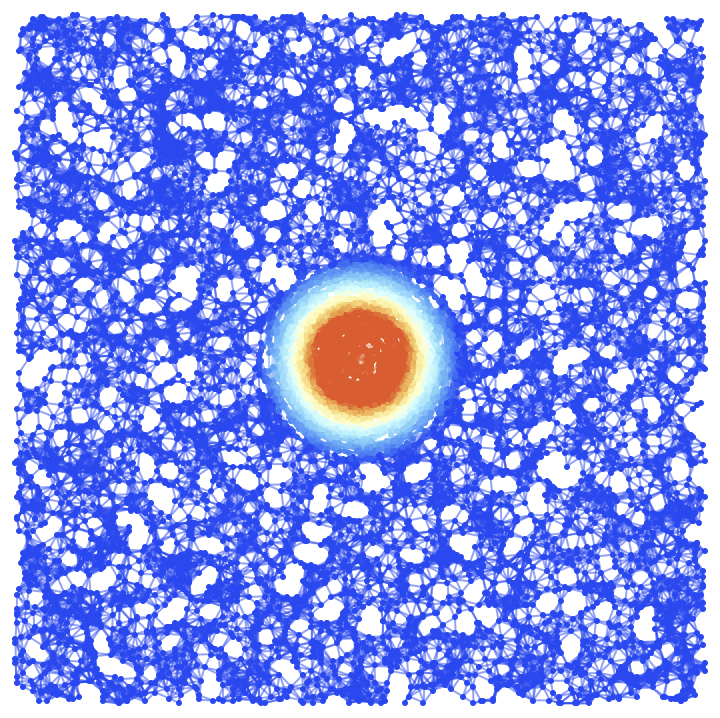}
\caption{On the left, the initial (${t = 0}$) hypersurface configuration of the mildly-relativistic blast wave problem, embedded within a spherically-symmetric Minkowski geometry in spherical polar coordinates ${\left( t, r, \theta, \phi \right)}$, with a resolution of 10,000 hypergraph vertices (colored based on fluid density), and with spatial coordinate information assigned to the vertices. On the right, the initial (${t = 0}$) hypersurface configuration of the mildly-relativistic blast wave problem, embedded within a rectangular Minkowski geometry in Cartesian coordinates ${\left( t, x, y, z \right)}$, with a resolution of 10,000 hypergraph vertices (colored based on fluid density), and with spatial coordinate information assigned to the vertices.}
\label{fig:Figure1}
\end{figure}

\begin{figure}[ht]
\centering
\includegraphics[width=0.295\textwidth]{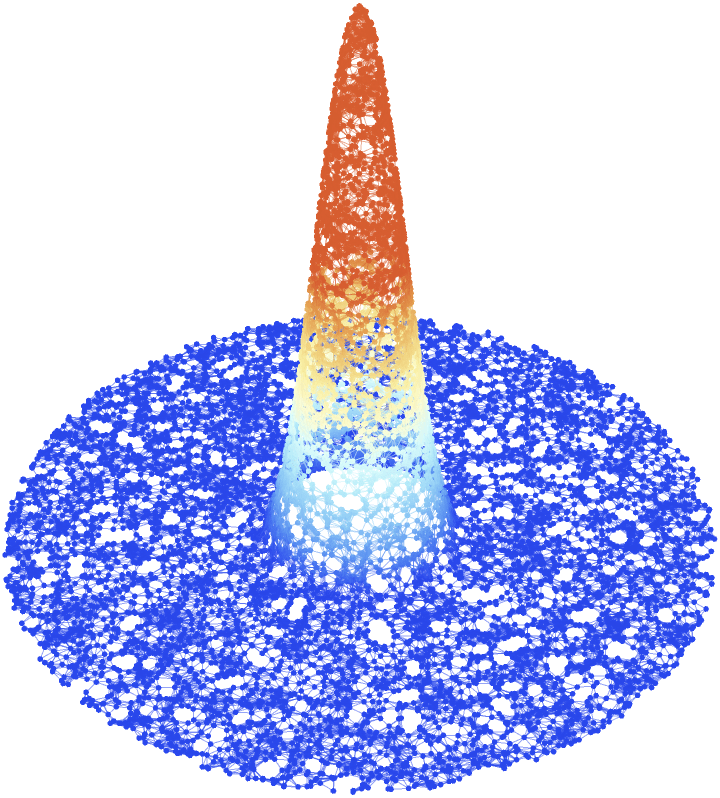}\hspace{0.2\textwidth}
\includegraphics[width=0.295\textwidth]{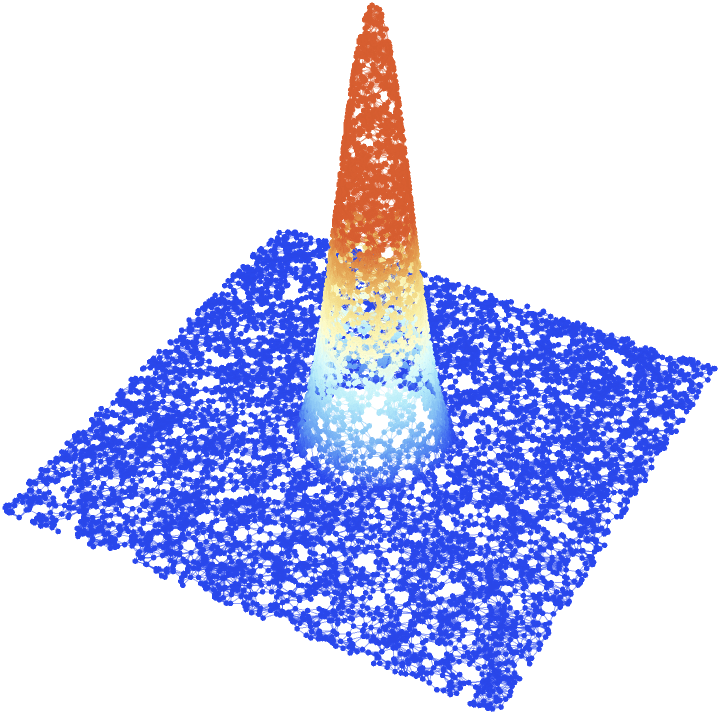}
\caption{On the left, the initial (${t = 0}$) hypersurface configuration of the mildly-relativistic blast wave problem, embedded within a spherically-symmetric Minkowski geometry in spherical polar coordinates ${\left( t, r, \theta, \phi \right)}$, with a resolution of 10,000 hypergraph vertices (colored based on fluid density), and with both spatial coordinate and fluid density coordinate information assigned to the vertices. On the right, the initial (${t = 0}$) hypersurface configuration of the mildly-relativistic blast wave problem, embedded within a rectangular Minkowski geometry in Cartesian coordinates ${\left( t, x, y, z \right)}$, with a resolution of 10,000 hypergraph vertices (colored based on fluid density), and with both spatial coordinate and fluid density coordinate information assigned to the vertices.}
\label{fig:Figure2}
\end{figure}

\begin{figure}[ht]
\centering
\includegraphics[width=0.295\textwidth]{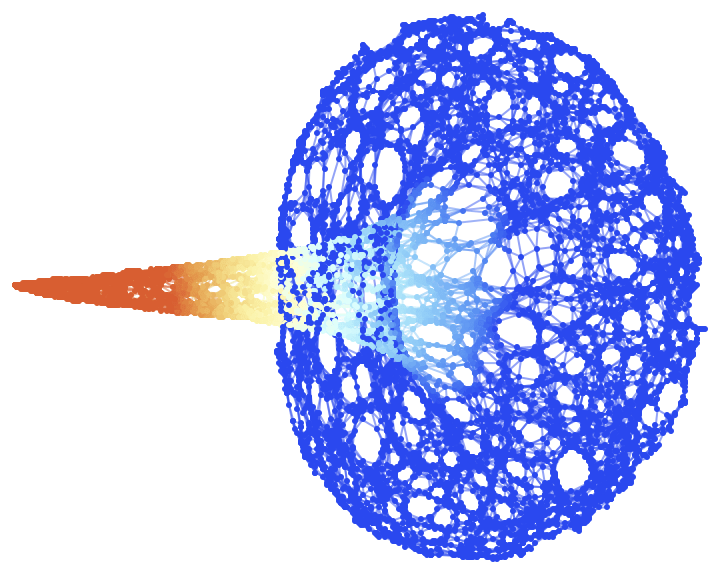}\hspace{0.2\textwidth}
\includegraphics[width=0.295\textwidth]{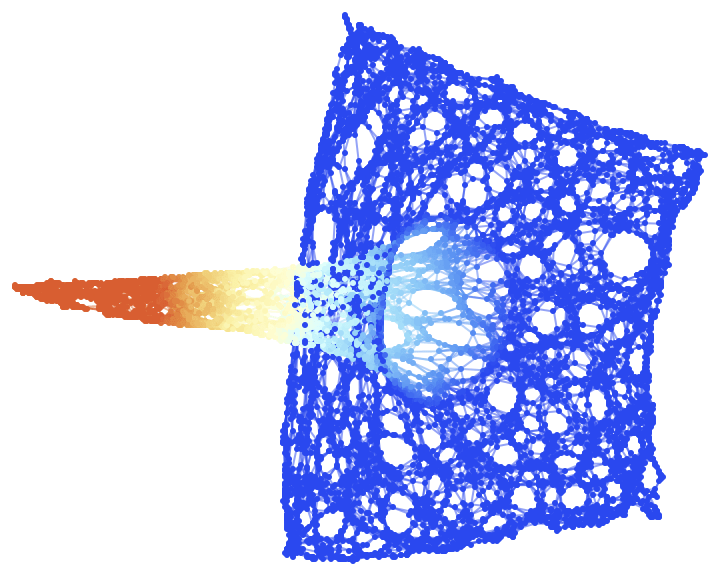}
\caption{On the left, the initial (${t = 0}$) hypersurface configuration of the mildly-relativistic blast wave problem, embedded within a spherically-symmetric Minkowski geometry in spherical polar coordinates ${\left( t, r, \theta, \phi \right)}$, with a resolution of 10,000 hypergraph vertices (colored based on fluid density), and with no coordinate information assigned to the vertices. On the right, the initial (${t = 0}$) hypersurface configuration of the mildly-relativistic blast wave problem, embedded within a rectangular Minkowski geometry in Cartesian coordinates ${\left( t, x, y, z \right)}$, with a resolution of 10,000 hypergraph vertices (colored based on fluid density), and with no coordinate information assigned to the vertices.}
\label{fig:Figure3}
\end{figure}

\begin{figure}[ht]
\centering
\includegraphics[width=0.295\textwidth]{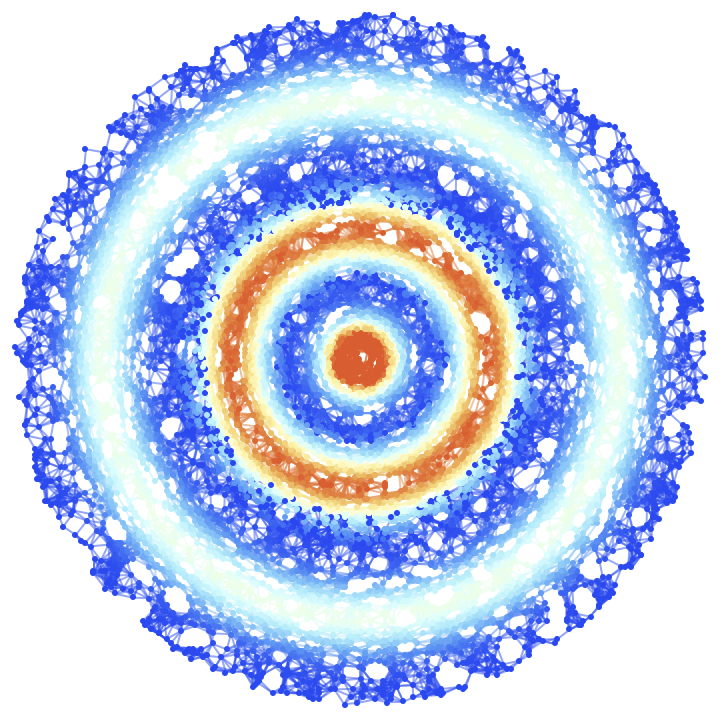}\hspace{0.2\textwidth}
\includegraphics[width=0.295\textwidth]{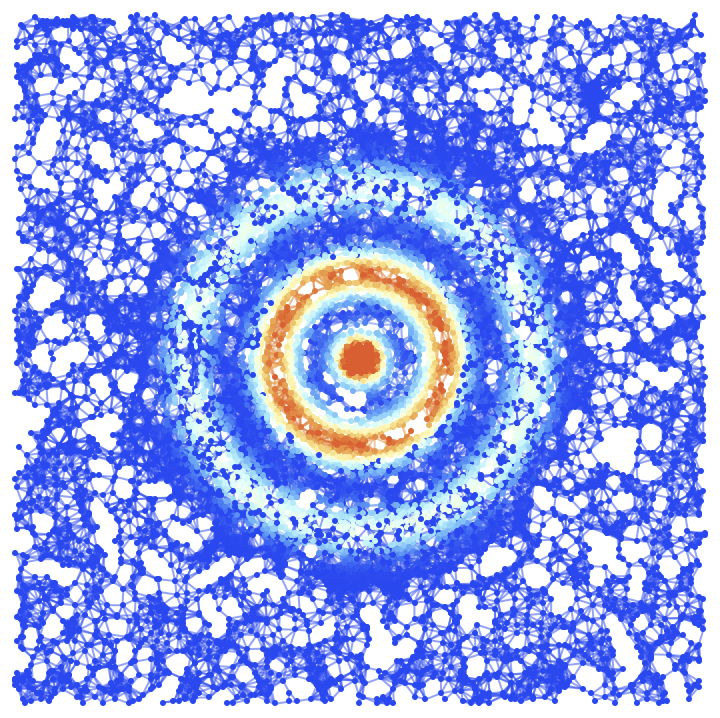}
\caption{On the left, the final (${t = 0.4}$) hypersurface configuration of the mildly-relativistic blast wave problem, embedded within a spherically-symmetric Minkowski geometry in spherical polar coordinates ${\left( t, r, \theta, \phi \right)}$, with a resolution of 10,000 hypergraph vertices (colored based on fluid density), and with spatial coordinate information assigned to the vertices. On the right, the final (${t = 0.4}$) hypersurface configuration of the mildly-relativistic blast wave problem, embedded within a rectangular Minkowski geometry in Cartesian coordinates ${\left( t, x, y, z \right)}$, with a resolution of 10,000 hypergraph vertices (colored based on fluid density), and with spatial coordinate information assigned to the vertices.}
\label{fig:Figure4}
\end{figure}

\begin{figure}[ht]
\centering
\includegraphics[width=0.295\textwidth]{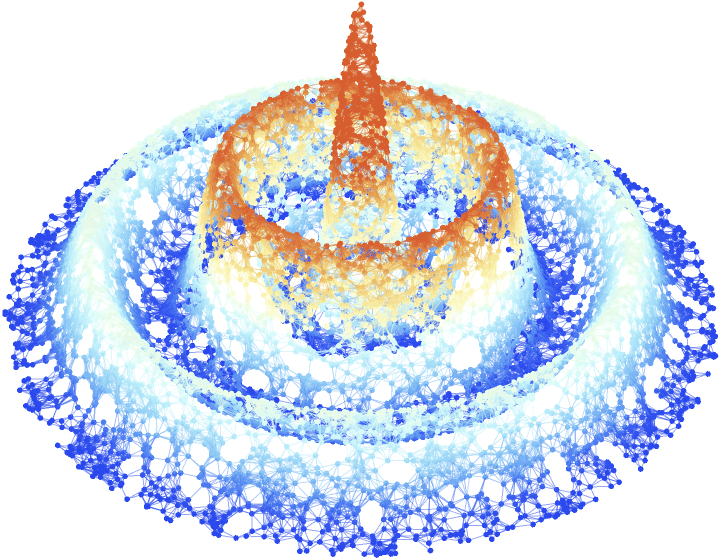}\hspace{0.2\textwidth}
\includegraphics[width=0.295\textwidth]{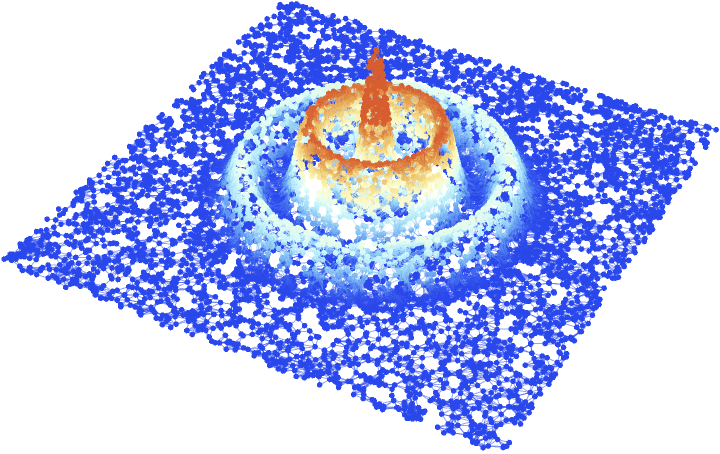}
\caption{On the left, the final (${t = 0.4}$) hypersurface configuration of the mildly-relativistic blast wave problem, embedded within a spherically-symmetric Minkowski geometry in spherical polar coordinates ${\left( t, r, \theta, \phi \right)}$, with a resolution of 10,000 hypergraph vertices (colored based on fluid density), and with both spatial coordinate and fluid density coordinate information assigned to the vertices. On the right, the final (${t = 0.4}$) hypersurface configuration of the mildly-relativistic blast wave problem, embedded within a rectangular Minkowski geometry in Cartesian coordinates ${\left( t, x, y, z \right)}$, with a resolution of 10,000 hypergraph vertices (colored based on fluid density), and with both spatial coordinate and fluid density coordinate information assigned to the vertices.}
\label{fig:Figure5}
\end{figure}

\begin{figure}[ht]
\centering
\includegraphics[width=0.295\textwidth]{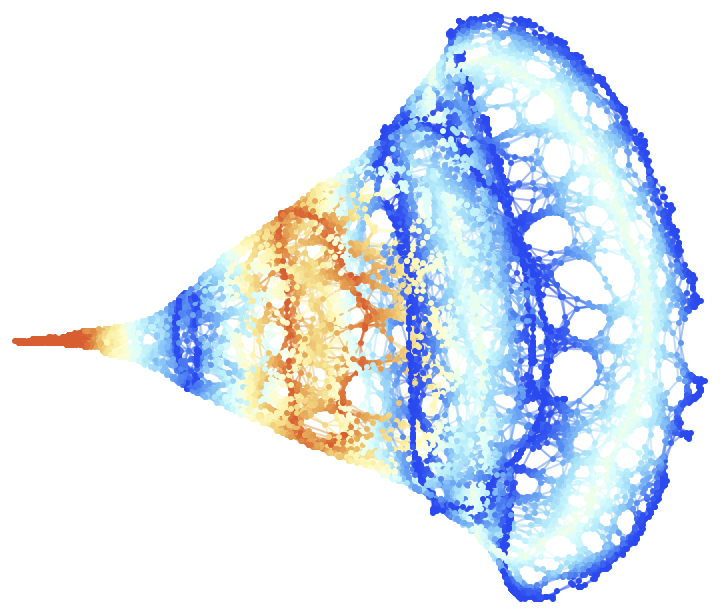}\hspace{0.2\textwidth}
\includegraphics[width=0.295\textwidth]{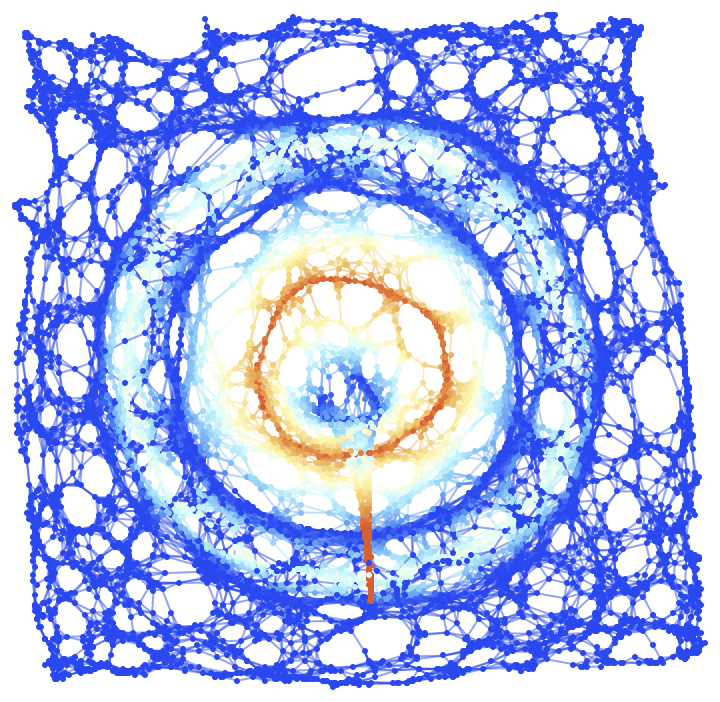}
\caption{On the left, the final (${t = 0.4}$) hypersurface configuration of the mildly-relativistic blast wave problem, embedded within a spherically-symmetric Minkowski geometry in spherical polar coordinates ${\left( t, r, \theta, \phi \right)}$, with a resolution of 10,000 hypergraph vertices (colored based on fluid density), and with no coordinate information assigned to the vertices. On the right, the final (${t = 0.4}$) hypersurface configuration of the mildly-relativistic blast wave problem, embedded within a rectangular Minkowski geometry in Cartesian coordinates ${\left( t, x, y, z \right)}$, with a resolution of 10,000 hypergraph vertices (colored based on fluid density), and with no coordinate information assigned to the vertices.}
\label{fig:Figure6}
\end{figure}

We validate against a standard one-dimensional special relativistic shock tube problem, namely the mildly-relativistic blast wave problem proposed by Donat, Font, Ib\'a\~nez and Marquina\cite{donat}, and we compare against the numerical solution of Del Zanna and Bucciantini\cite{delzanna}, and, since Riemann problems in one-dimensional special relativistic hydrodynamics admit exact solutions, we compare also against the exact solution derived by Pons, Mart\'i and M\"uller\cite{pons}. The fluid on the left-hand-side of the shock tube is of a high temperature and pressure, with ${\rho = 10}$ and ${P = 13.3}$; the fluid on the right-hand-side of the shock tube is of a low temperature and negligible pressure, with ${\rho = 1}$ and ${P = 0}$ (in some older papers using more unstable numerical methods, ${P = 10^{-6}}$ is chosen instead). Although this is a one-dimensional problem, since \textsc{Gravitas}'s hypergraph-based numerical algorithms work optimally in higher-dimensional geometries, we choose instead to evolve it as a three-dimensional problem in spherical symmetry, with the high temperature/high pressure fluid contained initially within a small spherical region in the center of the domain. It is then trivial to interpolate from a solution to this higher-dimensional spherically-symemtric problem back to a solution to the original one-dimensional Riemann problem. Since it will facilitate certain comparisons that we intend to perform later on in the case of general relativistic hydrodynamics in black hole spacetimes (which will be simulated in either spherical or axial symmetry), we choose to evolve this problem within both a spherically-symmetric Minkowksi spacetime parameterized by spherical polar coordinates ${\left( t, r, \theta, \phi \right)}$:

\begin{equation}
d s^2 = g_{\mu \nu} d x^{\mu} d x^{\nu} = - d t^2 + d r^2 + r^2 d \theta^2 + r^2 \sin^2 \left( \theta \right) d \phi^2,
\end{equation}
with the initial hypersurface geometry:

\begin{equation}
d l^2 = \gamma_{\mu \nu} d x^{\mu} d x^{\nu} = d r^2 + r^2 d \theta^2 + r^2 \sin^2 \left( \theta \right) d \phi^2,
\end{equation}
as well as within a standard rectangular Minkowski spacetime parameterized by Cartesian coordinates ${\left( t, x, y, z \right)}$:

\begin{equation}
d s^2 = g_{\mu \nu} d x^{\mu} d x^{\nu} = - d t^2 + d x^2 + d y^2 + d z^2, \qquad \text{ with } \qquad d l^2 = \gamma_{\mu \nu} d x^{\mu} dx^{\nu} = d x^2 + d y^2 + d z^2.
\end{equation}
We run all simulations with a hypergraph resolution of 10,000 vertices. Internally, \textsc{Gravitas} uses an adaptive fourth-order Runge-Kutta algorithm\cite{gorard8}\cite{gorard9} with a hypergraph rewriting/canonicalization algorithm based on \cite{gorard11}. An ideal gas equation of state with adiabatic exponent ${\Gamma = \frac{5}{3}}$ is assumed throughout. The initial (${t = 0}$) configurations of the domains in both cases are shown in Figure \ref{fig:Figure1}, with vertices colored based on fluid density and with vertex coordinates assigned using a two-dimensional projection of the spatial coordinates, yielding two-dimensional visualizations of the respective simulation domains. We can construct three-dimensional visualizations of the domains by also assigning a third vertex coordinate based on the fluid density, as shown in Figure \ref{fig:Figure2}. Finally, to give an indication of the hypergraph topology produced by \textsc{Gravitas}'s adaptive hypergraph refinement algorithm (applied here as a preconditioning step for the initial data), we show the initial hypergraphs without any vertex coordinate information assigned in Figure \ref{fig:Figure3}. We shall make use of all three modes of visualization throughout the remainder of this article. The exact solution to the mildly-relativistic blast wave Riemann problem is known to consist of three waves: a slow-moving rarefaction wave, a contact discontinuity, and a fast-moving shock wave. In Figures \ref{fig:Figure4}, \ref{fig:Figure5} and \ref{fig:Figure6} (showing the solution at coordinate time ${t = 0.4}$ with two-dimensional spatial coordinates, three-dimensional spatial and fluid density coordinates, and no vertex coordinates, respectively), we see that all three waves are resolved correctly in both the spherically-symmetric and rectangular cases.

\clearpage

\section{Black Hole Accretion in Discrete Spacetime}
\label{sec:Section4}

Following the recent historical exposition of Aguayo-Ortiz, Tejeda, Sarbach and L\'opez-C\'amara\cite{aguayoortiz}, we note that Bondi\cite{bondi2} originally analyzed the case of an infinite, spherically-symmetric distribution of ideal gas with adiabatic exponent ${\Gamma}$, with initially uniform density ${\rho}$ and pressure $P$, accreting radially onto a compact object of mass $M$ in Newtonian gravity. Under these assumptions, the steady-state accretion flow in spherical polar coordinates ${\left( t, r, \theta, \phi \right)}$ is governed by the continuity and radial Euler equations, namely:

\begin{equation}
\frac{1}{r^2} \frac{d}{d r} \left( r^2 \rho v_r \right) = 0, \qquad \text{ and } \qquad v_r \frac{d v_r}{d r} + \frac{1}{\rho} \frac{d P}{d r} + \frac{M}{r^2} = 0,
\end{equation}
respectively, where ${v_r}$ designates the radial component of the fluid velocity, i.e:

\begin{equation}
v_r = \left\lvert \frac{d r}{d t} \right\rvert.
\end{equation}
The initial conditions of the fluid distribution are given in terms of its density and pressure at infinite radial distance, i.e. ${\rho_{\infty}}$ and ${P_{\infty}}$, respectively, with the fluid assumed to be at rest at infinite radial distance (i.e. ${\left( v_r \right)_{\infty} = 0}$) and with the (local) adiabatic speed of sound ${c_s}$ given by the following partial derivative assuming fixed internal energy ${\varepsilon}$:

\begin{equation}
c_s = \sqrt{\left. \frac{\partial P}{\partial \rho} \right\rvert_{\varepsilon}}.
\end{equation}
Subject to the additional assumption that the flow is transonic, i.e. that there exists a radius ${r = r_{trans}}$ such that the radial velocity ${v_r = v_{trans}}$ is equal to the (local) adiabatic speed of sound, which can be calculated to be:

\begin{equation}
r_{trans} = \frac{M}{2 v_{trans}^2{}}, \qquad \text{ where } \qquad v_{trans} = c_s = c_{\infty} \sqrt{\frac{2}{5 - 3 \Gamma}},
\end{equation}
and where ${c_{\infty}}$ is the adiabatic speed of sound at infinite radial distance, Bondi showed that there exists a unique analytic solution to the continuity and radial Euler equations which maximizes the rate at which mass is accreted onto the compact object, namely:

\begin{equation}
\frac{d M}{d t} = \pi \left( \frac{2}{5 - 3 \Gamma} \right)^{\frac{5 - 3 \Gamma}{2 \left( \Gamma - 1 \right)}} M^2 \frac{\rho_{\infty}}{c_{\infty}^{3}}.
\end{equation}

Bondi's purely Newtonian analysis was subsequently extended by Michel\cite{michel} to the case of an infinite, spherically-symmetric distribution of ideal gas, again with an initially uniform density ${\rho}$ and pressure $P$, accreting radially onto a static, uncharged and non-rotating black hole (as described by a Schwarzschild geometry) of mass $M$ in general relativity. The Schwarzschild metric is taken to be given in Schwarzschild/spherical polar coordinates ${\left( t, r, \theta, \phi \right)}$ by\cite{schwarzschild}\cite{droste}:

\begin{equation}
d s^2 = g_{\mu \nu} d x^{\mu} d x^{\nu} = - \left( 1 - \frac{2 M}{r} \right) d t^2 + \left( 1 - \frac{2 M}{r} \right)^{-1} d r^2 + r^2 \left( d \theta^2 + \sin^2 \left( \theta \right) d \phi^2 \right),
\end{equation}
and, within such a spacetime, the spherical symmetry of the problem allows us to reformulate the energy-momentum conservation equations as a single (radial) ordinary differential equation:

\begin{equation}
\nabla_{\nu} T^{\mu \nu} = \frac{\partial}{\partial x^{\nu}} \left( T^{\mu \nu} \right) + \Gamma_{\nu \sigma}^{\mu} T^{\sigma \nu} + \Gamma_{\nu \sigma}^{\nu} T^{\mu \sigma} = 0, \qquad \implies \qquad \frac{d}{d r} \left( r^2 \rho h u_t u^r \right) = 0
\end{equation}
and the baryon number continuity equation may be reformulated similarly:

\begin{equation}
\nabla_{\mu} \left( \rho u^{\mu} \right) = \frac{\partial}{\partial x^{\mu}} \left( \rho u^{\mu} \right) + \Gamma_{\mu \sigma}^{\mu} \left( \rho u^{\sigma} \right) = 0, \qquad \implies \qquad \frac{d}{d r} \left( r^2 \rho u^r \right) = 0,
\end{equation}
where $h$ is, as usual, the specific relativistic enthalpy of the fluid:

\begin{equation}
h = 1 + \varepsilon \left( \rho, P \right) + \frac{P}{\rho}.
\end{equation}
In all other respects, the specification of the initial conditions of the fluid in terms of its density ${\rho_{\infty}}$, pressure ${P_{\infty}}$ and specific relativistic enthalpy ${h_{\infty}}$ at infinite radial distance (with the fluid again assumed to be at rest at this point, i.e. ${\left( u_r \right)_{\infty} = 0}$) is directly analogous to the Newtonian case, with the (local) adiabatic speed of sound ${c_s}$ now given by the following partial derivative, assuming a fixed hydrostatic pressure $P$:

\begin{equation}
c_s = \sqrt{\frac{\rho}{h} \left( \left. \frac{\partial h}{\partial \rho} \right) \right\rvert_{P}}.
\end{equation}
Once again, we make the assumption that the flow is transonic, with the radial velocity ${u_r = \left( u_r \right)_{trans}}$ at the transonic radius ${r = r_{trans}}$ now being such that the norm of the spatial velocity of the fluid is measured by any local static observer as being equal to the (local) adiabatic speed of sound:

\begin{equation}
r_{trans} = \frac{M}{2 \left( u_r \right)_{trans}}, \qquad \text{ where } \qquad \left( u_r \right)_{trans} = \sqrt{\frac{\frac{1}{3} \left( \frac{h_{trans}^{2}}{h_{\infty}^{2}} - 1 \right)}{\left( \frac{h_{trans}^{2}}{h_{\infty}^{2}} \right)}},
\end{equation}
where the transonic value of the specific relativistic enthalpy ${h_{trans}}$ is calculated as:

\begin{equation}
h_{trans} = 2 h_{\infty} \sqrt{\Gamma - \frac{2}{3}} \sin \left( \frac{1}{3} \arccos \left( \frac{3 \left( \Gamma - 1 \right)}{2 h_{\infty} \left( \sqrt{\Gamma - \frac{2}{3}} \right)^3} \right) \right).
\end{equation}
Once again, there exists a unique analytic solution to the energy-momentum and baryonic number conservation equations within this setting, such that the flow satisfies the steady-state condition and regularity across the event horizon of the black hole is preserved (as proved by Chaverra and Sarbach\cite{chaverra}, and later Chaverra, Mach and Sarbach\cite{chaverra2}), with the accretion rate of mass onto the black hole given by:

\begin{equation}
\frac{d M}{d t} = \pi \left( \frac{h_{trans}}{h_{\infty}} \right)^{\frac{3 \Gamma - 2}{\Gamma - 1}} \left( \frac{\sqrt{\frac{1}{3} \left( \frac{h_{trans}^2}{h_{\infty}^{2}} - 1 \right)^2}}{c_{\infty}} \right)^{\frac{5 - 3 \Gamma}{\Gamma - 1}},
\end{equation}
where ${c_{\infty}}$ is, again, the adiabatic speed of sound at infinite radial distance.

\begin{figure}[ht]
\centering
\includegraphics[width=0.295\textwidth]{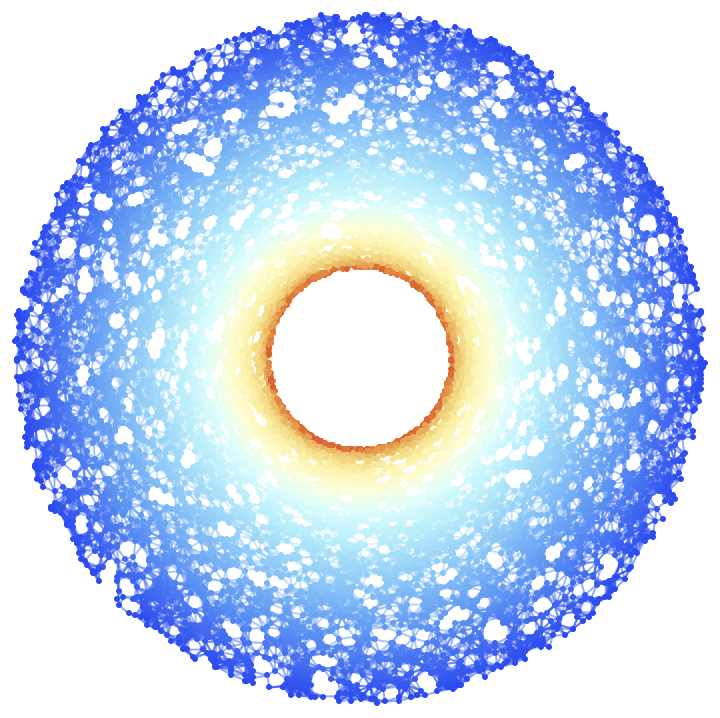}\hspace{0.2\textwidth}
\includegraphics[width=0.295\textwidth]{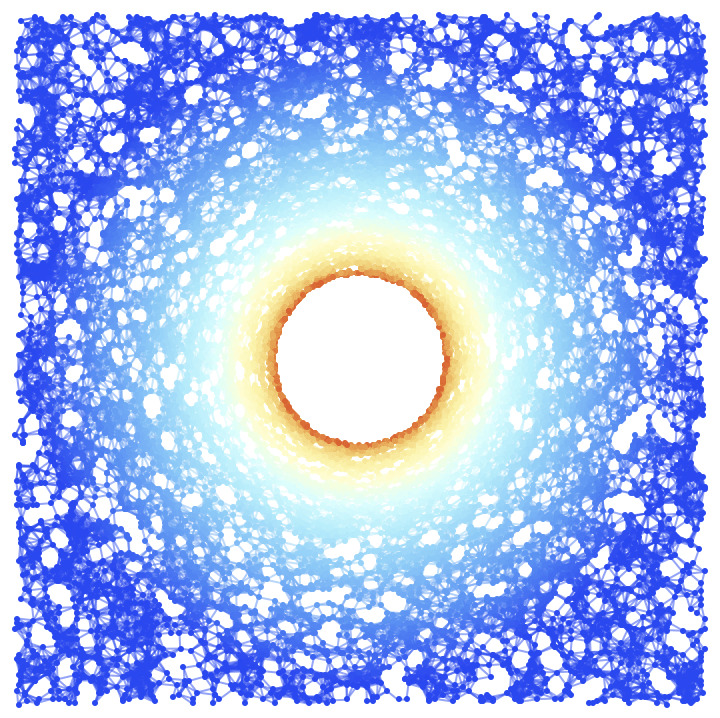}
\caption{On the left, the initial (${t = 0}$) hypersurface configuration for an uncharged, non-rotating black hole in Schwarzschild/spherical polar coordinates ${\left( t, r, \theta, \phi \right)}$, with a resolution of 10,000 hypergraph vertices (colored based on extrinsic curvature), and with spatial coordinate information assigned to the vertices. On the right, the initial (${t = 0}$) hypersurface configuration for an uncharged, non-rotating black hole in Kerr-Schild/Cartesian coordinates ${\left( t, x, y, z \right)}$, with a resolution of 10,000 hypergraph vertices (colored based on extrinsic curvature), and with spatial coordinate information assigned to the vertices.}
\label{fig:Figure7}
\end{figure}

\begin{figure}[ht]
\centering
\includegraphics[width=0.295\textwidth]{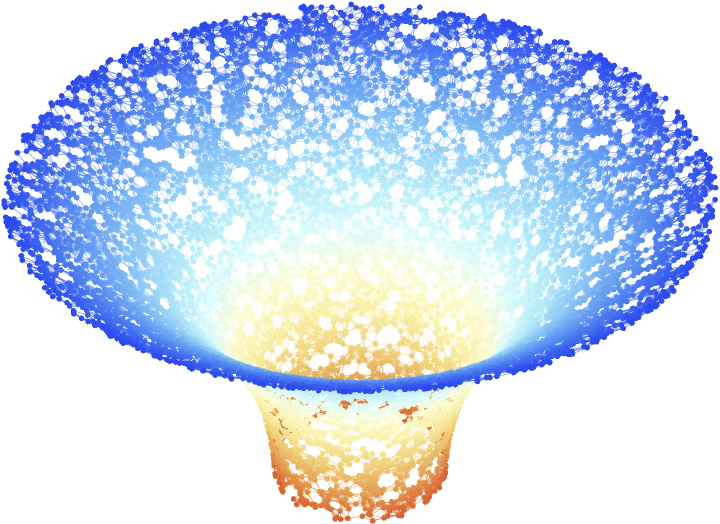}\hspace{0.2\textwidth}
\includegraphics[width=0.295\textwidth]{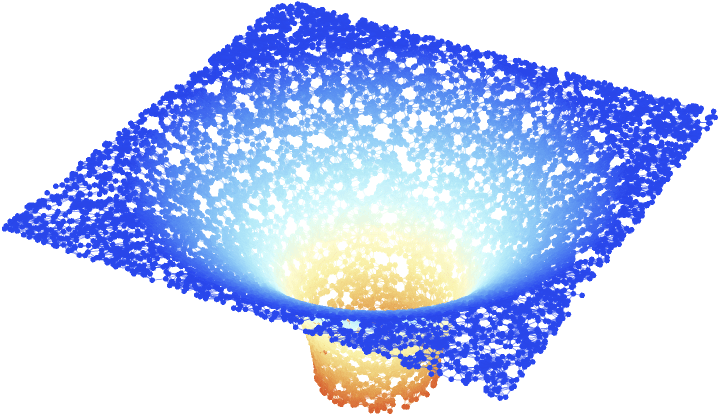}
\caption{On the left, the initial (${t = 0}$) hypersurface configuration for an uncharged, non-rotating black hole in Schwarzschild/spherical polar coordinates ${\left( t, r, \theta, \phi \right)}$, with a resolution of 10,000 hypergraph vertices (colored based on extrinsic curvature), and with both spatial coordinate and fluid density coordinate information assigned to the vertices. On the right, the initial (${t = 0}$) hypersurface configuration for an uncharged, non-rotating black hole in Kerr-Schild/Cartesian coordinates ${\left( t, x, y, z \right)}$, with a resolution of 10,000 hypergraph vertices (colored based on extrinsic curvature), and with both spatial coordinate and fluid density coordinate information assigned to the vertices.}
\label{fig:Figure8}
\end{figure}

\begin{figure}[ht]
\centering
\includegraphics[width=0.295\textwidth]{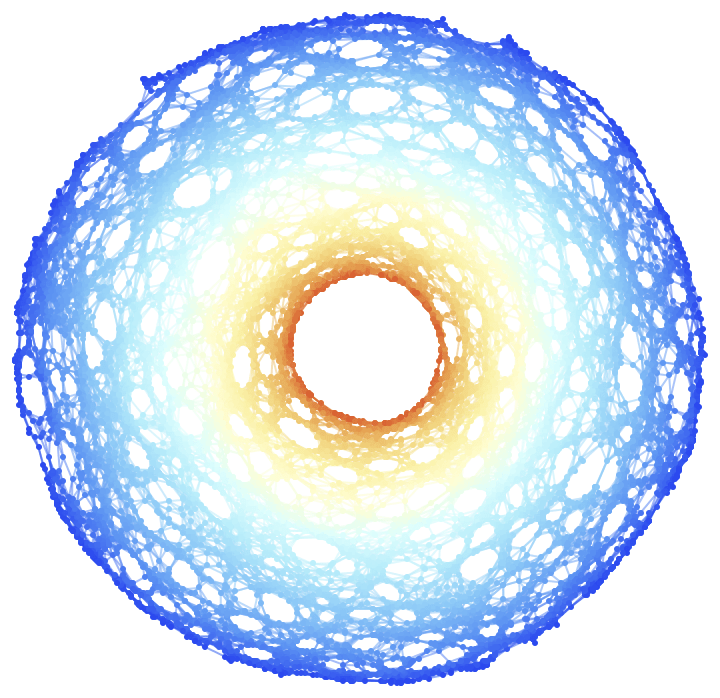}\hspace{0.2\textwidth}
\includegraphics[width=0.295\textwidth]{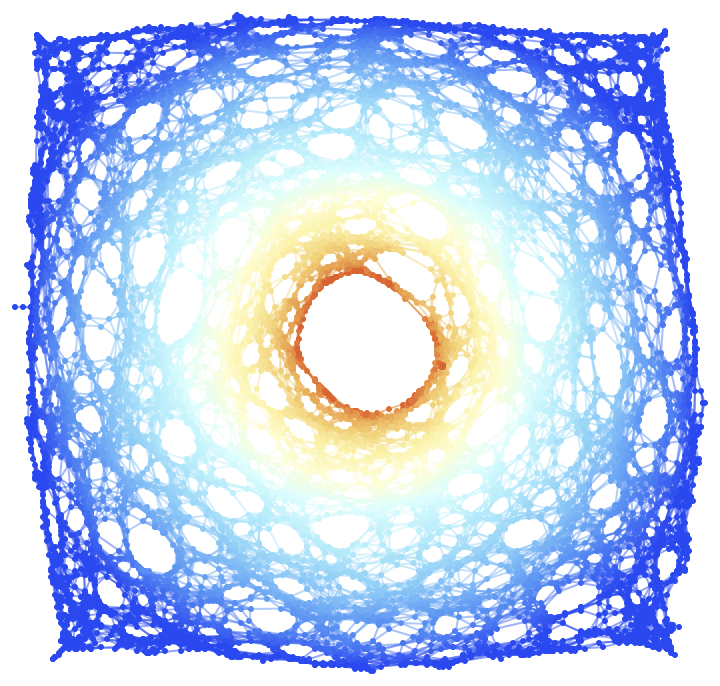}
\caption{On the left, the initial (${t = 0 }$) hypersurface configuration for an uncharged, non-rotating black hole in Schwarzschild/spherical polar coordinates ${\left( t, r, \theta, \phi \right)}$, with a resolution of 10,000 hypergraph vertices (colored based on extrinsic curvature), and with no coordinate information assigned to the vertices. On the right, the initial (${t = 0}$) hypersurface configuration for an uncharged, non-rotating black hole in Kerr-Schild/Cartesian coordinates ${\left( t, x, y, z \right)}$, with a resolution of 10,000 hypergraph vertices (colored based on extrinsic curvature), and with no coordinate information assigned to the vertices.}
\label{fig:Figure9}
\end{figure}

Font, Ib\'a\~nez and Papadopoulos\cite{font5} have stressed the importance of using ``horizon-adapted'' coordinates, i.e. coordinate systems (such as Kerr-Schild coordinates) which remain regular at the black hole event horizon, when performing black hole accretion studies, so as to avoid unphysical fluid behavior near the horizon resulting from coordinate divergences. For this reason, we shall perform our radial accretion simulations onto static, uncharged, non-rotating black holes (i.e. Schwarzschild black holes) expressed in both the Schwarzschild/spherical polar coordinate system ${\left( t, r, \theta, \phi \right)}$, which is not horizon-adapted, with the metric of the initial spacelike hypersurface given by:

\begin{equation}
d l^2 = \gamma_{\mu \nu} d x^{\mu} d x^{\nu} = \left( 1 - \frac{2 M}{r} \right)^{-1} d r^2 + r^2 \left( d \theta^2 + sin^2 \left( \theta \right) d \phi^2 \right),
\end{equation}
and the Kerr-Schild/Cartesian coordinate system ${\left( t, x, y, z \right)}$\cite{kerr}, which \textit{is} horizon-adapted, with the metric of the initial spacelike hypersurface given by:

\begin{equation}
d l^2 = \gamma_{\mu \nu} d x^{\mu} d x^{\nu} = d x^2 + d y^2 + d z^2 + \mathcal{F} \left( l_{\mu} d x^{\mu} \right)^2,
\end{equation}
where:

\begin{equation}
\mathcal{F} = \frac{2 M}{r}, \qquad \text{ and } \qquad l_{\mu} d x^{\mu} = \frac{x}{r} d x + \frac{y}{r} d y + \frac{z}{r} d z,
\end{equation}
and with $r$ being the usual radial coordinate:

\begin{equation}
r = \sqrt{x^2 + y^2 + z^2},
\end{equation}
in order that we be able to compare the hydrodynamic results obtained across the two coordinate schemes. As in the special relativistic hydrodynamics cases from before, we run all simulations presented in this section with a hypergraph resolution of 10,000 vertices, using an ideal gas equation of state:

\begin{equation}
h = 1 + \frac{P}{\rho} \left( \frac{\Gamma}{\Gamma - 1} \right), \qquad \text{ and } \qquad c_s = \sqrt{\frac{\Gamma P}{\rho \left( 1 + \left( \frac{P}{\rho} \right) \left( \frac{\Gamma}{\Gamma - 1} \right) \right)}},
\end{equation}
with adiabatic exponent ${\Gamma = \frac{5}{3}}$. The geometries of the initial (${t = 0}$) hypersurface configurations in both the Schwarzschild/spherical polar coordinate system and the Kerr-Schild/Cartesian coordinate system are shown in Figure \ref{fig:Figure7}, with vertices colored based on the value of the extrinsic curvature tensor, and with vertex coordinates assigned using a two-dimensional spatial projection. Three-dimensional visualizations are shown in Figure \ref{fig:Figure8}, with the third vertex coordinate assigned based on extrinsic curvature. Finally, coordinate-free representations of the pure hypergraph topologies are shown in Figure \ref{fig:Figure9}.

\begin{figure}[ht]
\centering
\includegraphics[width=0.295\textwidth]{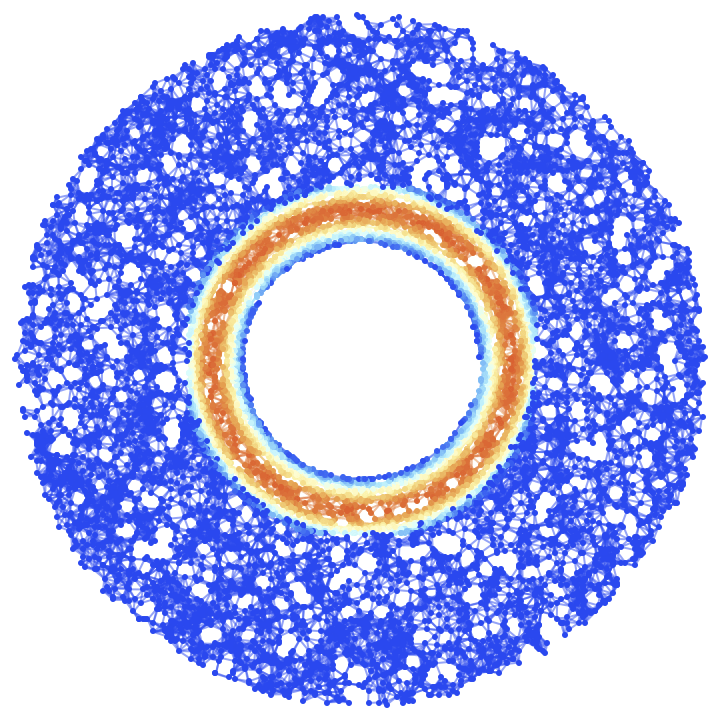}\hspace{0.2\textwidth}
\includegraphics[width=0.295\textwidth]{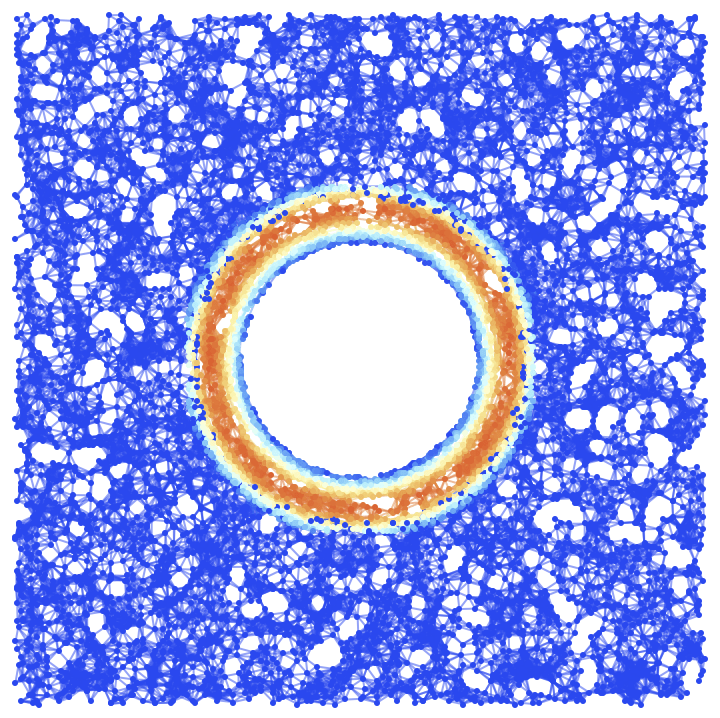}
\caption{On the left, the final (${t = 100 M}$) hypersurface configuration for the radial accretion of an initially spherically-symmetric fluid distribution onto an uncharged, non-rotating black hole in Schwarzschild/spherical polar coordinates ${\left( t, r, \theta, \phi \right)}$, with a resolution of 10,000 hypergraph vertices (colored based on fluid density), and with spatial coordinate information assigned to the vertices. On the right, the final (${t = 100 M}$) hypersurface configuration for the radial accretion of an initially spherically-symmetric fluid distribution onto an uncharged, non-rotating black hole in Kerr-Schild/Cartesian coordinates ${\left( t, x, y, z \right)}$, with a resolution of 10,000 hypergraph vertices (colored based on fluid density), and with spatial coordinate information assigned to the vertices.}
\label{fig:Figure10}
\end{figure}

\begin{figure}[ht]
\centering
\includegraphics[width=0.295\textwidth]{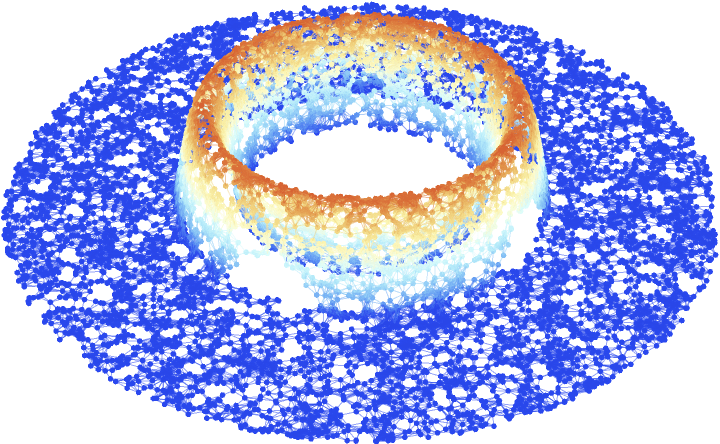}\hspace{0.2\textwidth}
\includegraphics[width=0.295\textwidth]{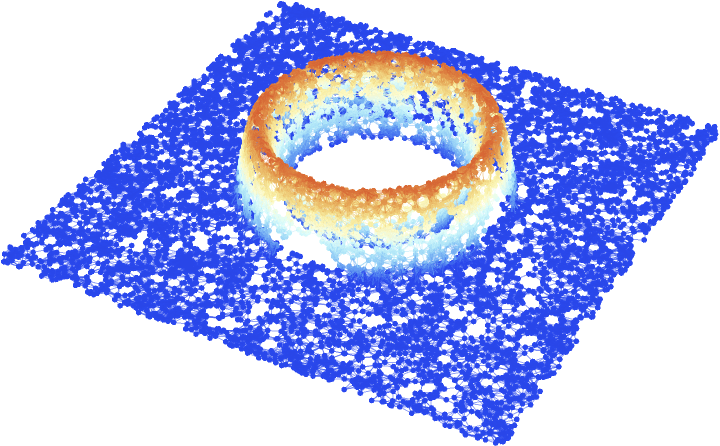}
\caption{On the left, the final (${t = 100 M}$) hypersurface configuration for the radial accretion of an initially spherically-symmetric fluid distribution onto an uncharged, non-rotating black hole in Schwarzschild/spherical polar coordinates ${\left( t, r, \theta, \phi \right)}$, with a resolution of 10,000 hypergraph vertices (colored based on fluid density), and with both spatial coordinate and fluid density coordinate information assigned to the vertices. On the right, the final (${t = 100 M}$) hypersurface configuration for the radial accretion of an initially spherically-symmetric fluid distribution onto an uncharged, non-rotating black hole in Kerr-Schild/Cartesian coordinates ${\left( t, x, y, z \right)}$, with a resolution of 10,000 hypergraph vertices (colored based on fluid density), and with both spatial coordinate and fluid density coordinate information assigned to the vertices.}
\label{fig:Figure11}
\end{figure}

\begin{figure}[ht]
\centering
\includegraphics[width=0.295\textwidth]{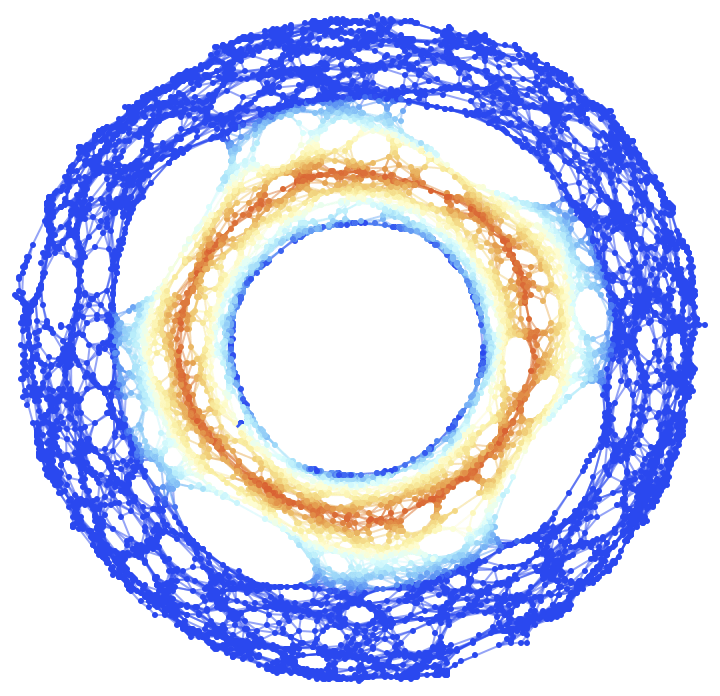}\hspace{0.2\textwidth}
\includegraphics[width=0.295\textwidth]{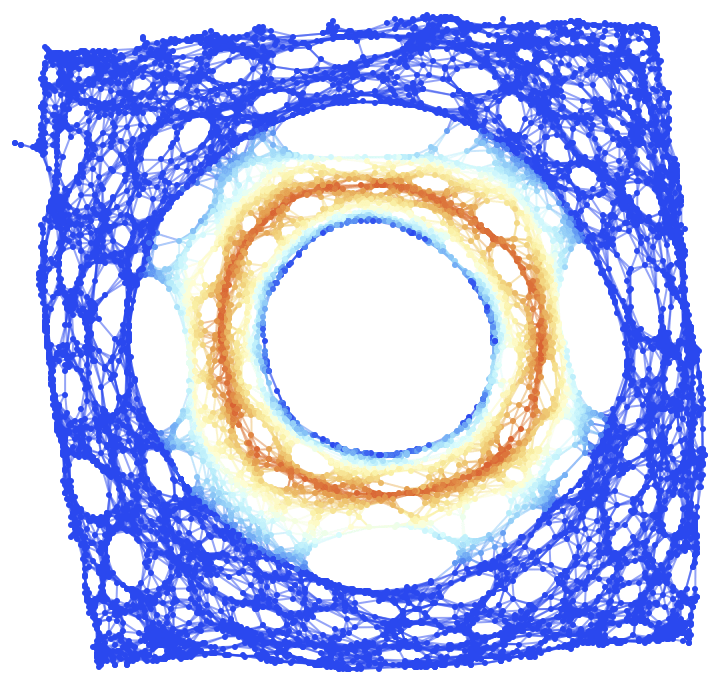}
\caption{On the left, the final (${t = 100 M}$) hypersurface configuration for the radial accretion of an initially spherically-symmetric fluid distribution onto an uncharged, non-rotating black hole in Schwarzschild/spherical polar coordinates ${\left( t, r, \theta, \phi \right)}$, with a resolution of 10,000 hypergraph vertices (colored based on fluid density), and with no coordinate information assigned to the vertices. On the right, the final (${t = 100 M}$) hypersurface configuration for the radial accretion of an initially spherically-symmetric fluid distribution onto an uncharged, non-rotating black hole in Kerr-Schild/Cartesian coordinates ${\left( t, x, y, z \right)}$, with a resolution of 10,000 hypergraph vertices (colored based on fluid density), and with no coordinate information assigned to the vertices.}
\label{fig:Figure12}
\end{figure}

In order to evolve this spatial metric (together with the fluid variables defined on top of it) forwards in time using \textsc{Gravitas}, we must first select an appropriate set of gauge conditions. For the lapse function ${\alpha}$, we choose to use the maximal slicing condition initially developed by Lichnerowicz\cite{lichnerowicz} and later developed into a directly usable form by York\cite{york}:

\begin{equation}
{}^{\left( 3 \right)} \Delta \alpha = \alpha K^{\mu \nu} K_{\mu \nu} - \frac{\partial}{\partial t} \left( K \right),
\end{equation}
where ${{}^{\left( 3 \right)} \Delta}$ denotes the connection Laplacian on spacelike hypersurfaces, defined for arbitrary scalar fields ${\phi}$ as:

\begin{equation}
{}^{\left( 3 \right)} \Delta \phi = {}^{\left( 3 \right)} \nabla^{\mu} \left( {}^{\left( 3 \right)} \nabla_{\mu} \phi \right) = \gamma^{\mu \sigma} {}^{\left( 3 \right)} \nabla_{\sigma} \left( {}^{\left( 3 \right)} \nabla_{\mu} \phi \right) = \gamma^{\mu \sigma} \left( \frac{\partial}{\partial x^{\sigma}} \left( \frac{\partial}{\partial x^{\mu}} \left( \phi \right) \right) - {}^{\left( 3 \right)} \Gamma_{\sigma \mu}^{\lambda} \left( \frac{\partial}{\partial x^{\lambda}} \left( \phi \right) \right) \right),
\end{equation}
which, using the contraction properties of the (spatial) Christoffel symbols ${{}^{\left( 3 \right)} \Gamma_{\sigma \mu}^{\lambda}}$, becomes:

\begin{equation}
{}^{\left( 3 \right)} \Delta \phi = \frac{1}{\sqrt{\det \left( \gamma_{\mu \nu} \right)}} \left( \frac{\partial}{\partial x^{\mu}} \left( \sqrt{\det \left( \gamma_{\mu \nu} \right)} \left( \gamma^{\mu \nu} \frac{\partial}{\partial x^{\nu}} \left( \phi \right) \right) \right) \right),
\end{equation}
leading to the following explicit form of the maximal slicing condition, with the lapse function ${\alpha}$ being treated as a scalar field defined over spacelike hypersurfaces:

\begin{equation}
\frac{1}{\sqrt{\det \left( \gamma_{\mu \nu} \right)}} \left( \frac{\partial}{\partial x^{\mu}} \left( \sqrt{\det \left( \gamma_{\mu \nu} \right)} \left( \gamma^{\mu \nu} \frac{\partial}{\partial x^{\mu}} \left( \alpha \right) \right) \right) \right) = \alpha K^{\mu \nu} K_{\mu \nu} - \frac{\partial}{\partial t} \left( K \right).
\end{equation}
Note, as before, that the indices of the extrinsic curvature tensor ${K_{\mu \nu}}$ are raised and lowered using the spatial metric tensor ${\gamma_{\mu \nu}}$, and so, in particular, for the contravariant form ${K^{\mu \nu}}$, one has:

\begin{equation}
K^{\mu \nu} = \gamma^{\mu \sigma} K_{\sigma}^{\nu} = \gamma^{\sigma \nu} K_{\sigma}^{\mu} = \gamma^{\mu \sigma} \gamma^{\lambda \nu} K_{\sigma \lambda}.
\end{equation}
The maximal slicing condition seeks to maximize the spatial volume of each spacelike hypersurface by reducing the evolution rate in high-curvature regions and increasing it in low-curvature regions, thus equipping it with highly favorable singularity-avoidance properties that make it ideal for simulating black hole spacetimes. For the shift vector ${\boldsymbol\beta}$, we choose to use the minimal distortion coordinate conditions of Smarr and York\cite{smarr}, subsequently adapted into the form employed here by Brady, Creighton and Thorne\cite{brady}:

\begin{multline}
{}^{\left( 3 \right)} \nabla^{\mu} \left( {}^{\left( 3 \right)} \nabla_{\mu} \beta^{\nu} \right) + {}^{\left( 3 \right)} \nabla^{\nu} \left( {}^{\left( 3 \right)} \nabla_{\mu} \beta^{\mu} \right) - 2 {}^{\left( 3 \right)} \nabla_{\mu} \left( \alpha K^{\mu \nu} \right)\\
= \gamma^{\mu \sigma} {}^{\left( 3 \right)} \nabla_{\sigma} \left( {}^{\left( 3 \right)} \nabla_{\mu} \beta^{\nu} \right) + \gamma^{\nu \sigma} {}^{\left( 3 \right)} \nabla_{\sigma} \left( {}^{\left( 3 \right)} \nabla_{\mu} \beta^{\mu} \right) - 2 {}^{\left( 3 \right)} \nabla_{\mu} \left( \alpha K^{\mu \nu} \right) = 0,
\end{multline}
which expands out to give:

\begin{multline}
\gamma^{\mu \sigma} \left( \frac{\partial}{\partial x^{\sigma}} \left( D_{\mu}^{\nu} \right) + {}^{\left( 3 \right)} \Gamma_{\sigma \lambda}^{\nu} D_{\mu}^{\lambda} - {}^{\left( 3 \right)} \Gamma_{\sigma \mu}^{\lambda} D_{\lambda}^{\mu} \right) + \gamma^{\nu \sigma} \left( \frac{\partial}{\partial x^{\sigma}} \left( D_{\mu}^{\mu} \right) + {}^{\left( 3 \right)} \Gamma_{\sigma \lambda}^{\mu} D_{\mu}^{\lambda} - {}^{\left( 3 \right)} \Gamma_{\sigma \mu}^{\lambda} D_{\lambda}^{\mu} \right)\\
- 2 \left( \frac{\partial}{\partial x^{\mu}} \left( \alpha K^{\mu \nu} \right) + {}^{\left( 3 \right)} \Gamma_{\mu \sigma}^{\mu} \left( \alpha K^{\sigma \nu} \right) + {}^{\left( 3 \right)} \Gamma_{\mu \sigma}^{\nu} \left( \alpha K^{\mu \sigma} \right) \right) = 0,
\end{multline}
with the rank-2 tensor ${D_{\mu}^{\nu}}$ consisting of (spatial) covariant derivatives of the shift vector components ${\beta^{\nu}}$:

\begin{equation}
D_{\mu}^{\nu} = {}^{\left( 3 \right)} \nabla_{\nu} \beta^{\mu} = \frac{\partial}{\partial x^{\mu}} \left( \beta^{\nu} \right) + {}^{\left( 3 \right)} \Gamma_{\mu \sigma}^{\nu} \beta^{\sigma}.
\end{equation}
The minimal distortion coordinate conditions seek to minimize the distortion (or ``strain'') in the spatial coordinates as one evolves from one hypersurface to the next, which makes it preferable for the case of hydrodynamics simulations, in which one generally wishes for the spatial coordinate system within which the fluid is evolved to remain as consistent as possible between time steps. In all of the above, ${\mu, \nu, \sigma, \lambda}$ range across spatial coordinate indices ${\left\lbrace 0, \dots, n - 2 \right\rbrace}$ only. We initialize our simulation with the (non-dimensional) gas temperature at infinite radial distance set to be ${\Theta_{\infty} = 0.1}$, from which the density and pressure at infinite radial distance (i.e. ${\rho_{\infty}}$ and ${P_{\infty}}$, respectively) can be computed using the relation ${\Theta_{\infty} = \frac{P_{\infty}}{\rho_{\infty}}}$. In Figures \ref{fig:Figure10}, \ref{fig:Figure11} and \ref{fig:Figure12} (showing the solution at coordinate time ${t = 100 M}$ with two-dimensional spatial coordinates, three-dimensional spatial and fluid density coordinates, and no vertex coordinates, respectively), we see that the fluid eventually evolves to a steady-state configuration with a high-density spherical accretion region surrounding the black hole event horizon, with no substantive difference between the solutions seen in the Schwarzschild/spherical polar and Kerr-Schild/Cartesian coordinate systems. The rates of mass/energy accretion onto the black hole are found to be slightly lower than in the analytic solution of Michel\cite{michel}, with this discrepancy vanishing in the limit as the discretization scale goes to zero.

The analyses of Bondi\cite{bondi2} and Michel\cite{michel} described above both made extensive use of the spherical symmetry of the accretion problem. However, the treatment of an infinite, initially spherically-symmetric distribution of ideal gas, with initially uniform density ${\rho}$ and pressure $P$, accreting radially onto an uncharged but spinning black hole (as described by a Kerr geometry) of mass $M$ and spin $J$ in full general relativity requires breaking this spherical symmetry, and replacing it with a more general axially-symmetric spacetime geometry. The Kerr metric\cite{kerr2} is taken to be given in Boyer-Lindquist/oblate spheroidal coordinates\cite{boyer} ${\left( t, r, \theta, \phi \right)}$ by:

\begin{multline}
d s^2 = g_{\mu \nu} d x^{\mu} d x^{\nu} = - \left( 1 - \frac{2 M}{\left( r^2 + \left( \frac{J}{M} \right)^2 \cos^2 \left( \theta \right) \right)} \right) d t^2 + \left( \frac{r^2 + \left( \frac{J}{M} \right)^2 \cos^2 \left( \theta \right)}{r^2 - 2 M + \left( \frac{J}{M} \right)^2} \right) d r^2\\
+ \left( r^2 + \left( \frac{J}{M} \right)^2 \cos^2 \left( \theta \right) \right) d \theta^2 + \left( r^2 + \left( \frac{J}{M} \right)^2 + \frac{2 J^2 \sin^2 \left( \theta \right)}{M \left( r^2 + \left( \frac{J}{M} \right)^2 \cos^2 \left( \theta \right) \right)} \right) \sin^2 \left( \theta \right) d \phi^2\\
- \left( \frac{4 J \sin^2 \left( \theta \right)}{r^2 + \left( \frac{J}{M} \right)^2 \cos^2 \left( \theta \right)} \right) d t d \phi,
\end{multline}
and, following Petrich, Shapiro and Teukolsky\cite{petrich}, we assume an ultra-relativistic equation of state in which the rest-mass energy of the fluid is negligible when compared to its internal energy:

\begin{equation}
P = \left( \Gamma - 1 \right) \rho,
\end{equation}
and we assume, moreover, that the fluid is \textit{stiff} in the sense that ${\Gamma = 2}$ and therefore ${P = \rho}$ identically. Subject to the additional assumption that the flow obtains a steady-state configuration that is non-rotational, the resulting fluid equations can be expressed purely in terms of the gradient of a certain stream function (or scalar potential) ${\Phi}$:

\begin{equation}
h u_{\mu} = \frac{\partial}{\partial x^{\mu}} \left( \Phi \right),
\end{equation}
which, due to the normalization convention ${u_{\mu} u^{\mu} = 1}$ for the spacetime velocity vector ${\mathbf{u}}$, implies that the specific relativistic enthalpy $h$ can be written purely in terms of the stream function gradient:

\begin{equation}
h = \sqrt{- \frac{\partial}{\partial x^{\mu}} \left( \Phi \right) g^{\mu \sigma} \frac{\partial}{\partial x^{\sigma}} \left( \Phi \right)}.
\end{equation}
Substituting the stream function gradient equation into the baryon number conservation equation:

\begin{equation}
{}^{\left( 4 \right)} \nabla_{\mu} \left( \rho u^{\mu} \right) = \frac{\partial}{\partial x^{\mu}} \left( \rho u^{\mu} \right) + {}^{\left( 4 \right)} \Gamma_{\mu \sigma}^{\mu} \left( \rho u^{\sigma} \right) = 0,
\end{equation}
yields the following harmonic equation for ${\Phi}$:

\begin{equation}
{}^{\left( 4 \right)} \Delta \Phi = {}^{\left( 4 \right)} \nabla^{\mu} \left( {}^{\left( 4 \right)} \nabla_{\mu} \Phi \right) = g^{\mu \sigma} {}^{\left( 4 \right)} \nabla_{\sigma} \left( {}^{\left( 4 \right)} \nabla_{\mu} \Phi \right) = 0,
\end{equation}
where ${{}^{\left( 4 \right)} \Delta}$ denotes the connection Laplacian on spacetime, which expands out to give:

\begin{equation}
g^{\mu \sigma} \left( \frac{\partial}{\partial x^{\sigma}} \left( \frac{\partial}{\partial x^{\mu}} \left( \Phi \right) \right) - {}^{\left( 4 \right)} \Gamma_{\sigma \mu}^{\lambda} \left( \frac{\partial}{\partial x^{\lambda}} \left( \Phi \right) \right) \right) = 0,
\end{equation}
which, using the contraction properties of the spacetime Christoffel symbols ${{}^{\left( 4 \right)} \Gamma_{\sigma \mu}^{\lambda}}$, becomes:

\begin{equation}
\frac{1}{\sqrt{- \det \left( g_{\mu \nu} \right)}} \left( \frac{\partial}{\partial x^{\sigma}} \left( \sqrt{- \det \left( g_{\mu \nu} \right)} \left( g^{\mu \sigma} \frac{\partial}{\partial x^{\mu}} \left( \Phi \right) \right) \right) \right) = 0.
\end{equation}
In all of the above, ${\mu, \nu, \lambda, \sigma}$ range across all spacetime coordinate indices ${\left\lbrace 0, \dots, n - 1 \right\rbrace}$.

By imposing the boundary condition that the fluid should be at rest at infinite radial distance (i.e. ${v_{\infty}^{\mu} = 0}$), and also that the fluid distribution should be uniform at this distance (i.e. that ${\rho_{\infty}}$, ${P_{\infty}}$ and ${h_{\infty}}$ should all be constant), there exists an analytic solution to this equation for the stream function, as derived by Aguayo-Ortiz, Sarbach and Tejeda\cite{aguayoortiz2}, namely:

\begin{equation}
\Phi = h_{\infty} \left[ - t + 2 M \log \left( \frac{r - M + \sqrt{M^2 - \left( \frac{J}{M} \right)^2}}{2 \sqrt{M^2 - \left( \frac{J}{M} \right)^2}} \right) \right].
\end{equation}
This, in turn, yields the following values for the timelike, radial, polar and azimuthal projections of the spacetime velocity vector ${\mathbf{u}}$:

\begin{equation}
\frac{h}{h_{\infty}} u^t = 1 + \frac{2 M r}{r^2 + \left( \frac{J}{M} \right)^2 \cos^2 \left( \theta \right)} \left( \frac{r + M + \sqrt{M^2 - \left( \frac{J}{M} \right)^2}}{r - M + \sqrt{M^2 - \left( \frac{J}{M} \right)^2}} \right),
\end{equation}
\begin{equation}
\frac{h}{h_{\infty}} u^r = - \frac{2 M \left( M + \sqrt{M^2 - \left( \frac{J}{M} \right)^2} \right)}{r^2 + \left( \frac{J}{M} \right)^2 \cos^2 \left( \theta \right)},
\end{equation}
\begin{equation}
\frac{h}{h_{\infty}} u^{\theta} = 0,
\end{equation}
and:

\begin{equation}
\frac{h}{h_{\infty}} u^{\phi} = \frac{2 J r}{\left( r^2 + \left( \frac{J}{M} \right)^2 \cos^2 \left( \theta \right) \right) \left( r - M + \sqrt{M^2 - \left( \frac{J}{M} \right)^2} \right)},
\end{equation}
respectively, as well as the following relation for the fluid (rest) mass density ${\rho}$ and/or specific relativistic enthalpy $h$:

\begin{equation}
\frac{\rho}{\rho_{\infty}} = \frac{h}{h_{\infty}} = \sqrt{1 + \left( \frac{2 M}{r^2 + \left( \frac{J}{M} \right)^2 \cos^2 \left( \theta \right)} \right) \left( \frac{r \left( r + M + \sqrt{M^2 - \left( \frac{J}{M} \right)^2} \right) + 2 M \left( M + \sqrt{M^2 - \left( \frac{J}{M} \right)^2} \right)}{r - M + \sqrt{M^2 - \left( \frac{J}{M} \right)^2}} \right)}.
\end{equation}
One consequently recovers the analytic solution of Petrich, Shapiro and Teukolsky\cite{petrich} for the accretion rate of mass onto the spinning black hole:

\begin{equation}
\frac{d M}{d t} = 8 \pi M \left( M + \sqrt{M^2 - \left( \frac{J}{M} \right)^2} \right) \rho_{\infty} = 4 \pi \left( \left( M + \sqrt{M^2 - \left( \frac{J}{M} \right)^2} \right)^2 + \left( \frac{J}{M} \right)^2 \right) \rho_{\infty}.
\end{equation}

\begin{figure}[ht]
\centering
\includegraphics[width=0.295\textwidth]{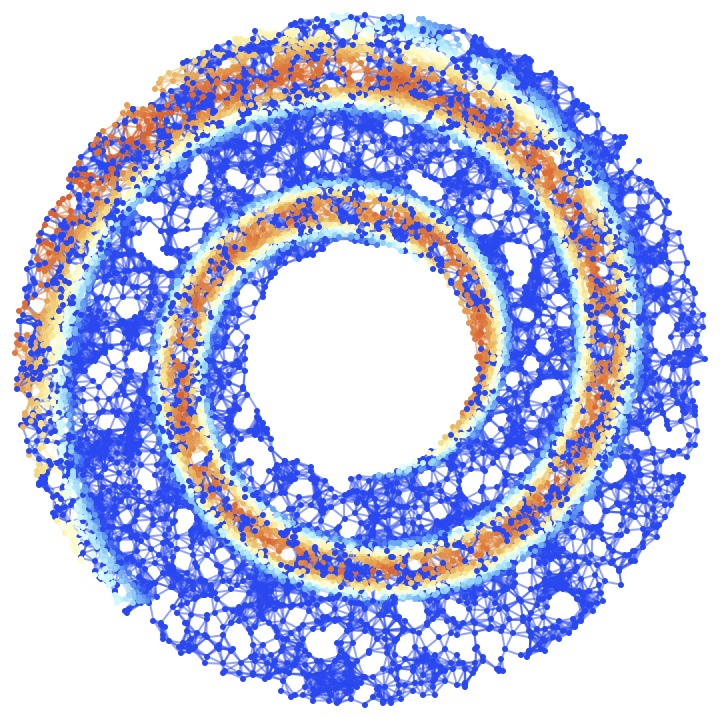}\hspace{0.2\textwidth}
\includegraphics[width=0.295\textwidth]{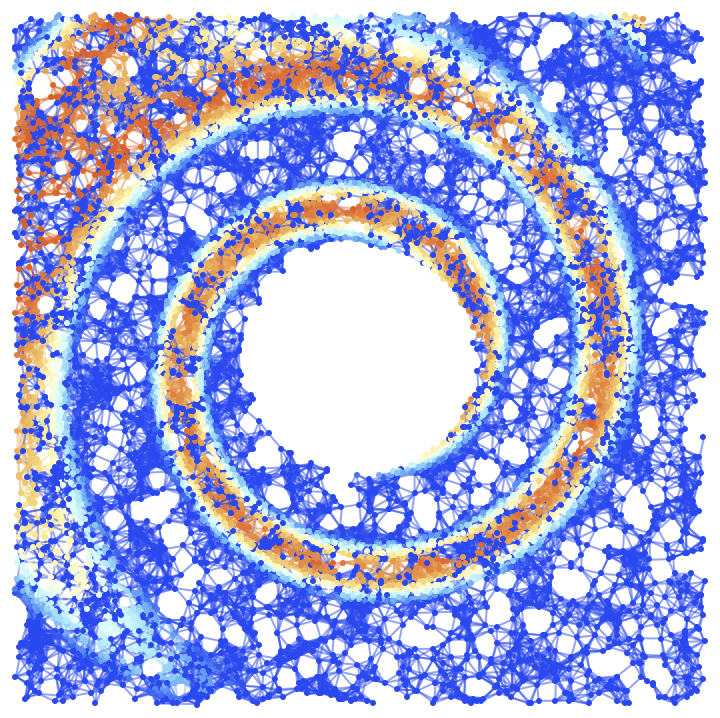}
\caption{On the left, the final (${t = 100 M}$) hypersurface configuration for the radial accretion of an initially spherically-symmetric fluid distribution onto an uncharged, spinning black hole with ${J = 0.6 M}$ in Boyer-Lindquist/oblate spheroidal coordinates ${\left( t, r, \theta, \phi \right)}$, with a resolution of 10,000 hypergraph vertices (colored based on fluid density), and with spatial coordinate information assigned to the vertices. On the right, the final (${t = 100 M}$) hypersurface configuration for the radial accretion of an initially spherically-symmetric fluid distribution onto an uncharged, spinning black hole with ${J = 0.6 M}$ in Kerr-Schild/Cartesian coordinates ${\left( t, x, y, z \right)}$, with a resolution of 10,000 hypergraph vertices (colored based on fluid density), and with spatial coordinate information assigned to the vertices.}
\label{fig:Figure13}
\end{figure}

\begin{figure}[ht]
\centering
\includegraphics[width=0.295\textwidth]{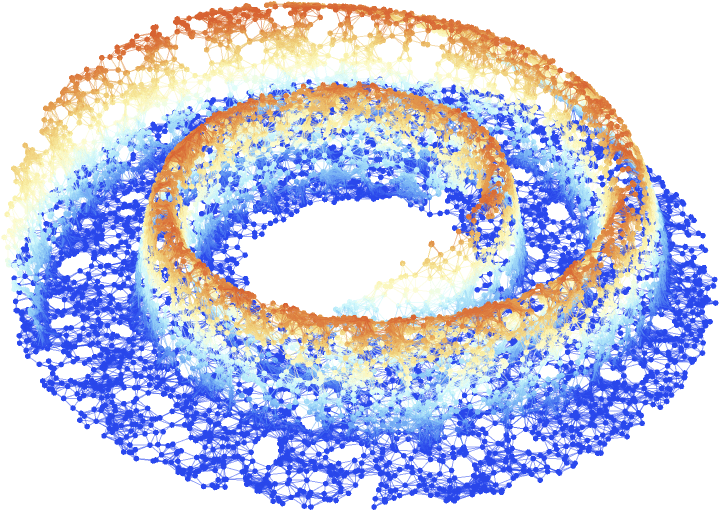}\hspace{0.2\textwidth}
\includegraphics[width=0.295\textwidth]{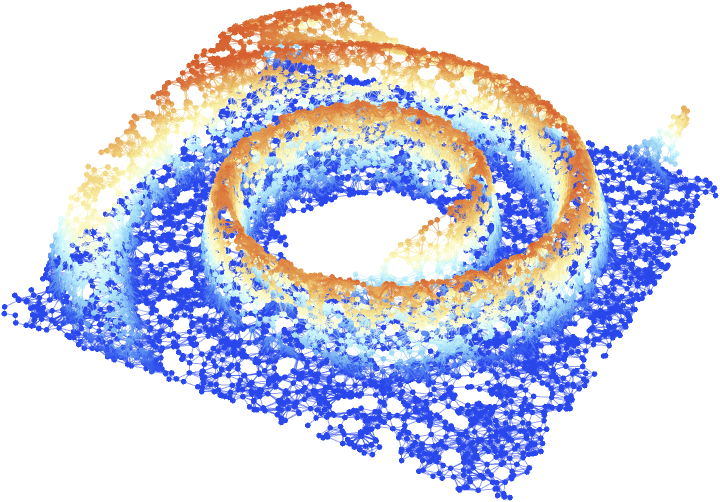}
\caption{On the left, the final (${t = 100 M}$) hypersurface configuration for the radial accretion of an initially spherically-symmetric fluid distribution onto an uncharged, spinning black hole with ${J = 0.6 M}$ in Boyer-Lindquist/oblate spheroidal coordinates ${\left( t, r, \theta, \phi \right)}$, with a resolution of 10,000 hypergraph vertices (colored based on fluid density), and with both spatial coordinate and fluid density coordinate information assigned to the vertices. On the right, the final (${t = 100 M}$) hypersurface configuration for the radial accretion of an initially spherically-symmetric fluid distribution onto an uncharged, spinning black hole with ${J = 0.6 M}$ in Kerr-Schild/Cartesian coordinates ${\left( t, x, y, z \right)}$, with a resolution of 10,000 hypergraph vertices (colored based on fluid density), and with both spatial coordinate and fluid density coordinate information assigned to the vertices.}
\label{fig:Figure14}
\end{figure}

\begin{figure}[ht]
\centering
\includegraphics[width=0.295\textwidth]{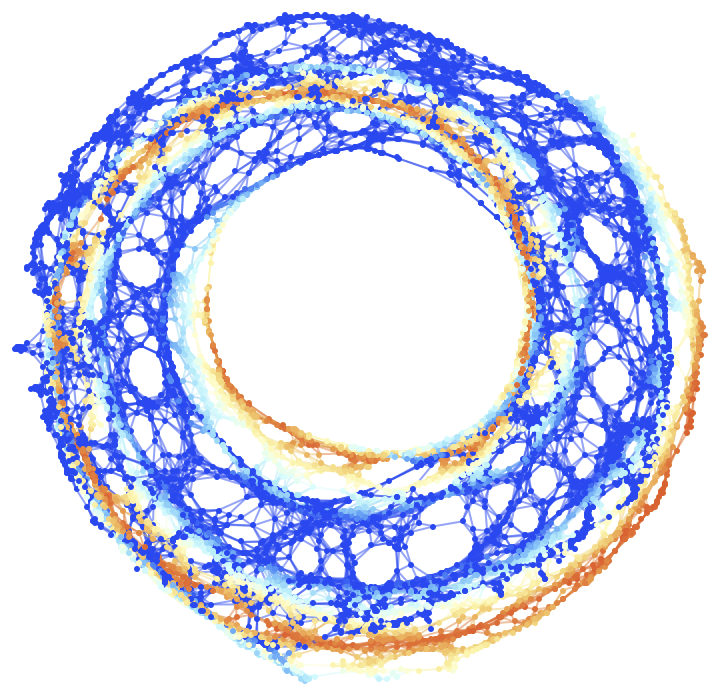}\hspace{0.2\textwidth}
\includegraphics[width=0.295\textwidth]{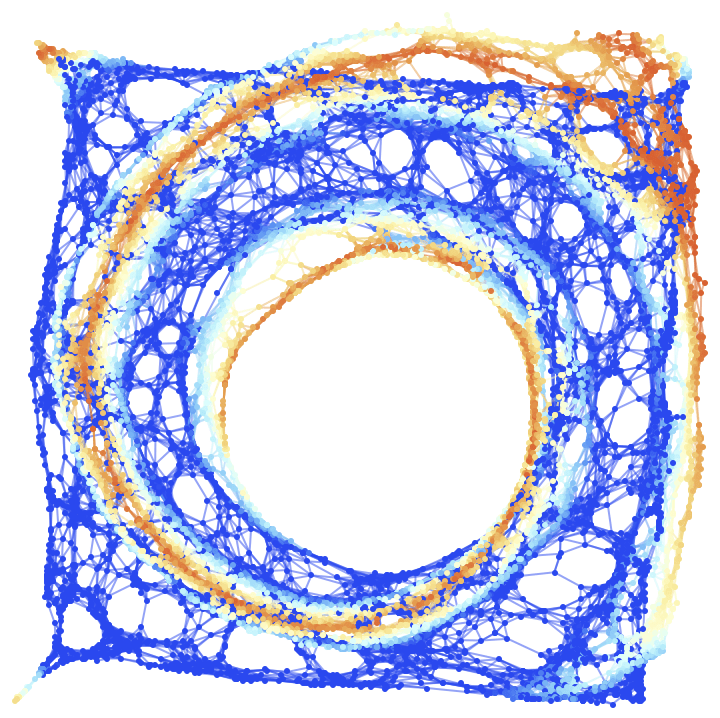}
\caption{On the left, the final (${t = 100 M}$) hypersurface configuration for the radial accretion of an initially spherically-symmetric fluid distribution onto an uncharged, spinning black hole with ${J = 0.6 M}$ in Boyer-Lindquist/oblate spheroidal coordinates ${\left( t, r, \theta, \phi \right)}$, with a resolution of 10,000 hypergraph vertices (colored based on fluid density), and with no coordinate information assigned to the vertices. On the right, the final (${t = 100 M}$) hypersurface configuration for the radial accretion of an initially spherically-symmetric fluid distribution onto an uncharged, spinning black hole with ${J = 0.6 M}$ in Kerr-Schild/Cartesian coordinates ${\left( t, x, y, z \right)}$, with a resolution of 10,000 hypergraph vertices (colored based on fluid density), and with no coordinate information assigned to the vertices.}
\label{fig:Figure15}
\end{figure}

\begin{figure}[ht]
\centering
\includegraphics[width=0.295\textwidth]{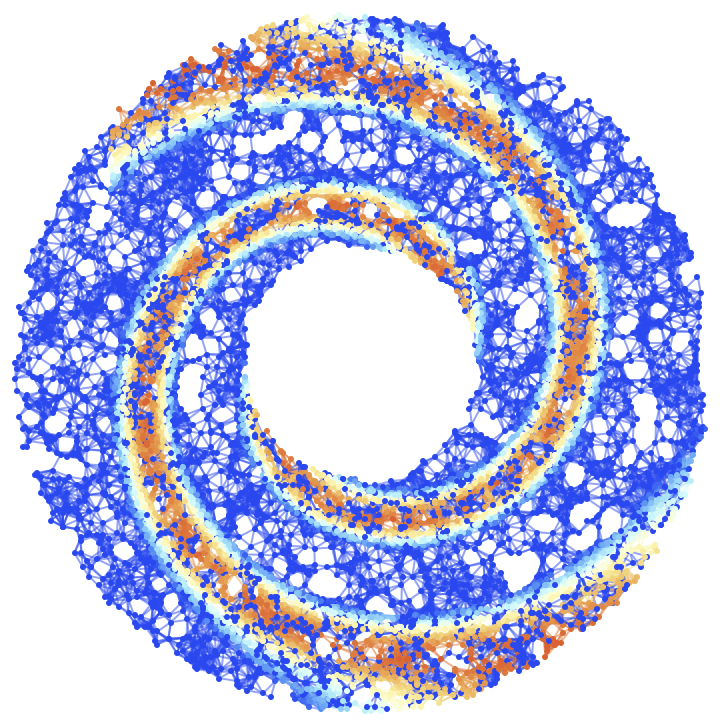}\hspace{0.2\textwidth}
\includegraphics[width=0.295\textwidth]{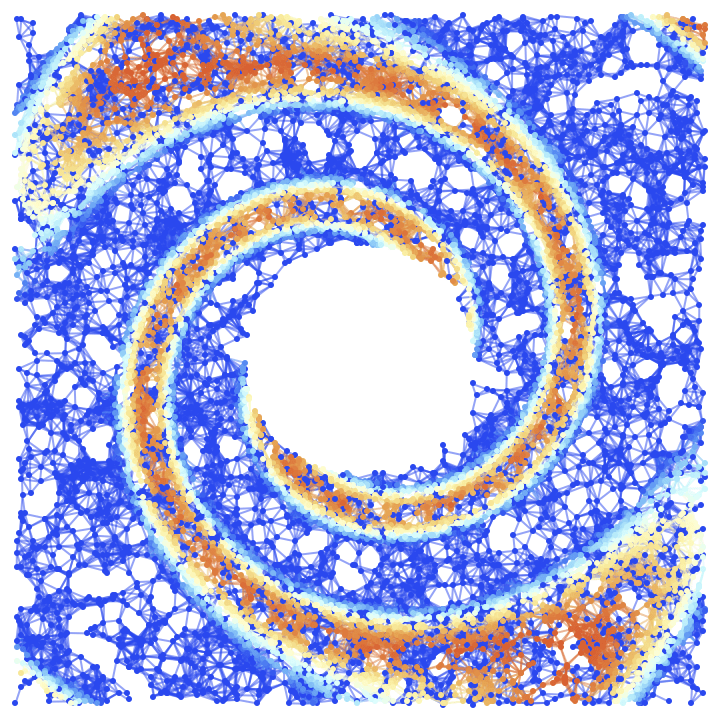}
\caption{On the left, the final (${t = 100 M}$) hypersurface configuration for the radial accretion of an initially spherically-symmetric fluid distribution onto an uncharged, rapidly-spinning black hole with ${J = 0.9 M}$ in Boyer-Lindquist/oblate spheroidal coordinates ${\left( t, r, \theta, \phi \right)}$, with a resolution of 10,000 hypergraph vertices (colored based on fluid density), and with spatial coordinate information assigned to the vertices. On the right, the final (${t = 100 M}$) hypersurface configuration for the radial accretion of an initially spherically-symmetric fluid distribution onto an uncharged, rapidly-spinning black hole with ${J = 0.9 M}$ in Kerr-Schild/Cartesian coordinates ${\left( t, x, y, z \right)}$, with a resolution of 10,000 hypergraph vertices (colored based on fluid density), and with spatial coordinate information assigned to the vertices.}
\label{fig:Figure16}
\end{figure}

\begin{figure}[ht]
\centering
\includegraphics[width=0.295\textwidth]{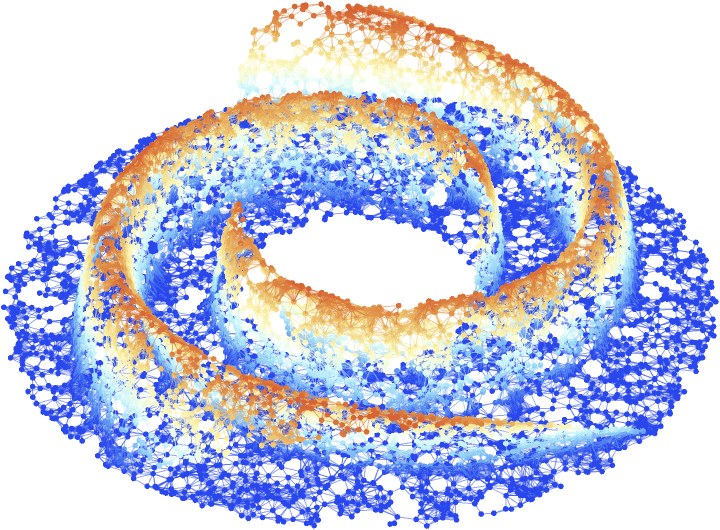}\hspace{0.2\textwidth}
\includegraphics[width=0.295\textwidth]{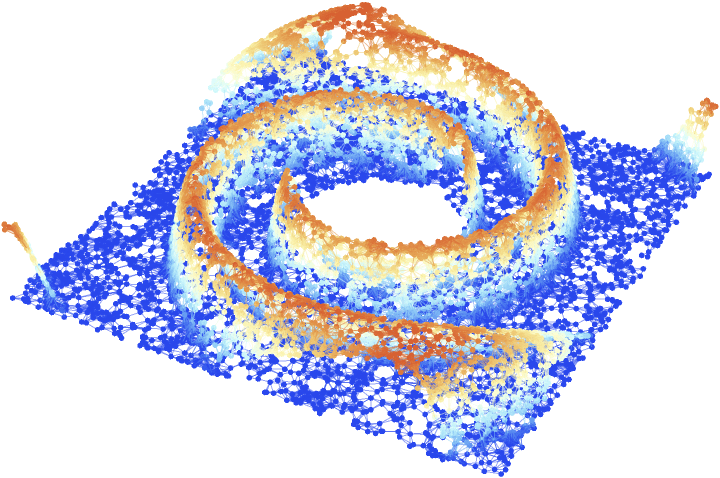}
\caption{On the left, the final (${t = 100 M}$) hypersurface configuration for the radial accretion of an initially spherically-symmetric fluid distribution onto an uncharged, rapidly-spinning black hole with ${J = 0.9 M}$ in Boyer-Lindquist/oblate spheroidal coordinates ${\left( t, r, \theta, \phi \right)}$, with a resolution of 10,000 hypergraph vertices (colored based on fluid density), and with both spatial coordinate and fluid density coordinate information assigned to the vertices. On the right, the final (${t = 100 M}$) hypersurface configuration for the radial accretion of an initially spherically-symmetric fluid distribution onto an uncharged, rapidly-spinning black hole with ${J = 0.9 M}$ in Kerr-Schild/Cartesian coordinates ${\left( t, x, y, z \right)}$, with a resolution of 10,000 hypergraph vertices (colored based on fluid density), and with both spatial coordinate and fluid density coordinate information assigned to the vertices.}
\label{fig:Figure17}
\end{figure}

\begin{figure}[ht]
\centering
\includegraphics[width=0.295\textwidth]{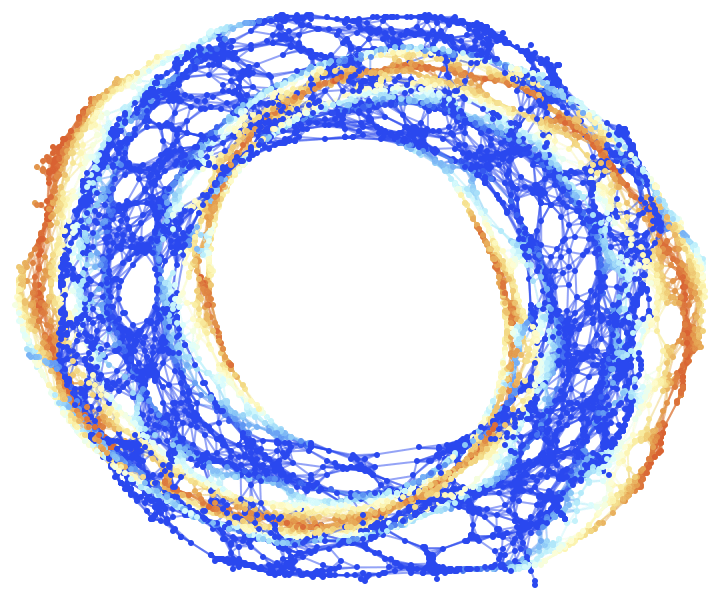}\hspace{0.2\textwidth}
\includegraphics[width=0.295\textwidth]{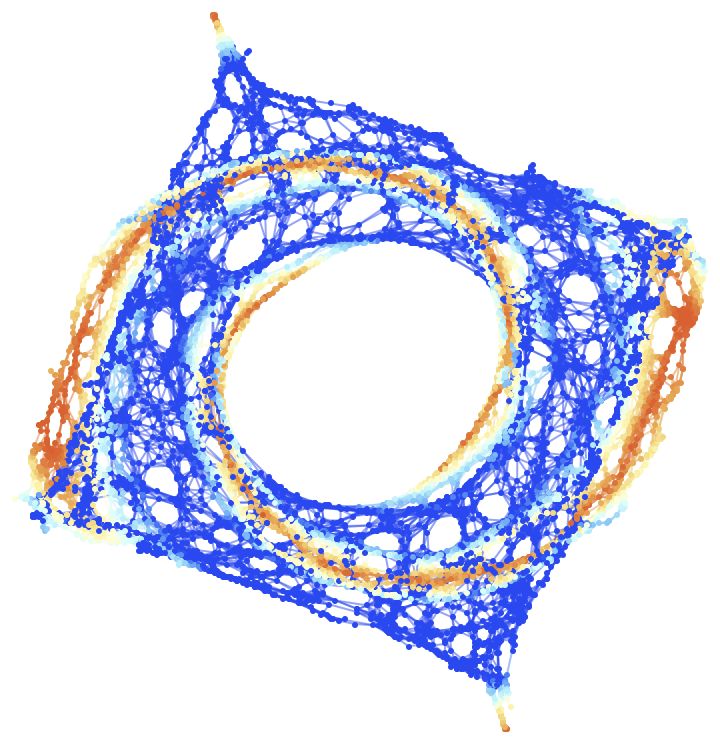}
\caption{On the left, the final (${t = 100 M}$) hypersurface configuration for the radial accretion of an initially spherically-symmetric fluid distribution onto an uncharged, rapidly-spinning black hole with ${J = 0.9 M}$ in Boyer-Lindquist/oblate spheroidal coordinates ${\left( t, r, \theta, \phi \right)}$, with a resolution of 10,000 hypergraph vertices (colored based on fluid density), and with no coordinate information assigned to the vertices. On the right, the final (${t = 100 M}$) hypersurface configuration for the radial accretion of an initially spherically-symmetric fluid distribution onto an uncharged, rapidly-spinning black hole with ${J = 0.9 M}$ in Kerr-Schild/Cartesian coordinates ${\left( t, x, y, z \right)}$, with a resolution of 10,000 hypergraph vertices (colored based on fluid density), and with both spatial coordinate and fluid density coordinate information assigned to the vertices.}
\label{fig:Figure18}
\end{figure}

\begin{figure}[ht]
\centering
\includegraphics[width=0.295\textwidth]{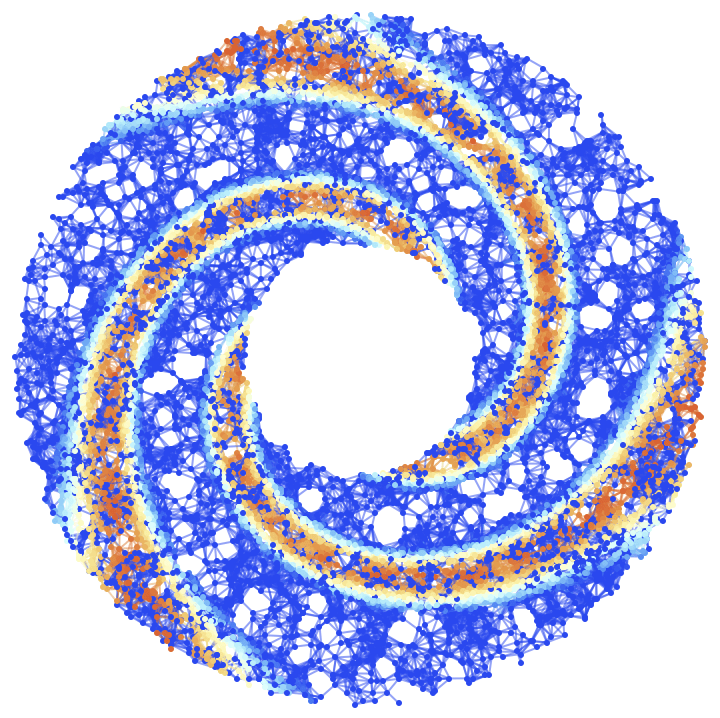}\hspace{0.2\textwidth}
\includegraphics[width=0.295\textwidth]{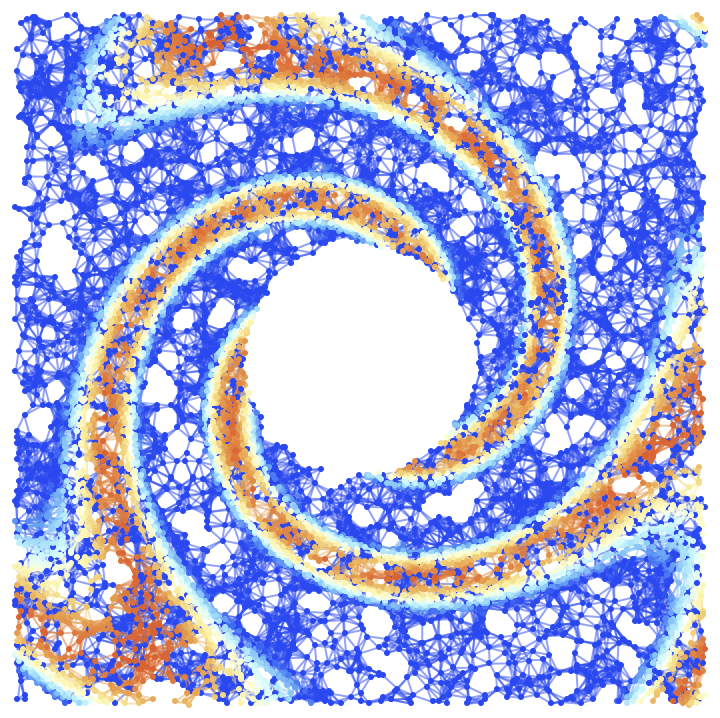}
\caption{On the left, the final (${t = 100 M}$) hypersurface configuration for the radial accretion of an initially spherically-symmetric fluid distribution onto an uncharged black hole spinning close to extremality with ${J = 0.99 M}$ in Boyer-Lindquist/oblate spheroidal coordinates ${\left( t, r, \theta, \phi \right)}$, with a resolution of 10,000 hypergraph vertices (colored based on fluid density), and with spatial coordinate information assigned to the vertices. On the right, the final (${t = 100 M}$) hypersurface configuration for the radial accretion of an initially spherically-symmetric fluid distribution onto an uncharged black hole spinning close to extremality with ${J = 0.99 M}$ in Kerr-Schild/Cartesian coordinates ${\left( t, x, y, z \right)}$, with a resolution of 10,000 hypergraph vertices (colored based on fluid density), and with spatial coordinate information assigned to the vertices.}
\label{fig:Figure19}
\end{figure}

\begin{figure}[ht]
\centering
\includegraphics[width=0.295\textwidth]{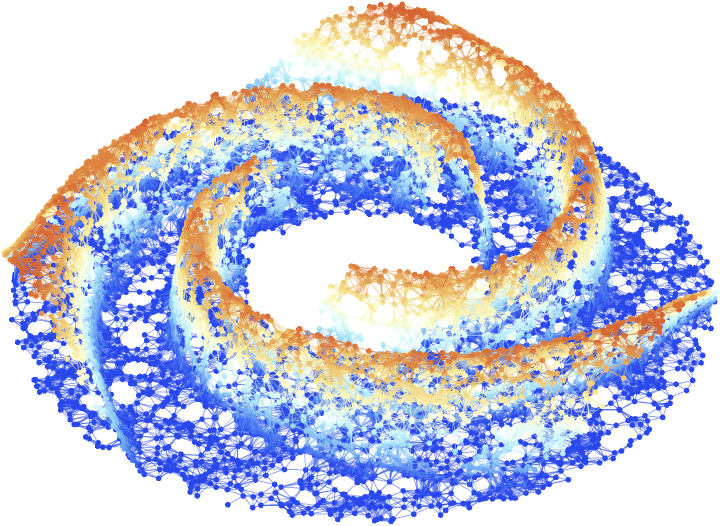}\hspace{0.2\textwidth}
\includegraphics[width=0.295\textwidth]{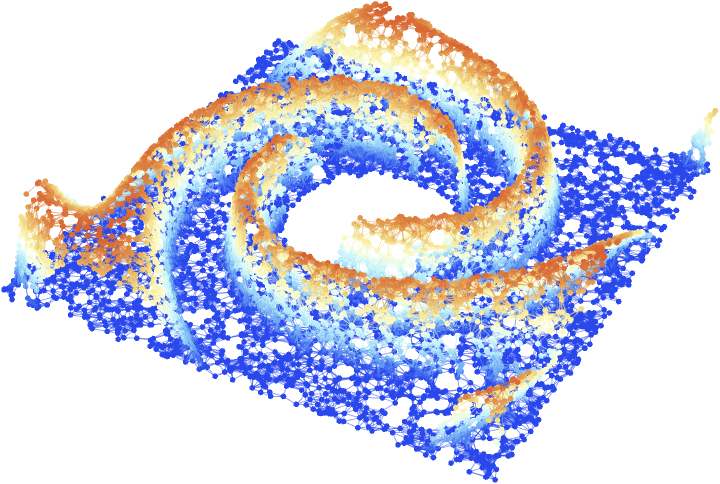}
\caption{On the left, the final (${t = 100 M}$) hypersurface configuration for the radial accretion of an initially spherically-symmetric fluid distribution onto an uncharged black hole spinning close to extremality with ${J = 0.99 M}$ in Boyer-Lindquist/oblate spheroidal coordinates ${\left( t, r, \theta, \phi \right)}$, with a resolution of 10,000 hypergraph vertices (colored based on fluid density), and with both spatial coordinate and fluid density coordinate information assigned to the vertices. On the right, the final (${t = 100 M}$) hypersurface configuration for the radial accretion of an initially spherically-symmetric fluid distribution onto an uncharged black hole spinning close to extremality with ${J = 0.99 M}$ in Kerr-Schild/Cartesian coordinates ${\left( t, x, y, z \right)}$, with a resolution of 10,000 hypergraph vertices (colored based on fluid density), and with both spatial coordinate and fluid density coordinate information assigned to the vertices.}
\label{fig:Figure20}
\end{figure}

\begin{figure}[ht]
\centering
\includegraphics[width=0.295\textwidth]{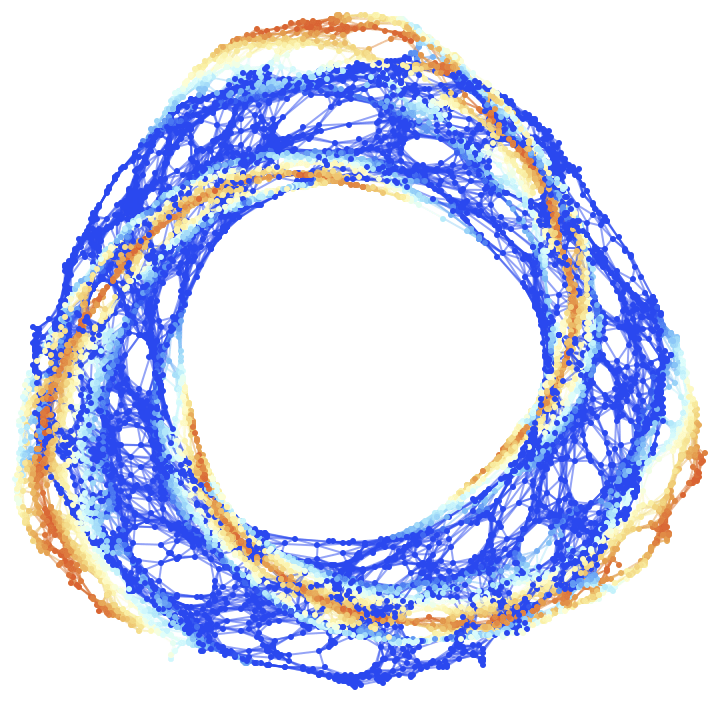}\hspace{0.2\textwidth}
\includegraphics[width=0.295\textwidth]{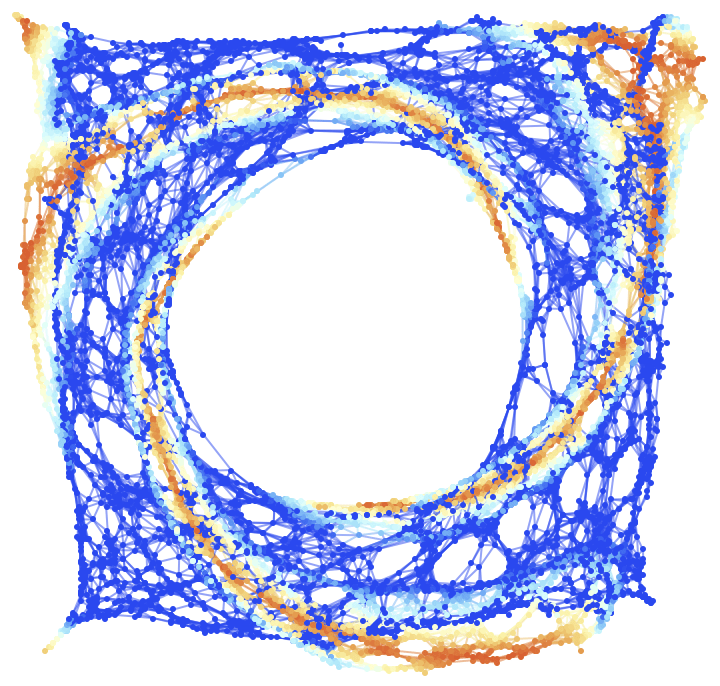}
\caption{On the left, the final (${t = 100 M}$) hypersurface configuration for the radial accretion of an initially spherically-symmetric fluid distribution onto an uncharged black hole spinning close to extremality with ${J = 0.99 M}$ in Boyer-Lindquist/oblate spheroidal coordinates ${\left( t, r, \theta, \phi \right)}$, with a resolution of 10,000 hypergraph vertices (colored based on fluid density), and with no coordinate information assigned to the vertices. On the right, the final (${t = 100 M}$) hypersurface configuration for the radial accretion of an initially spherically-symmetric fluid distribution onto an uncharged black hole spinning close to extremality with ${J = 0.99 M}$ in Kerr-Schild/Cartesian coordinates ${\left( t, x, y, z \right)}$, with a resolution of 10,000 hypergraph vertices (colored based on fluid density), and with no coordinate information assigned to the vertices.}
\label{fig:Figure21}
\end{figure}

As in the Schwarzschild case described previously, we perform our radial accretion simulations onto uncharged, spinning black holes (i.e. Kerr black holes) expressed in both the Boyer-Lindquist/oblate spheroidal coordinate system ${\left( t, r, \theta, \phi \right)}$, which is not horizon-adapted, with the metric of the initial spacelike hypersurface given by:

\begin{multline}
d l^2 = \gamma_{\mu \nu} d x^{\mu} d x^{\nu} = \left( \frac{r^2 + \left( \frac{J}{M} \right)^2 \cos^2 \left( \theta \right)}{r^2 - 2 M r + \left( \frac{J}{M} \right)^2} \right) d r^2 + \left( r^2 + \left( \frac{J}{M} \right)^2 \cos^2 \left( \theta \right) \right) d \theta^2\\
+ \left( r^2 + \left( \frac{J}{M} \right)^2 + \frac{2 J^2 \sin^2 \left( \theta \right)}{M \left( r^2 + \left( \frac{J}{M} \right)^2 \cos^2 \left( \theta \right) \right)} \right) \sin^2 \left( \theta \right) d \phi^2,
\end{multline}
and the Kerr-Schild/Cartesian coordinate system ${\left( t, x, y, z \right)}$, which \textit{is} horizon-adapted, with the metric of the initial spacelike hypersurface given by:

\begin{equation}
d l^2 = \gamma_{\mu \nu} d x^{\mu} d x^{\nu} = d x^2 + d y^2 + d z^2 + \mathcal{F} \left( l_{\mu} d x^{\mu} \right)^2,
\end{equation}
where:

\begin{equation}
\mathcal{F} = \frac{2 M r^2}{r^4 + \left( \frac{J}{M} \right)^2 z^2}, \qquad \text{ and } \qquad l_{\mu} dx^{\mu} = \frac{z}{r} d z + \frac{r}{r^2 + \left( \frac{J}{M} \right)^2} \left( x d x + y d y \right) - \frac{\left( \frac{J}{M} \right)^2}{r^2 + \left( \frac{J}{M} \right)^2} \left( x d y - y d x \right),
\end{equation}
with $r$ no longer being the usual radial coordinate, but rather being defined implicitly as a (positive, real) solution to the following algebraic equation:

\begin{equation}
\frac{x^2 + y^2}{r^2 + \left( \frac{J}{M} \right)^2} + \frac{z^2}{r^2} = 1.
\end{equation}
We initialize our simulations, as before, with an ideal gas equation of state with adiabatic exponent ${\Gamma = \frac{5}{3}}$, and the non-dimensional gas temperature at infinite radial distance (which determines both ${\rho_{\infty}}$ and ${P_{\infty}}$) set to be ${\Theta_{\infty} = 0.1}$. We begin by considering a black hole with only a modest spin value of ${J = 0.6 M}$, and in Figures \ref{fig:Figure13}, \ref{fig:Figure14} and \ref{fig:Figure15} (showing the solution at coordinate time ${t = 100 M}$ with, as before, two-dimensional spatial coordinates, three-dimensional spatial and fluid density coordinates, and no vertex coordinates, respectively), we see that the fluid in this case evolves to a steady-state configuration with a single high-density ``swirl'' surrounding the black hole horizon. For a rapidly-spinning black hole with a spin value of ${J = 0.9 M}$, as shown in Figures \ref{fig:Figure16}, \ref{fig:Figure17} and \ref{fig:Figure18}, we see that this high-density ``swirl'' effectively splits into two distinct ``arms'', while for a black hole spinning close to the threshold of extremality with ${J = 0.99 M}$, as shown in Figures \ref{fig:Figure19}, \ref{fig:Figure20} and \ref{fig:Figure21}, we see splitting of the ``swirl'' into three ``arms'' instead. We see evidence of some slight boundary effects around the edges and corners of the domain in Kerr-Schild/Cartesian coordinates, and no evidence of unphysical fluid behavior close to the horizon in Boyer-Lindquist/oblate spheroidal coordinates (which we attribute to our robust and singularity-avoiding choice of gauge). The rates of mass/energy accretion onto the spinning black holes are, again, found to be slightly lower than in the analytic solution of Petrich, Shapiro and Teukolsky\cite{petrich}, with this discrepancy vanishing in the limit as the discretization scale and the black hole spin both go to zero.

In order to conduct a more quantitative analysis of these simulation results, we begin by extracting the rates of mass/energy and momentum accretion onto the black hole in each case. Following the approach of Petrich, Shapiro, Stark and Teukolsky\cite{petrich2}, the total fluid (rest) mass $M$ contained within a given simply-connected spatial volume $V$ of dimension ${\left( n - 1 \right)}$, with a closed ${\left( n - 2 \right)}$-dimensional boundary ${\partial V}$, can be computed by means of the following integral of the fluid (rest) mass density:

\begin{equation}
M = \int_{V} \sqrt{\det \left( \gamma_{\mu \nu} \right)} \left( \frac{\rho}{\sqrt{1 - \gamma_{\mu \nu} v^{\mu} v^{\nu}}} \right) d V,
\end{equation}
which, upon application of the Leibniz integral rule, allows us to write the mass accretion rate within that volume as:

\begin{equation}
\frac{d M}{d t} = \int_{V} \frac{\partial}{\partial t} \left( \sqrt{\det \left( \gamma_{\mu \nu} \right)} \left( \frac{\rho}{\sqrt{1 - \gamma_{\mu \nu} v^{\mu} v^{\nu}}} \right) \right) d V,
\end{equation}
with ${\mu, \nu}$ in the above ranging across spatial coordinate indices ${\left\lbrace 0, \dots, n - 2 \right\rbrace}$ only. Since, by the definitions of the Lorentz factor ${\alpha u^0}$, the (rest) mass current vector ${J^{\mu}}$ and the spacetime metric determinant ${\det \left( g_{\mu \nu} \right)}$, we have:

\begin{equation}
\alpha u^0 = - u_{\left( \mu + 1 \right)} n^{\left( \mu + 1 \right)} = \frac{1}{\sqrt{1 - \gamma_{\mu \nu} v^{\mu} v^{\nu}}}, \qquad J^{\mu} = \rho u^{\mu}, \qquad \text{ and } \qquad \sqrt{- \det \left( g_{\mu \nu} \right)} = \alpha \sqrt{\det \left( \gamma_{\mu \nu} \right)},
\end{equation}
respectively, with ${\mu}$ ranging across all spacetime coordinate indices ${\left\lbrace 0, \dots, n - 1 \right\rbrace}$ in ${u_{\left( \mu + 1 \right)} n^{\left( \mu + 1 \right)}}$, ${J^{\mu}}$ and ${\rho u^{\mu}}$, and with ${\mu, \nu}$ ranging across spatial coordinate indices ${\left\lbrace 0, \dots, n - 2 \right\rbrace}$ only everywhere else, we can rewrite these two integrals as:

\begin{equation}
M = \int_{V} J^0 \sqrt{- \det \left( g_{\mu \nu} \right)} d V, \qquad \text{ and therefore } \qquad \frac{d M}{d t} = \int_{V} \frac{\partial}{\partial t} \left( J^0 \sqrt{- \det \left( g_{\mu \nu} \right)} \right) d V,
\end{equation}
respectively. The time derivative in the latter integral can now be rewritten as a difference between a spacetime covariant derivative and a spatial partial derivative, namely:

\begin{equation}
\int_{V} \frac{\partial}{\partial t} \left( J^0 \sqrt{- \det \left( g_{\mu \nu} \right)} \right) d V = \int_{V} \left[ {}^{\left( 4 \right)} \nabla_{\mu} \left( J^{\mu} \sqrt{- \det \left( g_{\mu \nu} \right)} \right) - \frac{\partial}{\partial x^{\nu}} \left( J^{\left( \nu + 1 \right)} \sqrt{- \det \left( g_{\mu \nu} \right)} \right) \right] d V,
\end{equation}
with ${\mu}$ ranging across all spacetime coordinate indices ${\left\lbrace 0, \dots, n - 1 \right\rbrace}$ and ${\nu}$ ranging across spatial coordinate indices ${\left\lbrace 0, \dots, n - 2 \right\rbrace}$ only. However, the covariant derivative term vanishes by virtue of the conservation of baryon number:

\begin{multline}
{}^{\left( 4 \right)} \nabla_{\mu} J^{\mu} = \frac{\partial}{\partial x^{\mu}} \left( J^{\mu} \right) + {}^{\left( 4 \right)} \Gamma_{\mu \sigma}^{\mu} J^{\sigma} = 0,\\
\implies {}^{\left( 4 \right)} \nabla_{\mu} \left( J^{\mu} \sqrt{- \det \left( g_{\mu \nu} \right)} \right) = \sqrt{- \det \left( g_{\mu \nu} \right)} \left( {}^{\left( 4 \right)} \nabla_{\mu} J^{\mu} \right) = \sqrt{- \det \left( g_{\mu \nu} \right)} \left( \frac{\partial}{\partial x^{\mu}} \left( J^{\mu} \right) + {}^{\left( 4 \right)} \Gamma_{\mu \sigma}^{\mu} J^{\sigma} \right) = 0,
\end{multline}
due to the metric compatibility of the spacetime covariant derivative operator (i.e. since ${{}^{\left( 4 \right)} \nabla_{\nu} g^{\mu \nu} = \mathbf{0}}$ identically), with ${\mu, \nu, \sigma}$ ranging across all spacetime coordinate indices ${\left\lbrace 0, \dots, n - 1 \right\rbrace}$. Thus, the mass accretion rate reduces to the following surface integral over the boundary ${\partial V}$, with surface element ${d S_{\mu}}$ and corresponding area element ${d A}$:

\begin{equation}
\frac{d M}{d t} = - \int_{\partial V} J^{\left( \mu + 1 \right)} \sqrt{- \det \left( g_{\mu \nu} \right)} d S_{\mu}, \qquad \text{ with } \qquad d S_{\mu} = \gamma_{\mu \nu} n^{\left( \nu + 1 \right)} d A,
\end{equation}
which, using the definitions of the (rest) mass current vector ${J^{\mu}}$ (i.e. ${J^{\mu} = \rho u^{\mu}}$) and the (spatial) fluid velocity vector ${v^{\mu}}$ perceived by an observer moving in the normal direction ${\mathbf{n}}$:

\begin{equation}
v^{\mu} = \frac{u^{\left( \mu + 1 \right)}}{\alpha u^0} + \frac{\beta^{\mu}}{\alpha} = \sqrt{1 - \gamma_{\mu \nu} v^{\mu} v^{\nu}} \left( u^{\left( \mu + 1 \right)} \right) + \frac{\beta^{\mu}}{\alpha},
\end{equation}
becomes:

\begin{multline}
- \int_{\partial V} J^{\left( \mu + 1 \right)} \sqrt{- \det \left( g_{\mu \nu} \right)} d S_{\mu} = - \int_{\partial V} \rho u^{\left( \mu + 1 \right)} \sqrt{- \det \left( g_{\mu \nu} \right)} d S_{\mu}\\
= - \int_{\partial V} \alpha \sqrt{\det \left( \gamma_{\mu \nu} \right)} \left( \frac{\rho}{\sqrt{1 - \gamma_{\mu \nu} v^{\mu} v^{\nu}}} \left( v^{\mu} - \frac{\beta^{\mu}}{\alpha} \right) \right) d S_{\mu},
\end{multline}
with ${\mu, \nu}$ in all of the above ranging across spatial coordinate indices ${\left\lbrace 0, \dots, n - 2 \right\rbrace}$ only. In order to evaluate this integral numerically, we exploit the radial nature of the accretion flow by sampling the integrand uniformly in the angular range ${\theta \in \left[ 0, \frac{\pi}{2} \right]}$ and then evaluating:

\begin{equation}
\frac{d M}{d t} = 4 \pi \int_{0}^{\frac{\pi}{2}} \alpha \sqrt{\det \left( \gamma_{\mu \nu} \right)} \left( \frac{\rho}{\sqrt{1 - \gamma_{\mu \nu} v^{\mu} v^{\nu}}} \left( v^r - \frac{\beta^r}{\alpha} \right) \right) d \theta,
\end{equation}
using standard methods of numerical quadrature. By means of a similar argument, we follow the approach of Petrich, Shapiro, Stark and Teukolsky\cite{petrich2} and approximate the linear momentum accretion rate (assuming only local gravitational effects, and assuming evaluation of the integral over an asymptotically-flat boundary of spacetime) as:

\begin{equation}
\frac{d P^{\mu}}{d t} = - \int_{\partial V} \alpha \sqrt{\det \left( \gamma_{\mu \nu} \right)} \left( T^{\left( \mu +  1 \right) \left( \nu + 1 \right)} \right) d S_{\nu},
\end{equation}
with ${\mu, \nu}$ ranging across spatial coordinate indices ${\left\lbrace 0, \dots, n - 2 \right\rbrace}$ only.

Next, we determine the rate of decrease in the spin value $J$ of the black hole due to dynamical/tidal friction effects (also known as the ``drag force'' of the fluid) exerted by the fluid onto the underlying spacetime geometry. Within a general Bondi-Hoyle-Lyttleton accretion setup, the asymmetry of the fluid pressure distribution surrounding a spinning black hole causes an increase in fluid pressure on the side of the black hole that is counter-rotating with the fluid and a decrease on the side that is co-rotating with it, which, in turn distorts the metric in such a way as to induce a gravitational field opposing the direction of spin of the black hole. In the purely radial (Bondi) accretion setup, this represents the general relativistic analog of the viscous dissipation of angular momentum due to tidal deformation that occurs in Newtonian gravity. In order to analyze this effect quantitatively, we follow the approach of Hawking and Hartle\cite{hawking}, and later Hartle\cite{hartle}, by exploiting the known relationship between the surface area $A$ of the horizon of a Kerr black hole and its corresponding spin value $J$:

\begin{equation}
A = 8 \pi M \left( M + \sqrt{M^2 - \left( \frac{J}{M} \right)^2} \right),
\end{equation}
thus implying that the rate of decrease in the spin of the black hole ${\frac{d J}{d t}}$ is related directly to the rate of decrease of the surface area of the black hole horizon ${\frac{d A}{d t}}$

\begin{equation}
\frac{d J}{d t} = - \left( \frac{\sqrt{M^2 - \left( \frac{J}{M} \right)^2}}{8 \pi \left( \frac{J}{M} \right)^2} \right) \left( \frac{d A}{d t} \right).
\end{equation}
If we now choose an orthonormal tetrad such that the vector ${\mathbf{l}}$ is normal to the horizon of the black hole, with:

\begin{equation}
\frac{\partial}{\partial x^{\mu}} \left( t \right) l^{\mu} = 1,
\end{equation}
and the vector ${\mathbf{m}}$ (along with its complex conjugate ${\overline{\mathbf{m}}}$) is such that:

\begin{equation}
m^{\mu} l_{\mu} = 0, \qquad \text{ and } \qquad m^{\mu} \overline{m}_{\mu} = -1,
\end{equation}
then the rate of decrease of the surface area of the black hole can be rewritten as:

\begin{equation}
\frac{d A}{d t} = - 2 \int_{S} \left( {}^{\left( 4 \right)} \nabla_{\nu} l_{\mu} \right) m^{\mu} \overline{m}^{\nu} d A = -2 \int_{S} \left( \frac{\partial}{\partial x^{\nu}} \left( l_{\mu} \right) - {}^{\left( 4 \right)} \Gamma_{\nu \mu}^{\sigma} l_{\sigma} \right) m^{\mu} \overline{m}^{\nu} d A,
\end{equation}
where the ${\left( n - 2 \right)}$-dimensional surface $S$ represents the intersection of the event horizon of the black hole with a constant-time hypersurface, and the integrand represents the convergence of null geodesic generators of the event horizon in the Newman-Penrose formalism\cite{newman}. Hence, the overall rate of decrease in the spin of the black hole is given by:

\begin{equation}
\frac{d J}{d t} = \left( \frac{\sqrt{M^2 - \left( \frac{J}{M} \right)^2}}{4 \pi \left( \frac{J}{M} \right)^2} \right) \left( \int_{S} \left( \frac{\partial}{\partial x^{\nu}} \left( l_{\mu} \right) - {}^{\left( 4 \right)} \Gamma_{\nu \mu}^{\sigma} l_{\sigma} \right) m^{\mu} \overline{m}^{\nu} d A \right).
\end{equation}
Equivalently, the rate of decrease can be calculated as:

\begin{equation}
\frac{d \left( \frac{J}{M} \right)}{d t} = - \frac{A}{4 J} \int_{S} \left\lvert \left( {}^{\left( 4 \right)} \nabla_{\nu} l_{\mu} \right) m^{\mu} m^{\nu} \right\rvert^2 d A = - \frac{A}{4 J} \int_{S} \left\lvert \left( \frac{\partial}{\partial x^{\nu}} \left( l_{\mu} \right) - {}^{\left( 4 \right)} \Gamma_{\nu \mu}^{\sigma} l_{\sigma} \right) m^{\mu} m^{\nu} \right\rvert^2 d A,
\end{equation}
where now the integrand represents the square of the shear of null geodesic generators of the \textit{perturbed} black hole event horizon, evaluated to the first-order of perturbation theory. In all of the above, ${\mu, \nu, \sigma}$ range across all spacetime coordinate indices ${\left\lbrace 0, \dots, n - 1 \right\rbrace}$. Within our numerical validation tests, we did not find any substantive difference in the results obtained through these two mathematical approaches, so we generally opt to use the latter due to the reduced algorithmic complexity of its implementation within \textsc{Gravitas}. It is instructive to compare this \textit{angular} drag force experienced by the black hole in the purely radial/Bondi accretion case with the \textit{linear} drag force ${\mathbf{F}_{\infty}^{D}}$ experienced by the black hole in the general Bondi-Hoyle-Lyttleton accretion case, as calculated analytically by Ostriker\cite{ostriker}, yielding:

\begin{equation}
\mathbf{F}_{\infty}^{D} = - \left( \frac{1}{2} \log \left( \frac{1 + \mathcal{M}_{\infty}}{1 - \mathcal{M}_{\infty}} \right) - \mathcal{M}_{\infty} \right) \left( \frac{4 \pi M^2 \rho_{\infty}}{v_{\infty}^{2}} \right) \mathbf{\hat{v}}_{\infty},
\end{equation}
in the case of subsonic flow at radial infinity (i.e. ${\mathcal{M}_{\infty} < 1}$), and:

\begin{equation}
\mathbf{F}_{\infty}^{D} = - \left( \frac{1}{2} \log \left( \mathcal{M}_{\infty}^{2} - 1 \right) + \log \left( \Lambda \right) \right) \left( \frac{4 \pi M^2 \rho_{\infty}}{v_{\infty}^{2}} \right) \mathbf{\hat{v}}_{\infty},
\end{equation}
in the case of supersonic flow at radial infinity (i.e. ${\mathcal{M}_{\infty} > 1}$). In the above, ${\mathbf{\hat{v}}_{\infty}}$ is the unit vector representing the relative velocity between the fluid flow (at radial infinity) and the black hole, ${\mathcal{M}_{\infty}}$ designates the Mach number at radial infinity, ${\rho_{\infty}}$ and ${v_{\infty}}$ denote (as usual) the fluid density and velocity at radial infinity, respectively, and ${\log \left( \Lambda \right)}$ represents the Coulomb logarithm for particle collisions, with a typical numerical estimate (for instance derived by Chapon, Mayer and Teyssier\cite{chapon} using numerical simulations of supermassive black hole binary mergers) of ${\log \left( \Lambda \right) = 3.2}$.

We find, in each of the cases simulated, that the rates of mass/energy accretion ${\frac{d M}{d t}}$ and linear momentum accretion ${\frac{d P^{\mu}}{d t}}$ decrease monotonically as the discretization scale of the underlying spacetime increases. This effect becomes progressively more pronounced as one increases the black hole spin value $J$, the dimensionless gas temperature at radial infinity ${\Theta_{\infty} = \frac{P_{\infty}}{\rho_{\infty}}}$, and the adiabatic exponent ${\Gamma}$. For instance, the rate of decrease for a black hole spinning close to extremality, with spin parameter ${J = 0.99 M}$, experiences a mass/energy accretion rate decrease that is approximately six times as rapid (as a function of discretization scale) with a stiff equation of state (i.e. ${\Gamma = 2}$), and approximately two times as rapid with an ultra-relativistic equation of state (i.e. ${\Gamma = \frac{4}{3}}$), as the rate of decrease for a non-rotating black hole with the equivalent equations of state. These effects disappear in the non-relativistic limit as ${\Theta_{\infty} \to 0}$, and become divergent in the ultra-relativistic limit as ${\Theta \to \infty}$. We do not find any correspondingly systematic relationship between the discretization scale and the angular drag force exerted on a spinning black hole, although we do find evidence of an advective-acoustic instability within the fluid, which then propagates to become an instability in the underlying disrcete spacetime structure, that appears for certain critical values of the discretization scale, and becomes more pronounced at higher Mach numbers. This may be a purely numerical artefact (since similar such instabilities were found by Beckmann, Slyz and Devriendt\cite{beckmann} in simulations of supermassive black holes using the \textsc{RAMSES} code, and were discovered to be dependent upon numerical resolution), or may be due to some more physical ``inverse energy cascade'' effect caused by a truncation of fluid interactions at short length-scales. As a consequence, we treat this result as necessarily more tentative than the mass/energy and momentum accretion rate results, and believe that it warrants further and more systematic investigation.

\clearpage

\section{Concluding Remarks}
\label{sec:Section5}

In this article, we have derived and numerically validated a new formulation of the equations of general relativistic hydrodynamics that is amenable to analysis within arbitrary discrete spacetime settings, and have implemented the resulting formalism into the \textsc{Gravitas} computational general relativity framework. We then proceeded to simulate radial (Bondi-type) accretion of a perfect relativistic fluid obeying an ideal gas equation of state onto both static and spinning black holes, as described by the Schwarzschild and Kerr metrics respectively, with a variety of black hole spin parameters and in a variety of different coordinate systems. Our simulations suggest that there exists a fairly robust (namely an inverse, monotonic) relationship between the mass/energy and momentum accretion rates onto the black hole and the discretization scale of the underlying spacetime. We have also found preliminary evidence of a possible advective-acoustic instability in the angular drag force exerted on the black hole by the fluid, that becomes significantly more pronounced at certain key values of the discretization scale, although further numerical experiments will be required before the extent to which this corresponds to a physically realistic effect can be confidently determined. These results provide tentative evidence that there may exist astrophysically observable effects of the underlying discreteness of spacetime arising within certain quantum gravity models that are reflected in the dynamics of the fluid region close to radially-accreting black holes, especially those whose spin values are approaching extremality, and onto whom the accretion flow is ultra-relativistic. However, in order to render the analysis both mathematically and computationally tractable, we have had to make several strong assumptions which limit the physical reasonableness and generality of our results, including the assumption of a form of the ideal gas equation of state that was shown by Taub\cite{taub} only to be physical in either the strictly non-relativistic limit (i.e. with ${\Theta_{\infty} \to 0}$ and ${\Gamma = \frac{5}{3}}$) or in the strictly ultra-relativistic limit (i.e. with ${\Theta_{\infty} \to \infty}$ and ${\Gamma = \frac{4}{3}}$), and crucially not in the relativistic case with dimensionless gas temperatures on the order of unity (i.e. ${\Theta_{\infty} \sim 1}$). Therefore, extension of the simulation results presented within this article to the case of fluid accretion involving more physically reasonable equations of state, and from the highly idealized case of purely radial/Bondi-type accretion to the more astrophysically relevant case of non-radial/Bondi-Hoyle-Lyttleton-type accretion, remains a particular priority.

Many other directions exist for future research, including the inclusion of the effects arising from certain quantum gravitational, quantum field-theoretic and/or quantum information-theoretic properties of black hole event horizons in discrete spacetimes\cite{gorard12}\cite{shah}\cite{gorard13} (especially within discrete black holes spinning close to extremality) into simulations of the resulting accretion dynamics, as well as the effects of certain features of the global spacetime topology that are characteristic of discrete/emergent spacetime theories\cite{arsiwalla}\cite{arsiwalla2}. It is also highly likely that extending these simulations into more complex astrophysical and cosmological settings involving strong relativistic field dynamics, such as the accretion of matter onto a merging binary black hole system\cite{vanmeter}\cite{farris}\cite{farris2}, would reveal yet more intricate physics that is peculiar to the discrete spacetime setting, although such an analysis would require significant advances in the numerical algorithms employed within the \textsc{Gravitas} framework, certainly well beyond the capabilities of the algorithms used within this article. On the more observational side, a more complete and systematic survey of the parameter space of black hole spin values, discretization scales and equation of state parameters would be necessary in order to determine which (if any) of the effects discussed within this article might realistically be detectable within the X-ray emission spectra of black hole accretion regions in the near-term, as would the incorporation of electromagnetic effects (which would, in turn, facilitate investigations of the impact of spacetime discreteness on phenomena such as the Blandford-Znajek mechanism\cite{blandford} and the formation of astrophysical jets within active galactic nuclei\cite{penna}, for example) into the spacetime description. Finally, it would be particularly exciting to extend the mathematical and numerical methods developed here to other general relativistic scenarios involving the two-way interaction between perfect fluid matter and a discrete underlying spacetime in strong gravity, such as the inspiral and collision of binary neutron star systems\cite{baiotti}\cite{bernuzzi}.

\end{document}